\documentclass[twocolumn]{aastex631}

\usepackage{float,graphicx,amsmath,multirow}
\usepackage[version=4]{mhchem}
\usepackage{amsmath}
\usepackage{booktabs}
\usepackage{enumitem}
\usepackage{wrapfig}
\usepackage{threeparttable}
\usepackage{comment}
\usepackage{upgreek}
\usepackage{longtable}
\usepackage{float}
\usepackage{hyperref}
\usepackage{cleveref}
\usepackage{makecell}
\usepackage{needspace}
\usepackage{placeins} 
\usepackage{afterpage}
\usepackage[normalem]{ulem}
\usepackage{microtype}

\graphicspath{{./}{figures/}}

\begin{document}

\title{The impact of hydrogen atom tunneling on aromatic chemistry in TMC-1}

\correspondingauthor{Thomas H. Speak, Ilsa R. Cooke}
\email{speakt@gmail.com, icooke@chem.ubc.ca}

\author[0009-0002-6372-9926]{Reace H. J. Willis}
\affiliation{Department of Chemistry, University of British Columbia, Vancouver, BC, V6T 1Z1, Canada.
}

\author[0000-0001-8134-5681]{Thomas H. Speak}
\affiliation{Department of Chemistry, University of British Columbia, Vancouver, BC, V6T 1Z1, Canada.
}

\author[0000-0002-4593-518X]{Alex N. Byrne}
\affiliation{Department of Chemistry, Massachusetts Institute of Technology, Cambridge, MA, USA.}

\author[0000-0002-5171-7568]{Christopher N. Shingledecker}
\affiliation{Department of Chemistry, Virginia Military Institute, Lexington, VA, USA}

\author[0000-0002-0850-7426]{Ilsa R. Cooke}
\affiliation{Department of Chemistry, University of British Columbia, Vancouver, BC, V6T 1Z1, Canada.
}
 
\begin{abstract}
Hydrogen atom tunneling likely plays a substantial role in the gas-phase chemistry of astrochemical environments. To determine the potential effect that it has on the chemical modeling of aromatic molecules, we screened the kida.uva.2024 network, and our own expanded network to find reactions which could be significantly accelerated by hydrogen atom tunneling in the ISM. In total, 64 reactions were identified. The hydrogen abstraction reactions from H$_{2}$ to four key interstellar radicals (C$_{2}$H, OH, CN, and NH$_{2}$) were studied further using newly calculated potential energy surfaces and RRKM analyses to determine rate coefficients for a temperature of 10 K and a density of 2 $\times$ 10$^{4}$ cm$^{-3}$. Despite having low rate coefficients of 1.66 $\times$ 10$^{-15}$, 8.17 $\times$ 10$^{-16}$ and 3.15 $\times$ 10$^{-16}$ $\mathrm{cm^{3}\,s^{-1}}$ the C$_{2}$H, OH, and CN reactions are competitive in the ISM, due to large overall rates caused by the high abundance of molecular hydrogen. The calculated value for the NH$_{2}$ reaction, however, was much smaller and found to be inefficient at ISM conditions. The possible effects of all other considered reactions were studied with simulations using calculated collision limit rate coefficients. Upper and lower bounds were then placed on modeled aromatic abundances using the most significant reactions. Due to the dependence of calculated aromatic abundances on reactions involving c-C$_{6}$H$_{5}^{+}$ and the recent questions surrounding its reactivity, we also explored the abundance variations caused by reactions leading to or involving c-C$_{6}$H$_{5}^{+}$.
\end{abstract}


\section{Introduction} \label{sec:intro}
To date, more than 340 molecules have been detected in the interstellar medium (ISM) or circumstellar shells \citep{Endres2016,Muller2025}. To explain the presence and predict the fate of these molecules in the ISM, reaction networks and astrochemical models are necessary. However, current models and chemical networks are often unable to reproduce the measured abundances of many molecules, frequently failing by several orders of magnitude \citep{Millar2023b, Ruaud2016,Wakelam2024a}. Furthermore, most astrochemical models containing only bottom-up formation routes are unable to reproduce the observed abundances of aromatic molecules. A noteworthy case of this is the commonly under-predicted abundances of 1- and 2-cyanonaphthalene (1- and 2-CNN) within models compared to their observed abundances in the Taurus Molecular Cloud 1 (TMC-1; \citealt{McGuire2021,Byrne2024}). One implication of this is that much more CNN and naphthalene (\ce{C10H8}) could be inherited by dark molecular clouds than expected, or alternatively, that critical chemical routes leading to their formation are currently missing or misrepresented kinetically in chemical networks. This inability of models to reproduce some measured molecular abundances, such as in the case of the CNNs, creates additional uncertainty in the predicted abundances of undetected species like \ce{C10H8}; which models indicate the CNNs could be a proxy for in molecular clouds due to the typically barrierless addition-elimination reaction of CN with unsaturated hydrocarbons \citep{Carty2001a, Cooke2020}.

Kinetic rate coefficients remain unmeasured for many astrophysically-relevant reactions, especially at low temperatures, and have been shown to be a significant source of uncertainty in astrochemical modeling \citep{Wakelam2006,Vasyunin2004}. Measuring kinetic parameters at low temperature is experimentally challenging and often time and resource-intensive \citep{Cooke2019a}. As such, low-temperature reaction rate coefficients are commonly found through estimation \citep{Woon2009a, Loison2013} or extrapolation from higher temperatures \citep{Hebrard2009, Cooke2019a}, leading to unknown uncertainties for many modeled molecular abundances. This problem is exacerbated by the fact that rate coefficients for some reactions increase at low temperatures \citep{Heard2018}, and therefore may be substantially different from the extrapolated values; especially those where long-range attractive forces are significant. These issues have been recognized for some time and analyzed by several sensitivity analyses \citep{Vasyunin2004,Wakelam2005, Wakelam2006,Wakelam2009,Wakelam2010,Heyl2023,Byrne2024} of various astrochemical reaction networks such as the UMIST Database for Astrochemistry (UDfA; \citealt{Millar2023b}), osu-09-2008 \citep{Garrod2008}, and Kinetic Database for Astrochemistry (KIDA; \citealt{Wakelam2024a}).

Typically, if a reaction is endothermic or contains an emerged energy barrier, then it will not occur in the low-temperature environment of the ISM. However, barrier-containing reactions involving the exchange of a single hydrogen atom have been shown to become faster at low temperatures due to quantum mechanical tunneling \citep{Heard2018}. Additionally, the effect of tunneling becomes even more pronounced at low temperatures when a weakly bound complex is present prior to the barrier \citep{Sims2013b}, with rate coefficients sometimes even similar to those of barrierless reactions \citep{Shannon2013}. In an experimental study, \citet{Shannon2013} showed this occurs for the reaction of OH with methanol, which reported rate coefficients almost two orders of magnitude larger at 63 K than at $\sim$200 K. Electronic structure calculations conducted prior to this by \citet{Xu2007} revealed the possibility of a $\sim$20 kJ mol$^{-1}$ hydrogen-bonded complex preceding the activation barrier. Subsequent studies where OH was reacted with dimethyl ether or acetone \citep{J.Shannon2014}, ethanol or propan-2-ol \citep{Caravan2015}, and methyl formate \citep{Jimenez2016} also displayed negative temperature dependencies with notable rate coefficient increases at low temperatures; all of which are suspected to be enhanced by hydrogen atom tunneling, aided by the presence of a weakly bound hydrogen-bonded complex. Consequently, extrapolated rate coefficients for unstudied low-temperature reactions involving hydrogen atom transfer could be many orders of magnitude too low in astrochemical reaction networks \citep{Herbst1994}. Although parameterizations of existing experimental data naturally include the effect of hydrogen atom tunneling, these parameterizations are generally unable to predict the low-temperature behavior far from the fitted region. The full impact of hydrogen atom tunneling on astrochemical models is currently unknown, but the effects could be significant---given the large number of reactions that involve the transfer of a hydrogen atom. 

In this work, we report a list of neutral-neutral, gas-phase reactions from the kida.uva.2024 network for interstellar chemistry \citep{Wakelam2024a} that display the potential to be accelerated by low-temperature hydrogen atom tunneling. The calculated collision limit rate coefficients for 55 reactions in this list were then used to estimate an upper bound to the possible impact of each in astrochemical models. The reactions which were found to have the greatest potential impact on the modeling of interstellar aromatic chemistry are highlighted. Four reactions identified in this work were chosen to be studied extensively through a computational process combining \textit{ab-initio} theory, and master equation calculations to predict low-temperature and low-pressure rate coefficients relevant to the ISM. These rate coefficients were then implemented into a refined astrochemical network and a model of a standard dark molecular cloud. In Section \ref{Ab+MESMER-Results} the most notable effects caused by the use of the newly predicted rate coefficients at estimated TMC-1 relevant conditions (10 K and 2 $\times$ 10$^{4}$ $\mathrm{cm^{-3}}$) are discussed.

Recent work by \citet{Kocheril2025a}, and \citet{Loison2025} has called into question the role that phenylium (\ce{c-C6H5+}) plays in the formation of interstellar benzene (\ce{C6H6}). As such, the formation of \ce{C6H6} in the ISM has become more uncertain, and with it, many other polycyclic aromatic hydrocarbon formation pathways, since all such species detected in the ISM (1- and 2-CNN \citep{McGuire2021}, indene \citep{Burkhardt2021, Cernicharo2021}, 2-cyanoindene \citep{Sita2022}, 1-, 3-, 4-, and 5-cyanoacenaphthylene \citep{Cernicharo2024, Cernicharo2026a}, phenalene \citep{Cabezas2025a}, 1H-cyclopent[cd]indene \citep{Fuentetaja2026a}, 1-, 2- and 4-cyanopyrene \citep{Wenzel2024a,Wenzel2025}, and cyanocoronene \citep{Wenzel2025a}) likely evolve from \ce{C6H6} to some extent. Given the reliance of aromatic abundances in our model on reactions involving \ce{c-C6H5+}, we have investigated the effects of a reduced rate coefficient for the reaction of \ce{H2} and \ce{c-C6H5+}, as well as the removal of reactions present in our network which were reported by \citet{Loison2025} to be inefficient at low temperatures.
\pagebreak
\section{Methods}
\subsection{Astrochemical Modeling}\label{Astrochemical-Model}
The \texttt{NAUTILUS} astrochemical modeling code \citep{Ruaud2016} was used for all simulations, along with gas-phase and grain-phase chemical networks originally based on kida.uva.2014 (\citealt{Wakelam2015} and \citealt{Ruaud2015}, respectively). The adapted gas-phase network, which will be referred to as Network 1, is described in \citet{Byrne2024} and contains an additional number of chemical routes surrounding large carbon-chains and aromatic species that have been observed in TMC-1. In addition to Network 1, Network 2 is the same chemical network, but has a rate coefficient of 9 $\times$ $10^{-14}$ $\mathrm{cm^{3}\,s^{-1}}$ (suggested by Kocheril (private communication)) for the reaction
\begin{align}
\ce{c-C6H5+ + H2 \rightarrow C6H7+ + h\nu}   
\tag{1} \label{eq:5}
\end{align}
instead of the kida.uva.2014 value of 6 $\times$ $10^{-11}$ $\mathrm{cm^{3}\,s^{-1}}$. Physical conditions resembling a typical dark cloud were assumed, as in \citet{Byrne2024}. Specifically, a gas density ($n_{H_{2}}$) of 2 $\times$ $10^{4}$ $\mathrm{cm^{-3}}$ \citep{Snell1982}, visual extinction ($A_{v}$) of 10 \citep{Rodriguez-Baras2021}, cosmic-ray ionization rate ($\zeta_{\mathrm{CR}}$) of 1.3 $\times$ $10^{-17}$ $\mathrm{s^{-1}}$ \citep{spitzer1968}, kinetic temperature of 10 K for gas and dust grains \citep{Pratap1997}, chemical desorption efficiency of 1\% \citep{Garrod2007}, peak grain temperature of 70 K lasting 1 $\times$ $10^{-5}$ s due to cosmic-ray heating \citep{Hasegawa1993}, and a C/O ratio of 1.1 were used \citep{Loomis2021, Hincelin2011}. The initial state of the model contains only atoms and molecular hydrogen, where initial elemental abundances also correspond to those in \citet{Byrne2024}. 

\subsubsection{Reaction Screening}
\label{Sec:Screening}
Network 1 and the kida.uva.2024 chemical network were screened for reactions that have the potential to be accelerated by hydrogen atom tunneling at low temperatures. The criteria used to identify these reactions was largely based on those described by \citet{Meisner2019} in their screening of the 1299 reactions from \citet{Kamp2017}. In addition to their criteria, we have also excluded endothermic reactions and only consider neutral-neutral reactions. To be included in Table \ref{tab:Reactions}, a reaction must meet the following seven criteria: (1) be a neutral-neutral reaction, (2) involve the transfer of a single hydrogen atom, (3) form multiple products, (4) be exothermic, (5) have a $\gamma$ value that is 0 K $<$ $\gamma$ $<$ $10^{4}$ K, (6) have a $\beta$ value $\geq$ 0, and (7) have an existing network, low-temperature rate coefficient originating from extrapolation or estimation. We do not consider ion-neutral reactions to limit the length of this study and because we expect relatively few to fit our search criteria. For instance, \citet{Meisner2019} found four of their 47 identified reactions to be ion-neutral, with the rest being neutral-neutral. They also explain that other types of reactions can have a higher difficulty in discerning tunneling effects from other non-classical effects, hence the scope of this study is limited to neutral-neutrals. Reactions involving the transfer of a single hydrogen atom are only considered because the likelihood of quantum tunneling greatly decreases as more massive species are used. The formation of multiple products is required because excess energy contained by one singular product can lead to its redissociation soon after formation in the ISM. Endothermic reactions are excluded from this study because these reactions will not be affected by hydrogen atom tunneling in the low-temperature ISM. Enthalpies of formation were used to determine if a reaction is exothermic and were taken from Active Thermochemical Tables (ATcT), version 1.220 \citep{Ruscic2005}. If no such value existed, these were calculated from the enthalpies of other molecules and \textit{ab-initio} enthalpies of reactions. These enthalpies of reaction were calculated using structures and zero point energies at the B2PLYP-D3(BJ)/aug-cc-pVTZ level with single point energy corrections using CCSD(T)-F12/cc-pVTZ-F12, asides from the molecules \ce{C10}, and \ce{C10H}; for these molecules CCSD(T)-F12/cc-pVDZ-F12 was used instead. The $\gamma$ and $\beta$ values, along with $\alpha$ are fitting parameters included in the modified Arrhenius equation:
\begin{align}
k(T) = \alpha \left( \frac{T}{300\,\mathrm{K}} \right)^{\beta} \exp\left( -\frac{\gamma}{T} \right),
\label{Modified_Arrhenius_Eq}
\end{align}
These values correspond to the fitting of experimental data or computational calculations, often over specific temperature ranges, and allow for temperature dependent rate coefficients through the solving of Equation \ref{Modified_Arrhenius_Eq}. Reactions with a $\gamma$ value between 0 and $10^{4}$ K were used because $\gamma$ $>$ 0 K, suggests the presence of a potential energy barrier, and $\gamma$ $>$ $10^{4}$ K is an indication for a high energy barrier containing-reaction, those of which are unlikely to be accelerated by low-temperature hydrogen atom tunneling (these choices are described in further detail by \citealt{Meisner2019}). Additionally, a $\beta$ value $\geq$ 0 is needed because reactions with negative $\beta$ values will accelerate at low temperatures, and therefore the effect of hydrogen atom tunneling will be decreased. Two reactions with both a negative $\beta$ value and a $\gamma$ value that is 0 K $<$ $\gamma$ $<$ $10^{4}$ K were found, however both are already efficient at low temperatures and were not included in Table \ref{tab:Reactions}. Lastly, we only include reactions with rate coefficients based on estimation or extrapolated data because these are the most vulnerable to error.
\newpage
\subsubsection{Simulations}\label{Simulations}
\begin{figure}
    \centering
    \includegraphics[width=1\linewidth]{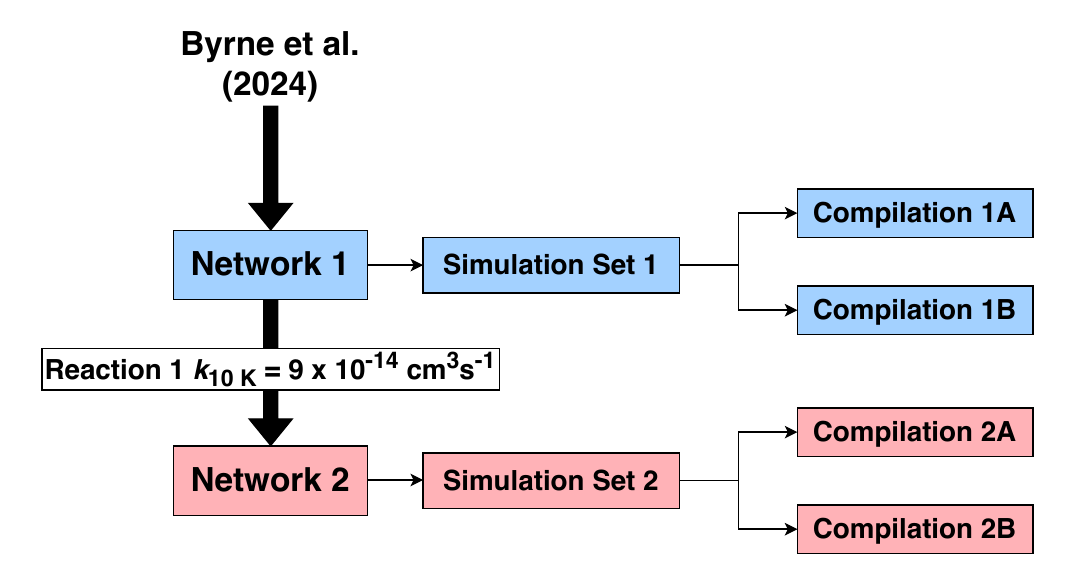}
    \caption{Shown is the workflow of the astrochemical modeling conducted in this study. Network 1 was taken from \citet{Byrne2024}, and Network 2 is Network 1, but with one modified reaction rate coefficient (Reaction 1). Both simulation sets were performed using \textit{k}\textsubscript{Coll, 10 K} values in Table \ref{tab:Reactions} as well as MESMER values. Reactions which increased the modeled abundances of species discussed in Section \ref{Simulations} were compiled together for one simulation (Compilation 1A and 2A), as were the ones which caused decreases (Compilation 1B and 2B).}
    \label{Network_FlowChart}
\end{figure}

Using Network 1, 59 simulations (Set 1) were performed to estimate the largest potential impact the reactions in Table \ref{tab:Reactions} could have on the modeled abundances of aromatic molecules in TMC-1. For each simulation, the calculated 10 K collision limit rate coefficient (\textit{k}\textsubscript{Coll, 10 K}: these calculations are described in Section \ref{Collision-Rate-Predictions}) of one reaction in Table \ref{tab:Reactions} was used in place of the $\alpha$, $\beta$, and $\gamma$ values from kida.uva.2024 asides from the reactions:
\begin{align}
\ce{C2H + H2 \rightarrow C2H2 + H}   
\tag{2} \label{eq:1} \\
\ce{OH + H2 \rightarrow H2O + H} 
\tag{3} \label{eq:2}\\
\ce{CN + H2 \rightarrow HCN + H}  
\tag{4} \label{eq:3} \\
\ce{NH2 + H2 \rightarrow NH3 + H} 
\tag{5} \label{eq:4}
\end{align}
where the rate coefficients for these were calculated using a more rigorous theoretical approach (described in Section \ref{Ab+ME}), providing improved estimations. These four reactions were chosen to be studied in detail due to their involvement of the very abundant \ce{H2} molecule and other important interstellar radicals. It was not feasible to use this treatment for each reaction in Table \ref{tab:Reactions}, however, we hope others will apply similar theory to that described in Section \ref{Ab+ME} to other reactions in Table \ref{tab:Reactions}. For the 59 simulations we conducted, each reaction found to increase or decrease the abundance of benzene (\ce{C6H6}), benzonitrile (\ce{C6H5CN}), phenylium (c-C$_{6}$H$_{5}$$^{+}$), benzenium (C$_{6}$H$_{7}$$^{+}$), phenyl (\ce{C6H5}), naphthalene (\ce{C10H8}), 1-cyanonaphthalene (\ce{1-C10H7CN}), 2-cyanonaphthalene (\ce{2-C10H7CN}), hydronaphthyl (C$_{10}$H$_{9}$), or dihydronaphthalene (C$_{10}$H$_{10}$) by a factor of 1.1 or greater—at any time point—were then compiled in two separate simulations. One of these simulations contained all of the \textit{k}\textsubscript{Coll, 10 K} values which increased aromatic abundances (Compilation 1A), and the other contained those which decreased them (Compilation 1B). In the few cases where a reaction provided similar increases and decreases, the net effect on the original abundance time profile—determined by integrating the original and the altered time profile—decided whether it was placed in Compilation 1A or 1B. Simulations using \textit{k}\textsubscript{Coll, 10 K} values for reaction numbers 17 and 25 from Table \ref{tab:Reactions} were not conducted due to the absence of HOCN, and HONC in the expanded network. HOCN and HONC chemical routes were not added to the network to perform these simulations in order to maintain the ability for comparisons to the aromatic sensitivity analysis conducted by \citet{Byrne2024} (Section \ref{subsec:SensitivityAnalysis}). 
In addition to Set 1, a second set of simulations (Set 2) was performed using Network 2, with the same methodology to compose two compilations of reactions with \textit{k}\textsubscript{Coll, 10 K} values which increase (Compilation 2A) and decrease (Compilation 2B) aromatic abundances. Figure \ref{Network_FlowChart} displays the relations of the various networks, sets of simulations, and compiled impactful reaction rate coefficients used in the astrochemical modeling of this work. This additional set of simulations was performed due to the substantial role that Reaction \ref{eq:5} plays in the formation of aromatics in the model. Experimental evidence from \citet{Kocheril2025a} suggests the rate coefficient of Reaction \ref{eq:5} is much lower than previously estimated, with an updated upper limit of 9 $\times$ $10^{-14}$ $\mathrm{cm^{3}\,s^{-1}}$ (Kocheril, private communication). The effects of a lower rate coefficient for Reaction \ref{eq:5} in our model are discussed in Section \ref{AromaticImpact}.

\subsubsection{Collision Rate Predictions}
\label{Collision-Rate-Predictions}
Within \texttt{NAUTILUS} simulations, \textit{k}\textsubscript{Coll, 10 K} values were used as an upper bound for the low-temperature and -pressure rate coefficients of the tunnelable reactions in Table \ref{tab:Reactions} (asides for Reactions \ref{eq:1}--\ref{eq:4} where  Master Equation Solver for Multi-Energy Well Reactions (MESMER) rate coefficients were used instead: see Section \ref{MESMER}). However, temperature-dependent \textit{k}\textsubscript{Coll} values were also required as input for MESMER calculations on Reactions \ref{eq:1}--\ref{eq:4}. All of these values were estimated using classical capture theory (CCT; \citealt{georgievskii2005,west2019a}):
\begin{displaymath} k_{\text {col }}(T)=\left[\pi\left(\frac{2 C_6}{k_B T}\right)^{1/3} 1.353\right]\left[\left(\frac{8 k_B T}{\pi \mu}\right)^{1/2}\right],\end{displaymath} 

In this equation, $k_B$ is the Boltzmann constant, $\mu$ is the reduced mass of the collision partners and $C_6$ is the sum of the attractive forces of the partners. For these CCT predictions, $C_6$ was calculated using the following equation:
\begin{displaymath}C_6=  \frac{2}{3}\left(\frac{\mu_1^2 \mu_2^2}{k_B T\left(4 \pi \epsilon_0\right)^2}\right)+\frac{\mu_1^2 \alpha_2+\mu_2^2 \alpha_1}{4 \pi \epsilon_0}  +\frac{3}{2} \alpha_1 \alpha_2\left(\frac{I_1 I_2}{I_1+I_2}\right) 
\end{displaymath}  

Where $\epsilon_0$ represents the permittivity of free space, $\mu_1$ and $\mu_2$ are the dipole moments of the reactants, $\alpha_1$/$\alpha_2$ are their polarizabilities and $I_1$/$I_2$ their ionization energies. For the barrierless formation of loose complexes included in the MESMER calculations, these values were available from the Computational Chemistry Comparison and Benchmark DataBase (CCCBDB; \citealt{CCCBDB-2022}). In the case of  the collision limit calculations for use in simulations, where these were not available in CCCBDB they were calculated in the manner described in Appendix Section \ref{CollisionLimit-RCs_SI}. 

The CCT approach has previously been shown to be accurate within a factor of 2 for the prediction of rate coefficients for barrierless neutral-neutral reactions \citep{west2019a}; which compares well with more rigorous (and resource intensive) methods such as full $E,J$-resolved microcanonical Variational Transition State Theory ($\mu$j-VTST) results \citep{Klippenstein2005} along with rotationally adiabatic capture theory (ACT) \citep{Stoecklin1991,Clary1993,Clary1994} and the Statistical Adiabatic Capture Model (SACM) \citep{Quack1974,Troe1985}. Previous studies have shown that $\mu$j-VTST can well reproduce experimentally determined reaction rate coefficients from \citet{west2019a} for the reaction of CH radicals with formaldehyde (\ce{CH2O}) using a pulsed Laval nozzle apparatus, typically within a factor of 2 to 5 \citep{Klippenstein2005}. Additionally, $\mu$j-VTST has been shown to be in good agreement with other approaches such as ACT \citep{Stoecklin1991,Clary1993,Clary1994} and SACM \citep{Quack1974,Troe1985}. ACT has been shown to reproduce experimentally observed collision limits (within a factor of 2) for the reactions of CN and CH with ammonia \citep{Stoecklin1995}. Although calculation of the potential energy surfaces required for these rigorous approaches is possible for the systems treated with MESMER in our work, CCT was chosen due to four main considerations. The first is that these reactions were far from the collision limit and the predicted product formation rate coefficient was relatively insensitive to this parameter at low temperatures. The second was that the experimental data available allowed for the floating of the parameters in an Arrhenius description of the collision limit for the formation of the loose entrance channel complex during data fitting. Thirdly, the intent of this analysis was to validate an approach that can be extended to the treatment of larger systems ($>10$ atoms) where resource intensive approaches such as $\mu$j-VTST would be beyond our capabilities. Finally, the use of CCT in the prediction of collision rates made practicable the screening of a large number of reactions for the potential impact of tunnelable reactions on the modeling of TMC-1. It should be noted, that the use of CCT for this last reason was to provide estimated upper limits to the actual reaction rate coefficients, where the true value is likely multiple orders of magnitude lower for many of these reactions in Table \ref{tab:Reactions}. However, this less-demanding approach allowed us to evaluate an estimation of the maximum effect that each reaction in Table \ref{tab:Reactions} could have on modeled aromatic molecule abundances.

\subsection{$\it{Ab-initio}$ and Master Equation Calculations}\label{Ab+ME}
\begin{figure}
    \centering
    \includegraphics[width=1\linewidth]{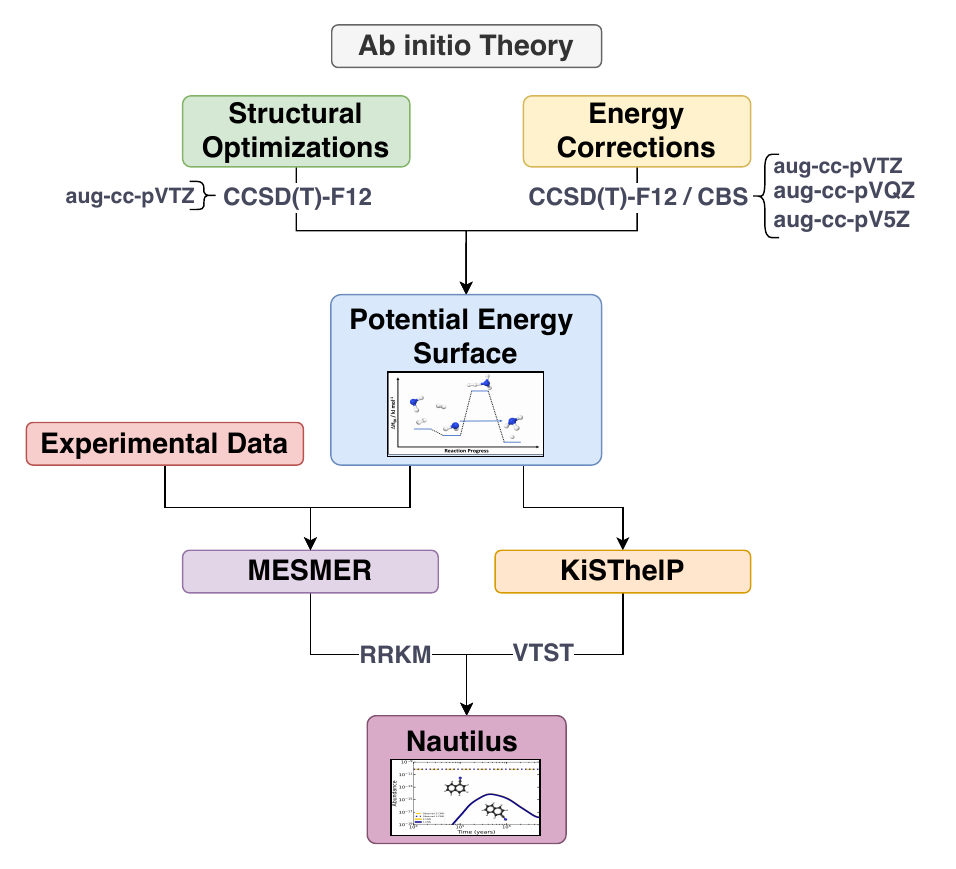}
    \caption{The general methodology for calculating rate coefficients and product branching ratios of low-temperature and -pressure reactions for implementation in astrochemical models in this work.}
    \label{Computational_Flow-Chart}
\end{figure}

Low-temperature and -pressure reaction rate coefficients for Reactions \ref{eq:1}--\ref{eq:4} were calculated through more resource-intensive calculations ($\it{ab-initio}$ and master equation calculations) following the general methodology outlined within Figure \ref{Computational_Flow-Chart}. A potential energy surface, rotational constants, and vibrational frequencies are calculated using $\it{ab-initio}$ theory and combined with experimental data as input for Master Equation Solver for Multi-Energy Well Reactions (MESMER) calculations. The calculated, experimentally fitted MESMER rate coefficients were then compared with those determined with Kinetic and Statistical Thermodynamical Package (KiSThelP) using transition state theory (TST), and the effects of each on a \texttt{NAUTILUS}-based astrochemical model were analyzed. Sections \ref{PESs}, \ref{kisthelp}, \ref{MESMER} and \ref{Astrochemical-Model} discuss the $\it{ab-initio}$, KiSThelP, and MESMER calculations, as well as the astrochemical modeling performed in this work in further detail.

\subsubsection{Potential Energy Surface Evaluations}\label{PESs}
Potential energy surfaces for the abstraction of a hydrogen atom from molecular hydrogen by C$_{2}$H, OH, CN, and NH$_{2}$ were primarily calculated with the explicitly-correlated coupled-cluster singles and doubles with perturbative triples method (CCSD(T)-F12; \citealt{Adler2007,Knizia2009,Rauhut2009}). For the reactions of OH and CN, CCSD(T)-F12b was combined with the augmented triple zeta Dunning type correlation consistent basis set aug-cc-pVTZ \citep{Dunning1989,Kendall1992} in Molpro 2010 \citep{Werner2012, MOLPROtest}. The reactions of C$_2$H and NH$_2$ used CCSD(T)-F12c with the correlation consistent basis set optimized for F12 calculations cc-pVTZ-F12 \citep{Peterson2008,ccREPO-2016} in ORCA 5.0.4 \citep{ORCA,ORCA5,ORCA2020}. The initial structures for the reactants and products were taken from the CCCBDB. These were then optimized and harmonic vibrational frequency analyses were performed. Structural visualization was performed in Avogadro 4.2.1 optimized for ORCA \citep{Avogadro2012}. 

Relaxed scans were performed where all but one coordinate were optimized at fixed steps of the scanned coordinate. The first set of scans took the separated reactants and scanned the X—H bond length (X = C$_2$H, OH, CN, NC, and NH$_2$), the second set of scans took the separated reactants and scanned the H—H bond length. For both sets of scans, the geometry corresponding to the energy maxima was used as the input structure for a transition-state optimization and the geometries corresponding to the energy minima on the surface were used as initial structures for loosely bound complexes on the entrance and exit channels. The energies of the optimized structures were further refined using CCSD(T) and CCSD(T)-F12 extrapolated to the complete basis set limit (CBS) using a series of correlation consistent basis sets (cc-pVTZ-F12 to cc-pV5Z-F12; \citealt{Peterson2008,ccREPO-2016}, and aug-cc-pVTZ to aug-cc-pV5Z; \citealt{Dunning1989,Kendall1992}) and a mixed Gaussian exponential scheme \citep{Peterson1994}. Intrinsic reaction coordinate scans (IRC) were performed from the optimized transition states to ensure that these geometries linked the reactants and products. The resultant IRC paths were also further refined for use in subsequent variational transition state theory (vTST) calculations; here, extrapolated single-point energy corrections were calculated for each point in the path along with projected frequencies. These surfaces were also calculated with CCSD(T)/CBS(aug-cc-pVTZ to aug-cc-pV5Z) corrected single-point energies on density functional theory structures (B2PLYP-D3(BJ)/aug-cc-pVTZ and M06-2X-D3/aug-cc-pVTZ). Here, both harmonic frequencies and anharmonic analysis at the VPT2 level \citep{Bloino2012} were calculated. Additionally, the surfaces were calculated at the CCSD(T)/CBS(aug-cc-pVTZ to aug-cc-pV5Z)//CCSD(T)/aug-cc-pVTZ level as well.

CCSD(T) is often described as the ``gold standard'' of computational chemistry, and the inclusion of explicit F12 treatment of the r12 terms eliminates some basis-set incompleteness error in the electron–electron cusp, and allows near CBS accuracy with triple zeta basis sets. \citet{Zhang2012} found CCSD(T)-F12 with the cc-pVTZ-F12 and aug-cc-pVTZ basis sets to generally be accurate to 1-2 kJ mol$^{-1}$ for barrier heights, with maximum errors typically below 5 kJ mol$^{-1}$. Our inclusion of single point energy corrections extrapolated to the CBS limit should improve this accuracy even further. The inclusion of core valence, scalar relativistic effects, higher order correlation terms, diagonal Born–Oppenheimer corrections and anharmonic vibrational analysis could improve the accuracies of our potential energy surface calculations even further. However, the calculations conducted on Reactions \ref{eq:1}--\ref{eq:4} were to be used in combination with MESMER for direct fitting of the potential energy surfaces to existing experimental data, so these higher-level corrections were not considered. The resultant parameters in the fitted potential energy surfaces were used along with their correlated uncertainties to determine the upper and lower bounds for the MESMER calculated low-temperature rate coefficients shown in Appendix Section \ref{RCC_SI}. Additionally, the deviation of the calculated 0 K enthalpies of reaction to those found in ATcT for Reactions \ref{eq:1}--\ref{eq:4} were all below 3 kJ mol$^{-1}$ (2.5, 0.1, 0.5, 0.9 and 0.7 kJ mol$^{-1}$ for the formation of \ce{C2H2} + H, \ce{H2O} + H, \ce{HCN} + H, \ce{HNC} + H and \ce{NH3} + H respectively). Torsional vibrations were replaced with hindered (or free) rotors based on relaxed scans of dihedrals mapping the atomic motions involved. Low frequency vibrational modes corresponding to translation of \ce{H2} were removed from the zero point energy of the Van der Waals complexes and is discussed in further detail in Appendix Section \ref{QCC_SI}.

\subsubsection{KiSThelP TST}\label{kisthelp}
TST and vTST rate coefficients for Reactions \ref{eq:1}--\ref{eq:4} were calculated in KiSThelP \citep{Canneaux2014}. Canonical TST calculates rates of reaction based on the equilibrium flux through a dividing surface as a function of temperature. The vTST calculations calculate the Gibbs free energy for each point along the reaction coordinate at each temperature simulated. This allows for the position of the transition state to shift from the maximum of the internal energy profile, as the entropy contribution at longer ranges can cause a single transition state model to underestimate the energy barrier. Quantum mechanical tunneling was accounted for using a multiplicative transmission coefficient, calculated for a one-dimensional asymmetric Eckart potential \citep{Miller1979} based on the imaginary frequency of the transition state and the relative energies of the reactants and products. 

\subsubsection{MESMER Modeling}\label{MESMER}
MESMER is an open-source, energy grain master equation calculator capable of calculating the time-dependent change in species population for a multi-energy well reaction. From this, MESMER can derive temperature and pressure dependent reaction rate coefficients describing the flux between energy wells. The calculated \textit{k}\textsubscript{Coll} values for Reactions \ref{eq:1}--\ref{eq:4} were incorporated into MESMER calculations as the high-pressure limit for the rate of formation of the Van der Waals complexes, while the microcanonical rate coefficients for the back dissociation from these were calculated by the inverse Laplace transformation (ILT) methodology \citep{davies1986}. Pressure dependent stabilization within energy wells was treated by an exponential down model where the probability of movement between two energy grains is linked to the energy gap, the properties of the bath gas, and the bath gas density. The subsequent reaction through well-defined transition states were treated using the Rice–Ramsperger–Kassel–Marcus (RRKM) method \citep{Robertson_1996,baer1996,lourderaj2009}. RRKM calculates energy-level specific (microcanonical) rate coefficients, $\it{k}$(E). In the absence of entrance channel complexes for bimolecular abstraction reactions, the Boltzmann averaged microcanonical rate coefficients should converge to the canonical TST results. Quantum mechanical tunneling rates for each energy grain were accounted for through a one dimensional Eckart barrier \citep{Miller1979} based on the imaginary frequency of the transition state. This imaginary frequency was floated within a small range during the data fitting procedure in MESMER. Where sufficient experimental data is available—over a range of temperatures such that barrier heights and tunneling effects are adequately uncorrelated—fitting the imaginary frequency can account for failings in the Eckart description of the tunneling potential. Although the MESMER rate coefficients calculated in this work for Reactions \ref{eq:1}--\ref{eq:4} are likely more accurate than the calculated \textit{k}\textsubscript{Coll, 10 K} values, it should be noted that they rely on this Eckart tunneling description which can overestimate low-temperature reaction rate coefficients by several orders of magnitude. The errors derived for the MESMER rate coefficients for Reactions \ref{eq:1}--\ref{eq:4} were found by varying the floated parameters (imaginary frequencies, barrier heights, well depths, as well as the pre-exponential factor (A, or $\alpha$ in Equation \ref{Modified_Arrhenius_Eq}) and temperature dependence of the pre-exponential factor (n, or $\beta$ in Equation \ref{Modified_Arrhenius_Eq}) within the modified Arrhenius equation) within their correlated uncertainties from the fits; the best fit parameters are provided in Appendix Section \ref{RCC_SI}. This varying of the imaginary frequency within the calculation of errors for these reaction rate coefficients propagates some of the uncertainty from the Eckart description of tunneling, and therefore some of this potential error caused by the Eckart description is reflected in our MESMER calculated rate coefficient uncertainties. Other more accurate descriptions of tunneling are available, however the Eckart description utilized here is a pragmatic choice when combined with fitting of experimental data. Additionally, it is also straightforward to apply to reactions involving larger chemical species in future work.

\Needspace{5\baselineskip} 
\section{Results \& Discussion} 
\subsection{Tunneling Reactions in KIDA\label{subsec:"Tunnellable Reactions"}}
In total, 61 reactions were found to have the potential to be significantly affected by hydrogen atom tunneling at low temperatures (Table \ref{tab:Reactions}). Of these, 59 were from the kida.uva.2024 chemical network, and two (15 and 16 in Table \ref{tab:Reactions}) from Network 1. These two are not included in kida.uva.2024, but are in kida.uva.2014, hence their presence in Network 1. Among the 61 reactions, 24 were first identified by \citet{Meisner2019} through the screening of 1299 reactions from \citet{Kamp2017}, and 37 were originally identified by this work. Three other endothermic, hydrogen atom transfer reactions were also found in kida.uva.2024, but their reverse reactions were absent from the network. The exothermic, reverse reactions of these three are displayed in Table \ref{tab:Reactions2}, and should possibly be added to the network.

When using \textit{k}\textsubscript{Coll, 10 K} values in both sets of simulations, reactions which involve the saturation of hydrocarbons via reaction with H$_{2}$ produced the greatest increases to aromatic abundances. Out of all the reactions in Table \ref{tab:Reactions}, reaction number 36 caused the most significant increase to any one aromatic molecule at any time point, where $\sim$3200$\times$ more 1-CNN was produced at a chemical age of $\sim$5 $\times$ 10$^{4}$ years using Network 2. However, saturating carbon chains of C$_{\text{x}}$ or C$_{\text{x}}$H, where x $\geq$ 6, proved to be less effective at increasing the abundances of these molecules. Among these x $\geq$ 6 reactions, reaction number 37 in Table \ref{tab:Reactions} provided the greatest abundance change to any aromatic molecule, producing slightly less than 3$\times$ more \ce{C6H7+} at $\sim$5 $\times$ 10$^{4}$ years, using Network 2. Network 1 also contained H$_{2}$ reactions with C$_{\text{y}}$ and C$_{\text{y}}$H, where y = 5, 7, and 9 which proved to provide increases—especially C$_{5}$ and C$_{5}$H—however, all of these were found to be endothermic. Additionally, the reaction of C$_{10}$H and H$_{2}$, which is contained within kida.uva.2024 and Network 1, was found to be endothermic as well. The reverse of these reactions are present in kida.uva.2024; however, these are not listed in Table \ref{tab:Reactions} as they have $\gamma$ values $>10$$^{4}$ K. 

Among the reactions in Table \ref{tab:Reactions}, those that aid the formation of either methane (10 and 13 in Table \ref{tab:Reactions}), acetylene (35 in Table \ref{tab:Reactions}), or ethylene (11 in Table \ref{tab:Reactions}) provided particularly notable aromatic increases when their \textit{k}\textsubscript{Coll, 10 K} values were used. The notable effects of these reactions is likely due to the direct involvement of methane, acetylene, and ethylene within major aromatic formation routes in the model \citep{Byrne2024}. The largest abundance increase to any singular aromatic molecule produced by these four reactions range from $\sim$3--1100$\times$ more, which were caused by reaction numbers 13 and 10 respectively, using Network 1. Although number 13 did not produce large increases to aromatic molecules at any particular point, it generated consistent $\sim$2--3$\times$ increases from the beginning of the simulation until $\sim$2 $\times$ 10$^{6}$ years. The saturation of C$_{4}$ and C$_{4}$H via reaction with H$_{2}$ (36 and 52 in Table \ref{tab:Reactions}) were significant as well, which is likely caused by the involvement of C$_{4}$H$_{2}$$^{+}$ and C$_{4}$H$_{3}$$^{+}$ in main aromatic formation routes. In this case however, the additional step of converting C$_{4}$H$_{2}$ to C$_{4}$H$_{2}$$^{+}$ or C$_{4}$H$_{3}$$^{+}$ is also necessary. Despite this additional requirement, these two reactions also provide similar increases to modeled aromatics. 

Interestingly, we also found propargyl (\ce{CH2CCH}) involving reactions (numbers 21 and 22 in Table \ref{tab:Reactions}) to provide $\sim$2$\times$ increases to \ce{C6H5}, \ce{C10H8}, 1-CNN, and 2-CNN when an even smaller rate coefficient of 1 $\times$ 10$^{-15}$ cm$^{3}$ s$^{-1}$ was used for Reaction \ref{eq:5}. However, when a rate coefficient of 6 $\times$ 10$^{-11}$ or 9 $\times$ 10$^{-14}$ cm$^{3}$ s$^{-1}$ were used, then reaction numbers 21 and 22 in Table \ref{tab:Reactions} provide negligible to very minor decreases to most aromatics. The reasoning for this is probably attributable to the decline of the pathway to aromatics via Reaction \ref{eq:5} when a smaller rate coefficient is used, causing the pathway to \ce{C6H5} through the reaction of two \ce{CH2CCH} radicals to become more important. Since the major route to \ce{C10H8} in the network is through the hydrogen abstraction, vinyl acetylene (\ce{C4H4}) addition (HAVA) mechanism—which requires \ce{C6H5} and \ce{C4H4}, and results in \ce{C10H8}—increases in \ce{C10H8} along with 1-CNN and 2-CNN were found as well. 

The only reaction which resulted in notable decreases to all aromatic abundances when \textit{k}\textsubscript{Coll, 10 K} values were used, was 57 in Table \ref{tab:Reactions} which were $\sim$2--3$\times$ decreases over the time period of $\sim$2 $\times$ 10$^{5}$--2 $\times$ 10$^{6}$ years. Its effect is likely due to the fact that it results in the destruction of both methane and the cyano radical, both of which are precursors to aromatics and their cyano derivatives. In addition to 57, changes in reaction 7 in Table \ref{tab:Reactions} resulted in significant decreases to \ce{C10H8}, 1-CNN and 2-CNN, with the largest effect being $\sim$80$\times$ less 1-CNN at $\sim$10$^{4}$ years. Since this reaction causes a decrease in \ce{C4H4}, the HAVA mechanism becomes less effective, and results in a decreased abundance of these three molecules. Although significant to \ce{C10H8} and its cyano derivatives, the impact of this reaction does not extend to any of the other aromatic molecules monitored in this work.

\startlongtable
\begin{deluxetable*}{cccccc}
\tablenum{1}
\tablecaption{Gas-phase reactions from kida.uva.2024, and Network 1 (our expanded network based on kida.uva.2014) with attributes indicating the ability to be significantly impacted by hydrogen atom tunneling at low temperatures.\label{tab:Reactions}}
\tablehead{\colhead{No.} & 
\colhead{Reaction} & \colhead{$\Delta$H$_{\text{Rxn}}$(0 K) (kJ mol$^{-1}$)}  & \colhead{\textit{k}\textsubscript{KIDA, 10 K} (cm$^{3}$ s$^{-1}$)} & \colhead{\textit{k}\textsubscript{Coll, 10 K} (cm$^{3}$ s$^{-1}$)}
}
\startdata
2* & \textbf{H$_{2}$ + CCH $\rightarrow$ H + C$_{2}$H$_{2}$} & $-118.83 \pm 0.11$ & 3.97 $\times$ 10$^{-68}$ & 5.79 $\times$ 10$^{-10}$ \\
3 & \textbf{H$_{2}$ + OH $\rightarrow$ H + H$_{2}$O} & $-60.1466 \pm 0.0011$ & 5.72 $\times$ 10$^{-58}$ & 4.93 $\times$ 10$^{-10}$ \\
4 & \textbf{CN + H$_{2}$ $\rightarrow$ H + HCN} & $-91.02 \pm 0.12$ & 5.69 $\times$ 10$^{-53}$ & 6.02 $\times$ 10$^{-10}$ \\
5 & \textbf{H$_{2}$ + NH$_{2}$ $\rightarrow$ H + NH$_{3}$} & $-11.46 \pm 0.11$ & 1.48 $\times$ 10$^{-173}$ & 4.78 $\times$ 10$^{-10}$ \\
\tableline
6* & CH + O$_{2}$H $\rightarrow$ O$_{2}$ + CH$_{2}$ & $-216.97 \pm 0.17$ & 6.88 $\times$ 10$^{-324}$ & 1.20 $\times$ 10$^{-09}$ \\
7* & H + C$_{4}$H$_{4}$ $\rightarrow$ H$_{2}$ + C$_{4}$H$_{3}$ & $-14.96 \pm 0.59 $ & 4.26 $\times$ 10$^{-309}$ & 9.08 $\times$ 10$^{-10}$ \\
8 & CH$_{2}$ + CH$_{4}$ $\rightarrow$ CH$_{3}$ + CH$_{3}$ & $-24.77 \pm 0.10$ & 3.42 $\times$ 10$^{-231}$ & 3.40 $\times$ 10$^{-10}$ \\
9 & CH$_{2}$ + CH$_{2}$ $\rightarrow$ CH + CH$_{3}$ & $-39.29 \pm 0.17$ & 2.85 $\times$ 10$^{-227}$ & 4.14 $\times$ 10$^{-10}$ \\
10 & H$_{2}$ + CH$_{2}$ $\rightarrow$ H + CH$_{3}$ & $-25.155 \pm 0.094$ & 1.58 $\times$ 10$^{-222}$ & 4.93 $\times$ 10$^{-10}$ \\
11* & H$_{2}$ + C$_{2}$H$_{3}$ $\rightarrow$ H + C$_{2}$H$_{4}$ & $-24.31 \pm 0.26$ & 1.02 $\times$ 10$^{-203}$ & 5.55 $\times$ 10$^{-10}$ \\
12* & H + C$_{2}$H$_{6}$ $\rightarrow$ H$_{2}$ + C$_{2}$H$_{5}$ & $-16.13 \pm 0.18$ & 1.26 $\times$ 10$^{-199}$ & 7.80 $\times$ 10$^{-10}$ \\
13 & H$_{2}$ + CH$_{3}$ $\rightarrow$ H + CH$_{4}$ & $-0.389 \pm 0.026$ & 3.64 $\times$ 10$^{-196}$ & 4.89 $\times$ 10$^{-10}$ \\
14 & C + NH $\rightarrow$ N + CH & $-6.57 \pm 0.17$ & 6.05 $\times$ 10$^{-186}$ & 2.97 $\times$ 10$^{-10}$ \\
15*{$\dag$} & S + CH $\rightarrow$ C + HS & $-15.61 \pm 0.15$ & 6.05 $\times$ 10$^{-186}$ & 3.18 $\times$ 10$^{-10}$ \\
16*{$\dag$} & S + NH $\rightarrow$ N + HS & $-22.18 \pm 0.21$ & 6.05 $\times$ 10$^{-186}$ & 2.80 $\times$ 10$^{-10}$ \\
17* & O + HONC $\rightarrow$ OH + CNO & $-55.5 \pm 1.1$ & 3.02 $\times$ 10$^{-166}$ & 2.46 $\times$ 10$^{-10}$ \\
18* & CH$_{2}$ + H$_{2}$CO $\rightarrow$ HCO + CH$_{3}$ & $-94.420 \pm 0.094$ & 3.19 $\times$ 10$^{-155}$ & 7.21 $\times$ 10$^{-10}$ \\
19 & OH + CH$_{2}$ $\rightarrow$ CH + H$_{2}$O & $-74.28 \pm 0.10$ & 1.35 $\times$ 10$^{-142}$ & 6.40 $\times$ 10$^{-10}$ \\
20 & OH + CH$_{2}$ $\rightarrow$ O + CH$_{3}$ & $-31.622 \pm 0.095$ & 1.35 $\times$ 10$^{-142}$ & 6.40 $\times$ 10$^{-10}$ \\
21* & H + CH$_{3}$CCH $\rightarrow$ H$_{2}$ + CH$_{2}$CCH & $-54.66 \pm 0.30$ & 1.81 $\times$ 10$^{-124}$ & 8.37 $\times$ 10$^{-10}$ \\
22* & O + CH$_{3}$CCH $\rightarrow$ OH + CH$_{2}$CCH & $-48.20 \pm 0.30$ & 2.38 $\times$ 10$^{-123}$ & 2.61 $\times$ 10$^{-10}$ \\
23* & O + C$_{2}$H$_{6}$ $\rightarrow$ OH + C$_{2}$H$_{5}$ & $-9.66 \pm 0.18$ & 1.64 $\times$ 10$^{-122}$ & 2.54 $\times$ 10$^{-10}$ \\
24 & CN + CH$_{2}$ $\rightarrow$ CH + HCN & $-105.15 \pm 0.15$ & 1.41 $\times$ 10$^{-120}$ & 5.74 $\times$ 10$^{-10}$ \\
25* & O + HOCN $\rightarrow$ OH + OCN & $-70.28 \pm 0.46$ & 1.79 $\times$ 10$^{-118}$ & 2.44 $\times$ 10$^{-10}$ \\
26 & H + NH $\rightarrow$ N + H$_{2}$ & $-104.19 \pm 0.16$ & 1.86 $\times$ 10$^{-116}$ & 5.88 $\times$ 10$^{-10}$ \\
27 & OH + CH$_{3}$ $\rightarrow$ O + CH$_{4}$ & $-6.855 \pm 0.034$ & 9.61 $\times$ 10$^{-115}$ & 2.71 $\times$ 10$^{-10}$ \\
28 & O + CH $\rightarrow$ C + OH & $-91.153 \pm 0.074$ & 1.09 $\times$ 10$^{-114}$ & 2.45 $\times$ 10$^{-10}$ \\
29 & H + NH$_{2}$ $\rightarrow$ H$_{2}$ + NH & $-46.20 \pm 0.17$ & 1.02 $\times$ 10$^{-108}$ & 6.12 $\times$ 10$^{-10}$ \\
30 & H + OH $\rightarrow$ O + H$_{2}$ & $-6.467 \pm 0.022$ & 1.03 $\times$ 10$^{-102}$ & 5.40 $\times$ 10$^{-10}$ \\
31* & H + HNC $\rightarrow$ H + HCN & $-62.38 \pm 0.30$ & 1.38 $\times$ 10$^{-97}$ & 7.32 $\times$ 10$^{-10}$ \\
32* & H + HOOH $\rightarrow$ H$_{2}$ + O$_{2}$H & $-71.55 \pm 0.14$ & 2.33 $\times$ 10$^{-94}$ & 6.42 $\times$ 10$^{-10}$ \\
33* & CH$_{3}$  + H$_{2}$CO $\rightarrow$ HCO + CH$_{4}$ & $-69.653 \pm 0.027$ & 2.78 $\times$ 10$^{-94}$ & 3.19 $\times$ 10$^{-10}$ \\
34* & O + \ce{H2S} $\rightarrow$ HS + OH & $-49.392 \pm 0.029$ & 6.19 $\times$ 10$^{-90}$ & 2.34 $\times$ 10$^{-10}$ \\
35* & C$_{2}$  + H$_{2}$ $\rightarrow$ H + CCH & $-40.19 \pm 0.13$ & 3.42 $\times$ 10$^{-72}$ & 6.17 $\times$ 10$^{-10}$ \\
36* & H$_{2}$  + C$_{4}$ $\rightarrow$ H + C$_{4}$H & $-34.3 \pm 1.1$ & 3.42 $\times$ 10$^{-72}$ & 7.41 $\times$ 10$^{-10}$ \\
37* & H$_{2}$ + C$_{6}$ $\rightarrow$ H + C$_{6}$H & $-48.6$ $\S$ & 3.42 $\times$ 10$^{-72}$ & 8.76 $\times$ 10$^{-10}$ \\
38* & H$_{2}$ + C$_{8}$ $\rightarrow$ H + C$_{8}$H & $-30.1$ $\S$ & 3.42 $\times$ 10$^{-72}$ & 1.02 $\times$ 10$^{-09}$ \\
39* & H$_{2}$ + C$_{10}$ $\rightarrow$ H + C$_{10}$H & $-21.4$ $\S$ & 3.42 $\times$ 10$^{-72}$ & 1.14 $\times$ 10$^{-09}$ \\
40 & OH  + CH$_{3}$ $\rightarrow$ CH$_{2}$ + H$_{2}$O & $-34.991 \pm 0.094$ & 1.90 $\times$ 10$^{-71}$ & 2.71 $\times$ 10$^{-10}$ \\
41* & O  + HOOH $\rightarrow$ OH + O$_{2}$H & $-65.08 \pm 0.14$ & 1.04 $\times$ 10$^{-70}$ & 2.05 $\times$ 10$^{-10}$ \\
42* & O + H$_{2}$CO $\rightarrow$ OH + HCO & $-62.797 \pm 0.023$ & 2.81 $\times$ 10$^{-67}$ & 2.28 $\times$ 10$^{-10}$ \\
43* & OH + CH$_{4}$ $\rightarrow$ H$_{2}$O + CH$_{3}$ & $-59.758 \pm 0.026$ & 4.20 $\times$ 10$^{-67}$ & 2.80 $\times$ 10$^{-10}$ \\
44* & H + H$_{2}$CO $\rightarrow$ H$_{2}$ + HCO & $-69.2641 \pm 0.0080$ & 3.98 $\times$ 10$^{-62}$ & 7.00 $\times$ 10$^{-10}$ \\
45* & \ce{H2S} + \ce{CH3} $\rightarrow$ HS + \ce{CH4} & $-56.247 \pm 0.032$ & 5.57 $\times$ 10$^{-61}$ & 3.26 $\times$ 10$^{-10}$ \\
46 & CN + NH $\rightarrow$ N + HCN & $-195.21 \pm 0.20$ & 2.00 $\times$ 10$^{-56}$ & 9.75 $\times$ 10$^{-10}$ \\
47* & N + HNO $\rightarrow$ NH + NO & $-131.16 \pm 0.18$ & 2.00 $\times$ 10$^{-56}$ & 2.39 $\times$ 10$^{-10}$ \\
48 & N + HCO $\rightarrow$ CO + NH & $-267.03 \pm 0.18$  & 3.88 $\times$ 10$^{-56}$ & 2.43 $\times$ 10$^{-10}$ \\
49 & CN + OH $\rightarrow$ O + HCN & $-97.49 \pm 0.12$ & 3.72 $\times$ 10$^{-55}$ & 3.70 $\times$ 10$^{-10}$ \\
50* & HCO + HNO $\rightarrow$ NO + H$_{2}$CO & $-166.081 \pm 0.093$ & 4.10 $\times$ 10$^{-55}$ & 8.83 $\times$ 10$^{-10}$ \\
51* & O + HS $\rightarrow$ S + OH & $-75.54 \pm 0.14$ & 5.40 $\times$ 10$^{-54}$ & 2.27 $\times$ 10$^{-10}$ \\
52* & H$_{2}$ + C$_{4}$H $\rightarrow$ H + C$_{4}$H$_{2}$ & $-127.46 \pm 0.89$ & 6.29 $\times$ 10$^{-53}$ & 7.60 $\times$ 10$^{-10}$ \\
53* & H$_{2}$ + C$_{6}$H $\rightarrow$ H + C$_{6}$H$_{2}$ & $-104.9 \pm 1.9$ & 6.29 $\times$ 10$^{-53}$ & 9.54 $\times$ 10$^{-10}$ \\
54* & H$_{2}$ + C$_{8}$H $\rightarrow$ H + C$_{8}$H$_{2}$ & $-131.6$ $\S$ & 6.29 $\times$ 10$^{-53}$ & 1.05 $\times$ 10$^{-09}$ \\
55* & H + H$_{2}$S $\rightarrow$ H$_{2}$ + HS & $-55.858 \pm 0.020$ & 6.71 $\times$ 10$^{-49}$ & 7.32 $\times$ 10$^{-10}$ \\
56* & C$_{4}$H + CH$_{4}$ $\rightarrow$ CH$_{3}$ + C$_{4}$H$_{2}$ & $-127.07 \pm 0.89$ & 5.15 $\times$ 10$^{-38}$ & 4.34 $\times$ 10$^{-10}$ \\
57 & CN + CH$_{4}$ $\rightarrow$ HCN + CH$_{3}$ & $-90.63 \pm 0.12$ & 2.23 $\times$ 10$^{-36}$ & 3.70 $\times$ 10$^{-10}$ \\
58 & CH + HCO $\rightarrow$ CO + CH$_{2}$ & $-357.09 \pm 0.14$ & 5.12 $\times$ 10$^{-35}$ & 1.05 $\times$ 10$^{-09}$ \\
59* & OH + HNO $\rightarrow$ NO + H$_{2}$O & $-295.491 \pm 0.093$ & 1.54 $\times$ 10$^{-32}$ & 1.04 $\times$ 10$^{-09}$ \\
60* & H + HNO $\rightarrow$ H$_{2}$ + NO & $-235.345 \pm 0.093$ & 1.15 $\times$ 10$^{-19}$ & 6.29 $\times$ 10$^{-10}$ \\
61 & OH + NH$_{3}$ $\rightarrow$ H$_{2}$O + NH$_{2}$ & $-48.69 \pm 0.11$ & 6.84 $\times$ 10$^{-17}$ & 1.09 $\times$ 10$^{-09}$ \\
62 & OH + OH $\rightarrow$ O + H$_{2}$O & $-66.613 \pm 0.022$ & 2.30 $\times$ 10$^{-16}$ & 1.18 $\times$ 10$^{-09}$ \\
\enddata
\tablecomments{Reactions \ref{eq:1}--\ref{eq:4} appear first, then all others are ordered by increasing \textit{k}\textsubscript{KIDA, 10 K} values. Those with an asterisk (*) were first identified as tunnelable reactions in this work, those without were first identified by \citet{Meisner2019} through the screening of 1299 reactions from \citet{Kamp2017}. Reactions with a dagger ($\dag$) are only present in kida.uva.2014, and ones in bold have reported, 10 K, 2 $\times$ 10$^{4}$ cm$^{-3}$, MESMER calculated rate coefficients in this work (Reactions 2--5). Enthalpy of reaction values without $\S$ were found from ATcT (version 1.220), and those with were calculated as described in Section \ref{Sec:Screening}.}
\end{deluxetable*}

\begin{figure*}
    \centering  \includegraphics[width=1\linewidth]{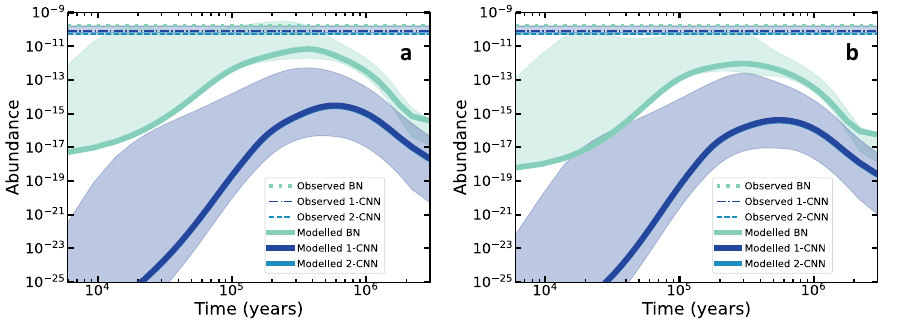}
    \caption{Displayed are the upper and lower modeled molecular abundance bounds for \ce{C6H5CN}, 1-CNN, and 2-CNN, which define an estimation of the uncertainty caused by dormant hydrogen atom transfer reactions in Network 1. \textbf{a}, The modeled abundances of \ce{C6H5CN} (thick green line), 1-CNN (thick dark blue line) and 2-CNN (thick light blue line) using Network 1. The top of each shaded region represents Compilation 1A, and the bottom represents Compilation 1B. The observed abundances of each species are denoted by dashed and/or dotted lines, with the horizontal shading above and below corresponding to the uncertainty. \textbf{b}, The same plot as \textbf{a}, but using Network 2, and Compilations 2A and 2B as the upper and lower bounds of the shaded regions. In both plots the modeled abundances of 1-CNN and 2-CNN are almost identical; since 1-CNN is plotted on top, the plot of 2-CNN is barely visible.} 
    \label{BN+CNN_Plot}
\end{figure*}
\vspace{-35 pt}
\subsection{Potential Tunneling Impacts on Aromatic Molecules\label{AromaticImpact}}
The formation of aromatic molecules are among the most uncertain in astrochemical models. Given that they are also among the most complex species in our model, and hence are indirectly linked to a large number of chemical processes, their abundances are more vulnerable to changes in the network than most species. Using Compilations 1A, 1B, 2A, and 2B for the two sets of simulations, an estimation of the upper and lower bounds for the modeled abundances of C$_{6}$H$_{6}$, C$_{6}$H$_{5}$CN, c-C$_{6}$H$_{5}$$^{+}$, C$_{6}$H$_{7}$$^{+}$, C$_{6}$H$_{5}$, C$_{10}$H$_{8}$, 1-CNN, 2-CNN, C$_{10}$H$_{9}$, and C$_{10}$H$_{10}$ were produced. These upper and lower bounds represent an estimation of the uncertainty found strictly from the presence of dormant, inactive hydrogen atom transfer reactions within Network 1. These reactions are currently inefficient in the model due to having very small rate coefficients. Figure \ref{BN+CNN_Plot} displays these upper and lower modeled molecular abundance bounds for \ce{C6H5CN}, 1-CNN, and 2-CNN, in relation to a simulation using the unmodified networks (Networks 1 and 2). Also displayed in this figure, are each molecule's observed abundances relative to H$_{2}$ ($1.65^{+0.09}_{-0.06}$ $\times$ 10$^{-10}$ for \ce{C6H5CN}, $8.0^{+8.1}_{-1.6}$ $\times$ 10$^{-11}$ for 1-CNN, and $5.3^{+1.4}_{-0.5}$ $\times$ 10$^{-11}$ for 2-CNN), which are taken from \citet{Xue2025}. Similar plots for all aromatic species analyzed in this work are shown and discussed in Appendix Section \ref{AAUAWT}. When compared to simulations conducted with the unmodified networks, the upper bound is much more pronounced than the lower bound for the monitored species at early times. In fact, the largest uncertainty for all molecules shown in Figure \ref{BN+CNN_Plot}, and Appendix Section \ref{AAUAWT} exist as increases, prior to $\sim$10$^{5}$ years, with the largest increase being $> 9$ orders of magnitude more 1-CNN at $\sim$2 $\times$ 10$^{4}$ years. Until times after $\sim$6 $\times$ 10$^{5}$ years the upper bound is much more pronounced for \ce{C10H8}, 1-CNN, and 2-CNN, while the lower bound for all other species never rivals or surpasses the upper bound in magnitude until after $\sim$10$^{6}$ years.
\begin{deluxetable}{ccc}
\tablenum{2}
\tablecaption{Exothermic, gas-phase, hydrogen atom transfer reactions absent from kida.uva.2024, which have endothermic, reverse reactions present in the network.}
\label{tab:Reactions2}
\tablehead{\colhead{No.} & 
\colhead{Reaction} & \colhead{$\Delta$H$_{\text{Rxn}}$ (0 K) (kJ mol$^{-1}$)}}
\startdata
63* & CN + H$_{2}$O $\rightarrow$ OH + HCN & $-30.87 \pm 0.12$ \\
64 & NH$_{2}$ + CH$_{4}$ $\rightarrow$ CH$_{3}$ + NH$_{3}$ & $-11.07 \pm 0.11$ \\
65* & HCO + HOOH $\rightarrow$ O$_{2}$H + H$_{2}$CO & $-2.28 \pm 0.14$
\enddata
\tablecomments{Reactions are ordered by increasing reaction enthalpies. Those with an asterisk (*) were first identified as tunnelable reactions in this work, those without were first identified by \citet{Meisner2019} through the screening of 1299 reactions from \citet{Kamp2017}. Enthalpies of reaction were found from ATcT (version 1.220).} \vspace{-25 pt}
\end{deluxetable}
At early time points, prior to 10$^{5}$ years, all molecules shown in Figures \ref{BN+CNN_Plot}, and Appendix Section \ref{AAUAWT} aside from \ce{C6H5} have an upper bound that is much greater than five orders of magnitude to that of the unmodified networks. The chemical age of the cyanopolyyne peak in TMC-1 is generally predicted to be within 1--5 $\times$ 10$^{5}$ years \citep{Majumdar2016,Hincelin2011, McGuire2018c,Loomis2021} based on the agreement of modeled carbon chains (with a high C/O ratio) to observations. Within this range, the upper bound of \ce{C6H5CN} reaches its observed abundance and surpasses it, with a minor lower bound using Set 1. The upper bounds of 1- and 2-CNN are also significantly closer to the observed abundances during this time frame, but each also has a more notable lower bound and a lessened upper bound after $\sim$10$^{5}$ years. During this time frame the upper bound of \ce{C6H5CN} ranges from $\sim$5--470$\times$ larger, and the lower bound ranges from having negligible change to being $\sim$4$\times$ smaller in Sets 1 and 2. For 1-CNN and 2-CNN the upper bound ranges from $\sim$130--100,000$\times$ larger, with the lower bound ranging from $\sim$20--70$\times$ smaller. Although there were more reactions from Table \ref{tab:Reactions} found to decrease the abundances of the studied species, only two were found in simulation Sets 1 and 2 to cause notable decreases, while, six were found in Sets 1 and 2 to provide significant increases (Section \ref{subsec:"Tunnellable Reactions"}). The major increases that these six reactions provide are the main reason for the imbalanced bounds.

\subsection{Recent Debates on Aromatic Formation Routes\label{PhenyliumImpacts}}
\begin{figure*}
    \centering  \includegraphics[width=1\linewidth]{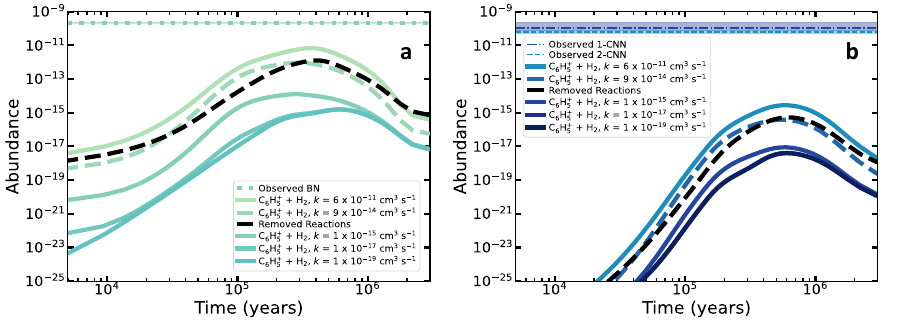}
    \caption{The modeled abundance of \ce{C6H5CN} (\textbf{a}), 1-CNN and 2-CNN (\textbf{b}) as different 10 K rate coefficients ($\it{k}$) are used for Reaction \ref{eq:5}, with the solid 6 $\times$ 10$^{-11}$ cm$^{3}$ s$^{-1}$ line being the value used in kida.uva.2024 and the dashed 9 $\times$ 10$^{-14}$ cm$^{3}$ s$^{-1}$ line being the experimentally determined upper limit by (Kocheril, private communication). The black dashed line corresponds to the change in the abundance of the three molecules when Reactions \ref{eq:6}--\ref{eq:8} are removed from Network 1—which were claimed to possibly be inefficient under low-temperature conditions by \citet{Loison2025}.} 
    \label{BN+CNN_C6H5++H2-Plot}
\end{figure*}

Changing the rate coefficient for Reaction \ref{eq:5} changes the modeled abundances of each species analyzed in this work. When 9 $\times$ 10$^{-14}$ cm$^{3}$ s$^{-1}$ is used as the rate coefficient in Network 2, instead of the kida.uva.2024 value of 6 $\times$ 10$^{-11}$ cm$^{3}$ s$^{-1}$ in Network 1, the abundance of c-C$_{6}$H$_{5}$$^{+}$ is increased, and all other aromatic abundances are decreased. From 1--5 $\times$ 10$^{5}$ years, the abundance increases of c-C$_{6}$H$_{5}$$^{+}$ range from $\sim$10--100$\times$ more, whereas the decreases for all other aromatic molecules during the same time period range from $\sim$3--10$\times$ less. As the rate coefficient for Reaction \ref{eq:5} is decreased, the impact to aromatic molecular abundances becomes more dramatic. For instance, when a rate coefficient of 1 $\times$ 10$^{-15}$ cm$^{3}$ s$^{-1}$ is used, the abundance of c-C$_{6}$H$_{5}$$^{+}$ is increased by $\sim$80--280$\times$, and all other aromatic abundances are decreased by $\sim$200--740$\times$ over the same time period, with the largest decreases to \ce{C6H6}, \ce{C6H5CN}, and C$_{6}$H$_{7}$$^{+}$. Figure \ref{BN+CNN_C6H5++H2-Plot} shows the decreases of \ce{C6H5CN}, 1-CNN, and 2-CNN as this rate coefficient is varied between 6 $\times$ 10$^{-11}$ and 1 $\times$ 10$^{-19}$ cm$^{3}$ s$^{-1}$. 

\begin{figure}
    \centering  \includegraphics[width=1\linewidth]{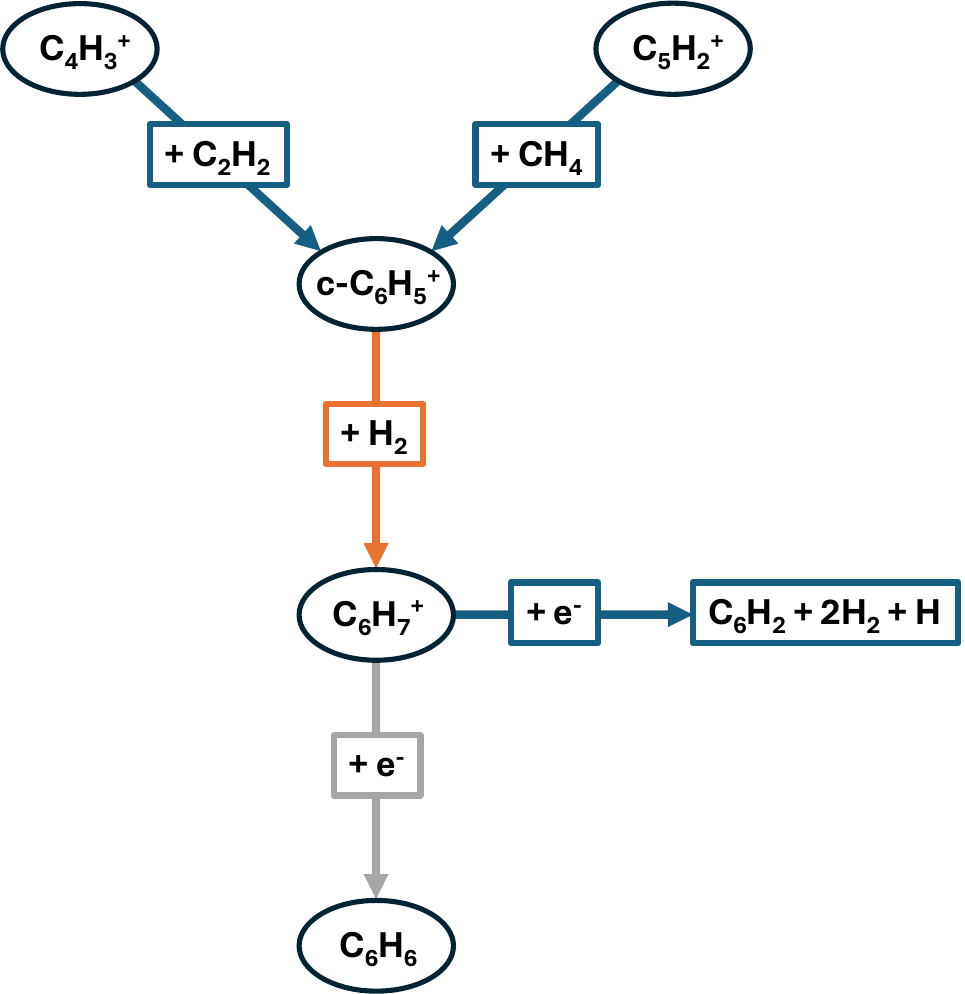}
    \caption{The involvement that Reactions \ref{eq:5}, and \ref{eq:6}--\ref{eq:9} have in the formation of interstellar benzene is displayed here. The non-aromatic species from these reactions are denoted by rectangles, with orange corresponding to Reaction \ref{eq:5} dark blue to Reactions \ref{eq:6}--\ref{eq:8}, and grey to Reaction \ref{eq:9}. All aromatic species are encased in black. Only the aromatic product is shown for each reaction aside from Reaction \ref{eq:7}, which has none.} 
    \label{Aromatic-Formation-Routes}
\end{figure}

In contrast to this evidence, recent work by \citet{Loison2025} has proposed that the reaction
\begin{align}
\ce{C2H2 + C4H3+ \rightarrow c-C6H5+ + hv}  \tag{66} \label{eq:6}
\end{align}
is instead inefficient in low-temperature environments and not Reaction \ref{eq:5}, and that reactions  
\begin{align}
\ce{C6H7+ + e- \rightarrow C6H2 + 2H2 + H} \tag{67} \label{eq:7} \\
\ce{C5H2+ + CH4 \rightarrow  c-C6H5+ + H}   \tag{68} \label{eq:8}
\end{align}
are also negligible at low temperatures. Experimental results from \citet{Loison2025} suggest that Reaction \ref{eq:6} may produce the linear form of C$_{6}$H$_{5}^{+}$, instead of its cyclic form. The linear isomer likely does not contribute to the formation of interstellar \ce{C6H6}, nor is it present in any chemical network in this work. Reaction \ref{eq:7} shows the products of one of two branches for this reaction in kida.uva.2024, whereas 
\begin{align}
\ce{C6H7+ + e- \rightarrow C6H6 + H} \tag{69} \label{eq:9}
\end{align}
 is the other outcome of the \ce{C6H7+ + e-} reaction. \citet{Loison2025} found Reaction \ref{eq:7} to be endothermic by 13 kJ mol$^{-1}$ through quantum chemical calculations. Inspecting this reaction via data from ATcT (version 1.220), suggests it is indeed endothermic, with a $\Delta$H$_{Rxn}$(0 K) = $10.4 \pm 1.6$ kJ mol$^{-1}$. Removing this reaction from our astrochemical network increases the abundance of \ce{C6H6} due to the only products for this reaction now being \ce{C6H6} + H. Furthermore, they also found an energy barrier of 9 kJ mol$^{-1}$ in the reaction of methane and C$_{5}$H$_{2}^{+}$ via quantum chemical calculations, suggesting that Reaction \ref{eq:8} could also be inefficient at low temperatures. Thus, removing Reactions \ref{eq:6} and \ref{eq:8} decrease modeled aromatic abundances. Figure \ref{Aromatic-Formation-Routes} displays Reactions \ref{eq:5}, and \ref{eq:6}--\ref{eq:9}, and displays how each are involved in the formation of interstellar benzene in our model. Between 1--5 $\times$ 10$^{5}$ years, the modeled abundance changes of c-C$_{6}$H$_{5}$$^{+}$, \ce{C6H6}, \ce{C6H5CN}, C$_{6}$H$_{7}$$^{+}$, \ce{C6H5}, \ce{C10H8}, 1-CNN, 2-CNN, C$_{10}$H$_{9}$, and C$_{10}$H$_{10}$ when Reactions \ref{eq:6}--\ref{eq:8} are removed, range from $\sim$4--20$\times$ less. The overall effect on \ce{C6H5CN} and 1- and 2-CNN when these three reactions are removed from Network 1 is shown in Figure \ref{BN+CNN_C6H5++H2-Plot}. 

\subsection{The Byrne et al. (2024) Sensitivity Analysis \label{subsec:SensitivityAnalysis}}
A recent study by \citet{Byrne2024} identified key reactions for the overall network (Network 1 in this work) and for 1-CNN using a Monte Carlo and a modified one-at-a-time approach. In this analysis, all gas-phase rate coefficients were varied within a factor of two of their nominal values. As shown in Table \ref{tab:Reactions}, the majority of the reactions studied here have very small 10 K rate coefficients in the kida.uva.2024 chemical network, and were thus found to play a negligible role in their study. 
However, we show consideration of quantum mechanical tunneling can increase the rate coefficients of these reactions by many orders of magnitude.
This demonstrates an important limitation of sensitivity analysis techniques; they can fail to identify important reactions when the actual rate coefficient uncertainty is significantly larger than estimated. In spite of this, there is agreement between their analysis and ours regarding the pathways important to aromatic chemistry. Both analyses find reactions involving methane, acetylene, ethylene, \ce{C4H2+}, and \ce{C4H3+} are important parts of the aromatic network, likely because they are involved in the formation of other semi-saturated aromatic precursors such as \ce{CH2CCH}, \ce{CH3CCH}, and \ce{C4H4}. 
Despite the weakness of sensitivity analyses when it comes to underestimated uncertainties, the determination of similar key species here as in \citet{Byrne2024} validates the importance of this technique in identifying incomplete areas of the network that should be further explored. 

\subsection{Ab initio and MESMER Results}
\label{Ab+MESMER-Results}
\subsubsection{C$_{2}$H + H$_{2}$}
\label{subsec:C2H+H2}
\begin{figure}
    \centering  \includegraphics[width=0.85\linewidth]{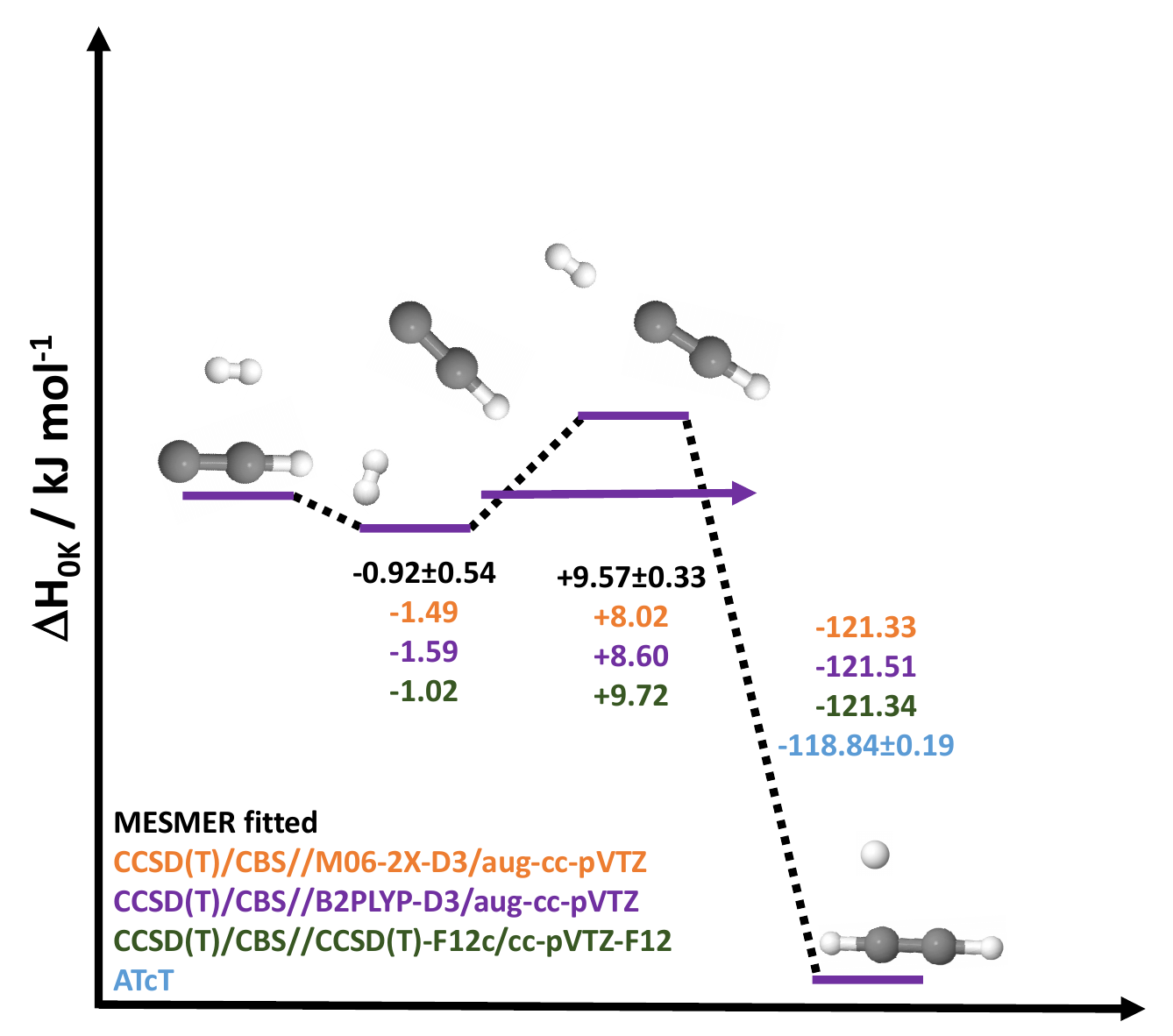}
    \caption{Potential energy surface of Reaction \ref{eq:1}. Calculated energies using the different methodologies in orange, purple, and green, with the ATcT (version 1.220) exothermicity of the reaction in blue, and MESMER fitted values in black. The geometries of the stationary points along the surface determined in the CCSD(T)/CBS//CCSD(T)-F12c/cc-pVTZ-F12 set of calculations are shown above each point.}
    \label{C2H+H2}
\end{figure}

Figure \ref{C2H+H2} shows the calculated potential energy surface of the C$_{2}$H + H$_{2}$ reaction. The MESMER fitted -1.82 ($\pm$ 0.27) kJ mol$^{-1}$ pre-reactive complex, 7.48 ($\pm$ 0.38) kJ mol$^{-1}$ barrier, along with the exothermicity (-118.84 kJ mol$^{-1}$) of the reaction are the lowest of the four studied reactions. The MESMER calculated rate coefficient of 1.66 $\times$ 10$^{-15}$ cm$^{3}$ s$^{-1}$ at 10 K and 2 $\times$ 10$^{4}$ cm$^{-3}$ for this reaction is the largest of those calculated for Reactions \ref{eq:1}--\ref{eq:4}, and is $\sim$53 orders of magnitude larger than the kida.uva.2024 value of 3.97 $\times$ 10$^{-68}$ cm$^{3}$ s$^{-1}$. The suggested 10 K rate coefficient of 2.2 $\times$ 10$^{-15}$ cm$^{3}$ s$^{-1}$ from \citet{Herbst1994} is similar to our prediction, with theirs being $\sim$1.3$\times$ larger. The inclusion of the updated rate coefficient of this reaction into Networks 1 and 2 does reduce the predicted abundance of the C$_{2}$H radical, and increases the abundance of \ce{C2H2} as well as saturated molecules in the model. This leads to minor increases of aromatic molecules, and thus makes the model more efficient at converting simple species into more complex ones as predicted by \citet{Herbst1994}.

For both the KiSThelP TST and MESMER RRKM simulations a hindered rotor model was used for a low frequency torsional vibrational mode in the transition state. The difference between a harmonic oscillator and hindered rotor treatment converges at low temperatures where thermal energy is below the torsional barrier. However, at elevated temperatures the density of states that results from the harmonic oscillator treatment differs markedly from that of a rotation.  Comparing the MESMER and TST prediction at 10 K, \textit{k}$_{TST,10 K}$ $=$ 3.27 $\times$ $10^{-18}$ $\mathrm{cm^{3}\,s^{-1}}$, the importance of the loose entrance channel complex and the impact of fitting the PES to the experimental data is highlighted. With the MESMER result being orders of magnitude larger than the TST rate coefficient. However, both results were approximately 50 orders of magnitude larger than the KIDA value. Highlighting, the need for the consideration of large deviations from the estimated rate coefficients in chemical reaction networks where tunneling could be significant.

The \ce{C2H} and \ce{H2} reaction has a much earlier (longer range) transition state than the other four \ce{H2}-radical reactions studied here. This leads to a smaller imaginary frequency corresponding to translation of the \ce{H2} towards the \ce{C2H} radical. Whereas, for the other reactions the transition state involves the movement of a single H atom. This results in a reduced propensity for tunneling through the barrier in the Eckart model for \ce{C2H} and \ce{H2} than might be expected given the exothermicity and low barrier height.  

\subsubsection{OH + H$_{2}$\label{subsec:OH+H2}}
\begin{figure}
    \centering  \includegraphics[width=0.85\linewidth]{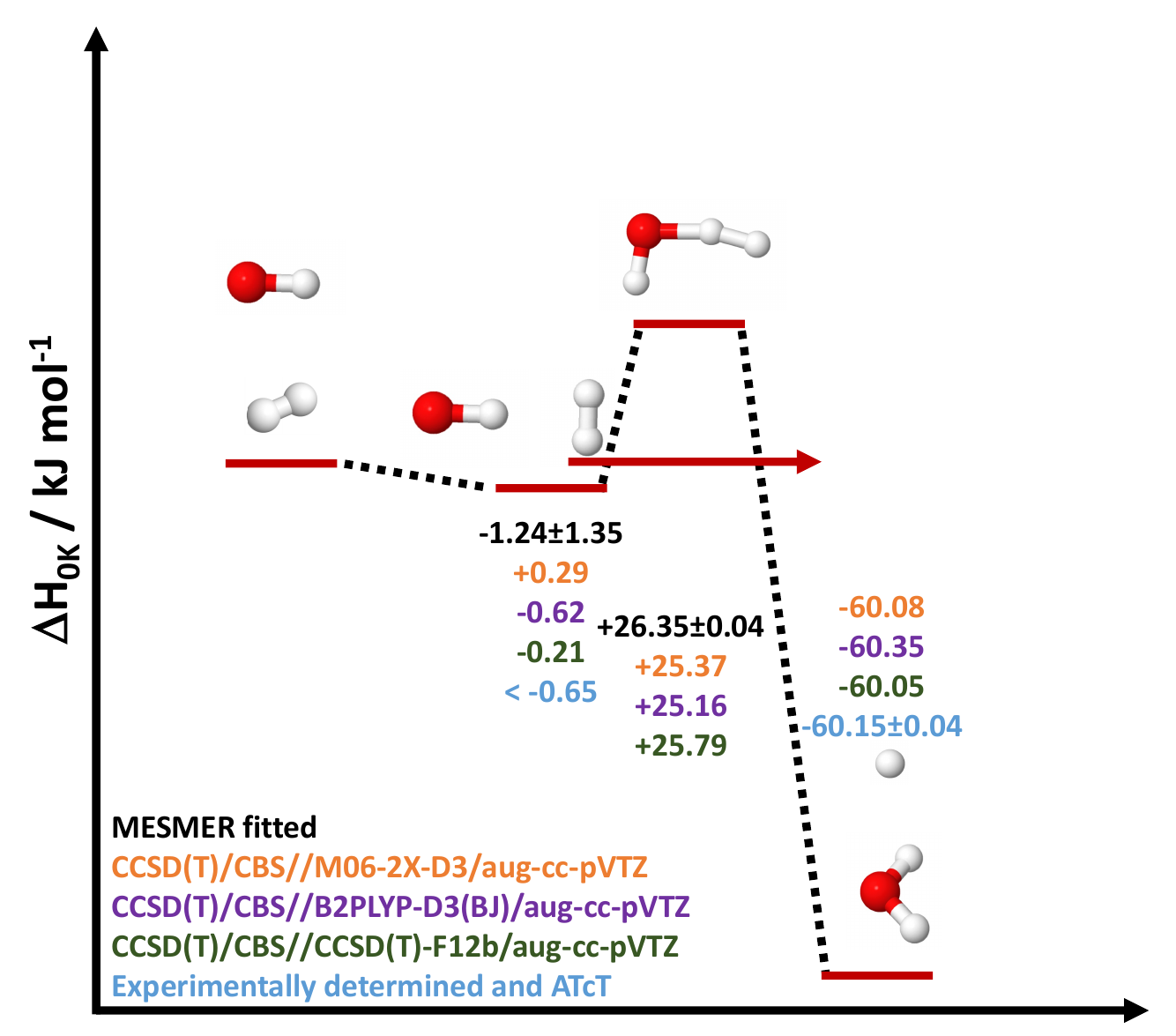}
    \caption{Potential energy surface of Reaction \ref{eq:2} calculated with various levels of theory (orange, purple, and green) including harmonic zero point energy corrections. MESMER fitted values are shown in black, and experimental values of the Van der Waals complex (taken from: \citet{Hernandez1995}, \citet{Loomis1996a}, \citet{Lester1997}, \citet{Loomis1997}, \citet{Schwartz1997a}, \citet{Anderson1998}, \citet{Krause1998}, and \citet{Wheeler1999}) along with the ATcT (version 1.220) exothermicity of the reaction are shown in blue. The geometries of the stationary points along the surface determined in the CCSD(T)/CBS//CCSD(T)-F12b/aug-cc-pVTZ set of calculations are shown above each point.}
    \label{OH+H2}
\end{figure}

The gas-phase reaction of OH + H$_{2}$ was found to have the second highest 10 K, 2 $\times$ 10$^{4}$ $\mathrm{cm^{-3}}$ rate coefficient of the four MESMER analyzed reactions. A MESMER calculated value of $8.17^{+9.17}_{-4.46}$ $\times$ 10$^{-16}$ $\mathrm{cm^{3}\,s^{-1}}$ was found, and the MESMER fitted potential energy surface is displayed in Figure \ref{OH+H2}. This rate coefficient is $\sim$42 orders of magnitude faster than the kida.uva.2024 value of 5.72 $\times$ 10$^{-58}$ $\mathrm{cm^{3}\,s^{-1}}$. Meanwhile, the calculated KiSThelP vTST prediction at 10 K for this reaction was \textit{k}$_{vTST,10 K}$ $=$ 1.39 $\times$ $10^{-17}$ $\mathrm{cm^{3}\,s^{-1}}$. Inclusion of the new MESMER value—along with the three other new rate coefficients for Reactions \ref{eq:1}--\ref{eq:4}—into Network 1 led to a maximal increase of $\sim$4$\times$ more water between 6--7 $\times$ 10$^{5}$ years. Changes to the abundances of other oxygenated species including \ce{OH}, \ce{CH2O}, \ce{CH3OH}, and \ce{C2H5OH} are shown in Appendix Section \ref{H2+Radical-modelling_SI}. This OH + \ce{H2} reaction, along with Reaction \ref{eq:3}, demonstrates that reactions involving molecular hydrogen as a reactant, even where bimolecular rate coefficients are below 10$^{-15}$ $\mathrm{cm^{3}\,s^{-1}}$, can still be influential in the chemistry of dark molecular clouds.

\subsubsection{CN + H$_{2}$} \label{subsec:CN+H2}
At low pressures (for example a gas density of 2 $\times$ 10$^{4}$ cm$^{-3}$), the reaction of CN and H$_{2}$ has two product channels. One of these channels produces HCN and a hydrogen atom, and the other producing HNC and a hydrogen atom. When the collision limit of this reaction was used in simulations, its potential impact on modeled aromatic abundances was the largest among all other reactions in Table \ref{tab:Reactions}. The 10 K reaction rate coefficient in kida.uva.2024 is 5.69 $\times$ 10$^{-53}$ $\mathrm{cm^{3}\,s^{-1}}$ and assumed that HCN + H was the only product channel. The MESMER simulated results for this reaction both prior to and after data fitting predict that HNC formation is not closed at low temperatures due to the influence of quantum mechanical tunneling on the predicted reaction rates. Following data fitting, the MESMER predicted rate coefficients of $3.02^{+0.57}_{-3.49}$ $\times$ 10$^{-16}$ and $1.34^{+13.21}_{-1.34}$ $\times$ 10$^{-17}$ $\mathrm{cm^{3}\,s^{-1}}$ were found for the HCN and HNC channels, respectively, and an overall value (consisting of both product channels) of $3.15^{+1.90}_{-3.05}$ $\times$ 10$^{-16}$ $\mathrm{cm^{3}\,s^{-1}}$ at 10 K and 2 $\times$ 10$^{4}$ cm$^{-3}$; making all three values over 36 orders of magnitude larger than the previously used kida.uva.2024 value. A large uncertainty exists for the HNC channel, with its reported uncertainty range being from $1.88$ $\times$ 10$^{-24}$ to $1.46$ $\times$ 10$^{-16}$ $\mathrm{cm^{3}\,s^{-1}}$. The KiSThelP vTST prediction at 10 K, for HCN and HNC were \textit{k}$_{vTST,10 K}$ $=$ 3.91 $\times$ $10^{-19}$ and 4.60 $\times$ $10^{-20}$ $\mathrm{cm^{3}\,s^{-1}}$. The large uncertainty associated with the HNC branching fraction is the result of its formation being the minor product channel for this reaction. In the MESMER results, less than 1\% of HNC is formed between 300 and 1000 K, which is the range of temperatures where most of the experimental data existed in the literature and that the calculated MESMER rate coefficients are fit to.

\begin{figure}
    \centering  \includegraphics[width=0.85\linewidth]{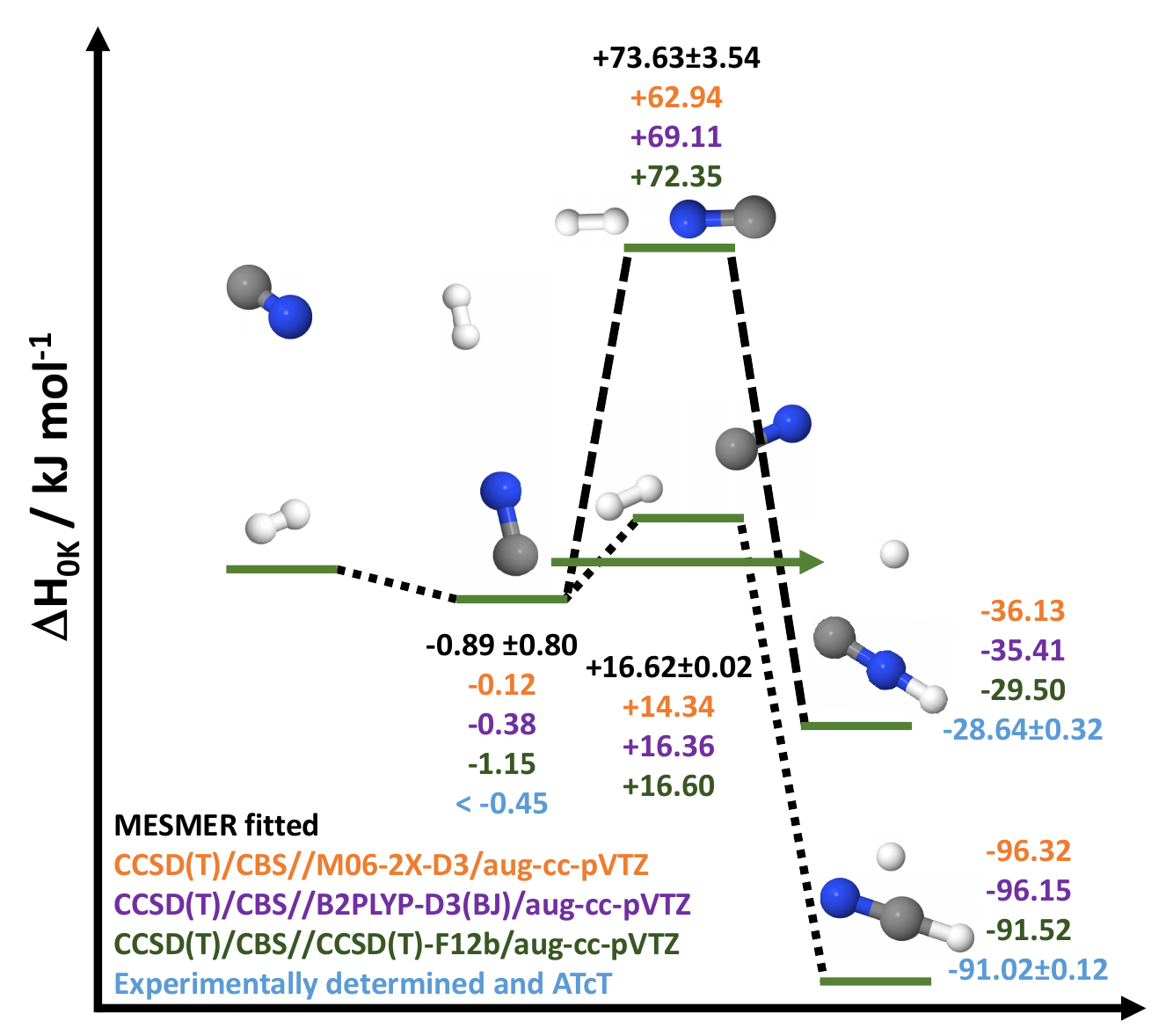}
    \caption{Potential energy surface of Reaction \ref{eq:3}, calculated with various levels of theory (orange, purple, and green values), including harmonic zero point energy corrections. MESMER fitted values are shown in black, and the experimental value of the Van der Waals complex (taken from \citet{Chen1998}) along with the ATcT (version 1.220) exothermicity of the reaction in blue. The two available product channels are shown, with the dashed line corresponding to the HNC channel, and the dotted line to HCN. The geometries of each stationary point along the surface determined in the CCSD(T)/CBS//CCSD(T)-F12b/aug-cc-pVTZ set of calculations are shown above each point.}
    \label{CN+H2}
\end{figure}

When monitoring CN, HCN, and HNC, with all four new rate coefficients for Reactions \ref{eq:1}--\ref{eq:4} implemented in Network 1, the maximal change was a $\sim$3$\times$ increase in the abundance of HCN for the time period of 1--5 $\times$ 10$^{5}$ years. As displayed in Figure \ref{CN+H2}, the HCN channel has a substantially lower energy barrier, while also producing more energetically favourable products than the HNC channel. When both barriers were floated and each channel was linked to the same pre-reaction complex the resulting fits were improved compared to when the barriers were floated concertedly ($\chi$$^{2}$ reduced by a factor of 2.5). The extrapolated MESMER results for the concerted fit fell within the prediction band for the independent fit due to the reduced uncertainty assigned to the HNC parameters due to their linking to the HCN channel. However, the uncertainty in the predicted HCN channel was reduced in the independent fit compared to the concerted fit due to a factor of 10 reduction in the assigned error in the HCN barrier height. Additionally, the HNC channel becomes competitive with the HCN channel at low temperatures, despite the higher energy barrier to the formation of HNC because it is offset by a larger imaginary frequency in the transition state. The HCN channel has a larger rate coefficient at 10 K; however, at this temperature we calculate a $\sim$4\% product branching fraction of HNC formation. Consequently, we suggest that the formation of HNC via hydrogen abstraction reactions involving CN should be included as a minor channel where HNC formation is exothermic, even where substantial barriers $>70$ kJ mol$^{-1}$ are present.
\subsubsection{$NH_{2}$ + H$_{2}$} \label{subsec:NH2+H2}
\begin{figure}
    \centering  \includegraphics[width=0.85\linewidth]{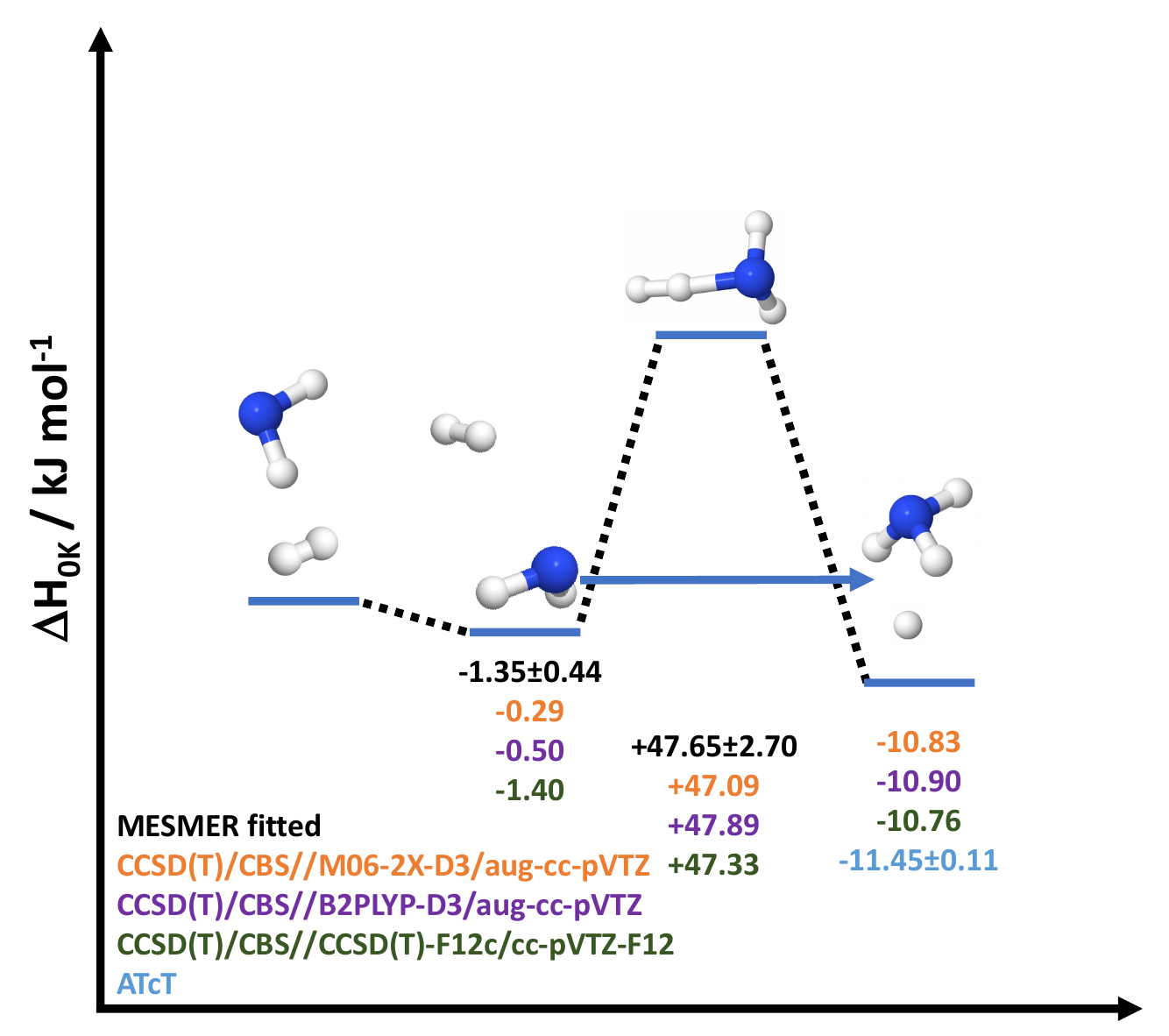}
    \caption{Potential energy surface of Reaction \ref{eq:4} calculated with various levels of theory (orange, purple, and green values), including harmonic zero point energy corrections. MESMER fitted values are shown in black, and the experimental ATcT (version 1.220) exothermicity of the reaction is shown in blue. The geometries of the stationary points along the surface determined in the CCSD(T)/CBS//CCSD(T)-F12c/cc-pVTZ-F12 set of calculations are shown above each point.}
    \label{NH2+H2}
\end{figure}

The formation of ammonia and a hydrogen atom through the reaction of NH$_{2}$ and H$_{2}$ was found to have the smallest MESMER predicted rate coefficient at 10 K and 2 $\times$ 10$^{4}$ cm$^{-3}$ among Reactions \ref{eq:1}--\ref{eq:4}. With a MESMER predicted value of $5.47^{+105}_{-5.36}$ $\times$ 10$^{-29}$ $\mathrm{cm^{3}\,s^{-1}}$, and a \textit{k}$_{vTST,10 K}$ value of 2.29 $\times$ $10^{-21}$  $\mathrm{cm^{3}\,s^{-1}}$, this reaction is non-competitive in the ISM. The predicted rate coefficient from the MESMER analysis at 10 K was $\sim$144 orders of magnitude larger than the 10 K kida.uva.2024 value previously included in our network: 1.48 $\times$ 10$^{-173}$ $\mathrm{cm^{3}\,s^{-1}}$. Consequently, the difference between the MESMER predicted rate coefficient, and its kida.uva.2024 value is the largest among Reactions \ref{eq:1}--\ref{eq:4}. The reaction between \ce{NH2} and \ce{H2} is the least exothermic reaction among Reactions \ref{eq:1}--\ref{eq:4}, as displayed in Figure \ref{NH2+H2}. The combination of its low exothermicity, as well as the presence of a substantial energy barrier cause the barrier to be wide at the entrance energy level, decreasing the probability of hydrogen atom tunneling, and likely being the main contributor to such a small rate coefficient for this reaction.
\vspace{-0.3em}
\section{Conclusion} \label{sec:Conclusion}
In this work, we estimate the degree to which the modeling of interstellar aromatic molecules is affected by reactions that involve hydrogen atom tunneling through an emerged energy barrier. A total of 64 reactions within Tables \ref{tab:Reactions} and \ref{tab:Reactions2} are highlighted to have the potential to be accelerated by low-temperature hydrogen atom tunneling. The potential affect of 59 of these reactions on aromatic abundances were determined. Reaction numbers 7, 10, 11, 13, 35, 36, 52, and 57 in Table \ref{tab:Reactions} were found to have the highest potential impact in altering the modeled molecular abundances of \ce{C6H6}, \ce{C6H5CN}, c-C$_{6}$H$_{5}$$^{+}$, C$_{6}$H$_{7}$$^{+}$, \ce{C6H5}, \ce{C10H8}, 1-CNN, 2-CNN, C$_{10}$H$_{9}$, and C$_{10}$H$_{10}$, with numbers 10, 35, and 36 having the three highest potential impacts. The potential impact on the abundance of these molecules from all 59 reactions range from negligible, to differing by $>9$ orders of magnitude, depending on the molecule and simulated time point.

The modeled abundances of \ce{C6H6}, \ce{C6H5CN}, C$_{6}$H$_{7}$$^{+}$, \ce{C6H5CN}, \ce{C10H8}, 1-CNN, 2-CNN, C$_{10}$H$_{9}$, and C$_{10}$H$_{10}$ decrease by $\sim$3--10$\times$ and c-C$_{6}$H$_{5}$$^{+}$ increases by $\sim$10--100$\times$ when the experimentally determined upper limit for the Reaction \ref{eq:5} rate coefficient of 9 $\times$ 10$^{-14}$ cm$^{3}$ s$^{-1}$ (Kocheril, private communication) instead of the KIDA value of 6 $\times$ 10$^{-11}$ cm$^{3}$ s$^{-1}$ is used during the time period of 1--5 $\times$ 10$^{5}$ years. As such, both rate coefficients were used in this work. However, a recent study from \citet{Loison2025} suggests that Reactions \ref{eq:6}--\ref{eq:8} should be removed in our chemical network rather than changing the Reaction \ref{eq:5} rate coefficient. Consequently, we find that the abundances of c-C$_{6}$H$_{5}$$^{+}$, \ce{C6H6}, \ce{C6H5CN}, C$_{6}$H$_{7}$$^{+}$, \ce{C6H5CN}, \ce{C10H8}, 1-CNN, 2-CNN, C$_{10}$H$_{9}$, and C$_{10}$H$_{10}$ decrease by $\sim$4--20$\times$ over the time period of 1--5 $\times$ 10$^{5}$ years when these reactions are removed (Figure \ref{BN+CNN_C6H5++H2-Plot}).

Four of the 64 hydrogen atom transfer reactions from Tables \ref{tab:Reactions} and \ref{tab:Reactions2}, were studied in detail to determine accurate low-temperature and -pressure rate coefficients. These four reactions involved the abstraction of a hydrogen atom from H$_{2}$ by either C$_{2}$H, OH, CN, or NH$_{2}$. 10 K and 2 $\times$ 10$^{4}$ cm$^{-3}$ rate coefficients of 1.66 $\times$ 10$^{-15}$, 8.17 $\times$ 10$^{-16}$ and 5.47 $\times$ 10$^{-29}$ cm$^{3}$ s$^{-1}$ were calculated for the C$_{2}$H, OH, and NH$_{2}$ involving reactions, with 3.02 $\times$ 10$^{-16}$ and 1.34 $\times$ 10$^{-17}$ cm$^{3}$ s$^{-1}$, corresponding to the respective HCN and HNC channels for the CN involving reaction. Even with rate coefficients smaller than 10$^{-14}$ cm$^{3}$ s$^{-1}$, the impacts of H$_{2}$ reactions persist in dark cloud models, given the high abundance of H$_{2}$; where the reactions of H$_{2}$ with C$_{2}$H, OH, and CN are evidence of this. 

Of the barrier possessing, hydrogen atom transfer reactions identified in this work, future studies should prioritize investigating H$_{2}$ involving reactions along with those found to be most impactful on the aromatic molecules studied here. Determinations of accurate, low-temperature and -pressure rate coefficients for these reactions will help astrochemical models better represent the reality of interstellar aromatic chemistry. This is especially true since low-temperature, neutral-neutral, gas-phase reactions with rate coefficients less than $\sim$10$^{-12}$ cm$^{3}$ s$^{-1}$ are often difficult to measure experimentally \citep{Cooke2019a}, and therefore necessitate computational studies such as the ones conducted here. Additionally, the significant potential effects of barrier possessing, hydrogen atom transfer reactions in chemical networks are likely not confined to aromatic molecules. As such, the possible influence of the reactions in Tables \ref{tab:Reactions} and \ref{tab:Reactions2} on other classes of molecules should be studied in order to further constrain the unclear impact of these reactions within astrochemical models.

\section*{Acknowledgements}
We thank Dr. Valentine Wakelam for the use of the \texttt{NAUTILUS} v1.1 code. R.H.J.W. acknowledges the support of the Walter C. Sumner Memorial Fellowship. I.R.C. acknowledges support from the Natural Sciences and Engineering Research Council of Canada (RGPIN-2022-04684), the Canada Foundation for Innovation and the B.C. Knowledge Development Fund. T.H.S. acknowledges the support of the Canadian Space Agency through grant 24AO3UBC14. A.N.B. acknowledges support from the National Science Foundation Grant Number 2141064. This research was made possible through the computational resources, services and support provided by Advanced Research Computing at the University of British Columbia and the University of Leeds. 

\software{ORCA 5.0.4, MESMER, KiSThelP, \texttt{NAUTILUS}, and Avogadro. All MESMER scripts and astrochemical reaction networks used in this work can be found at: 10.5281/zenodo.17298103}
\bibliographystyle{aasjournal}
\bibliography{tunnelling}

@article{Atkinson1975,
  title = {Rate Constants for the Reaction of the {{OH}} Radical with {{H2}} and {{NO}} ({{M}}={{Ar}} and {{N2}})},
  author = {Atkinson, R. and Hansen, D. A. and Pitts, J. N.},
  year = {1975},
  month = apr,
  journal = {J. Chem. Phys.},
  volume = {62},
  number = {8},
  pages = {3284--3288},
  issn = {0021-9606, 1089-7690},
  doi = {10.1063/1.430882},
  urldate = {2023-07-19},
  langid = {english}
}

@article{Burkhardt2021,
  title = {Discovery of the {{Pure Polycyclic Aromatic Hydrocarbon Indene}} (c-{{C9H8}}) with {{GOTHAM Observations}} of {{TMC-1}}},
  author = {Burkhardt, Andrew M. and Lee, Kin Long Kelvin and Changala, P. Bryan and Shingledecker, Christopher N. and Cooke, Ilsa R. and Loomis, Ryan A. and Wei, Hongji and Charnley, Steven B. and Herbst, Eric and McCarthy, Michael C. and McGuire, Brett A.},
  year = {2021},
  month = may,
  journal = {ApJL},
  volume = {913},
  number = {2},
  pages = {L18},
  publisher = {The American Astronomical Society},
  issn = {2041-8205},
  doi = {10.3847/2041-8213/abfd3a},
  urldate = {2023-09-12},
  langid = {english}
}

@article{Byrne2024,
  title = {Sensitivity Analysis of Aromatic Chemistry to Gas-Phase Kinetics in a Dark Molecular Cloud Model},
  author = {Byrne, Alex N. and Xue, Ci and Van Voorhis, Troy and McGuire, Brett A.},
  year = {2024},
  journal = {Phys. Chem. Chem. Phys.},
  pages = {26734-26747},
  issn = {1463-9076, 1463-9084},
  doi = {10.1039/D4CP03229B},
  urldate = {2024-10-29},
  langid = {english}
}

@article{Cernicharo2021,
  title = {Pure Hydrocarbon Cycles in {{TMC-1}}: {{Discovery}} of Ethynyl Cyclopropenylidene, Cyclopentadiene, and Indene},
  shorttitle = {Pure Hydrocarbon Cycles in {{TMC-1}}},
  author = {Cernicharo, J. and Ag{\'u}ndez, M. and Cabezas, C. and Tercero, B. and Marcelino, N. and Pardo, J. R. and De Vicente, P.},
  year = {2021},
  month = may,
  journal = {A\&A},
  volume = {649},
  pages = {L15},
  issn = {0004-6361, 1432-0746},
  doi = {10.1051/0004-6361/202141156},
  urldate = {2024-08-21},
  copyright = {https://www.edpsciences.org/en/authors/copyright-and-licensing},
  langid = {english}
}

@article{Cernicharo2024,
  title = {Discovery of Two Cyano Derivatives of Acenaphthylene ({{C12H8}}) in {{TMC-1}} with the {{QUIJOTE}} Line Survey},
  author = {Cernicharo, J. and Cabezas, C. and Fuentetaja, R. and Agundez, M. and Tercero, B. and Janeiro, J. and Juanes, M. and Kaiser, R.I. and Endo, Y. and Steber, A.L. and Perez, D. and P'erez, C. and Lesarri, A. and Marcelino, N. and De Vicente, P.},
  year = {2024},
  month = sep,
  journal = {A\&A},
  issn = {0004-6361, 1432-0746},
  doi = {10.1051/0004-6361/202452196},
  urldate = {2024-10-01},
  copyright = {https://www.edpsciences.org/en/authors/copyright-and-licensing},
  langid = {english}
}

@article{Cooke2019a,
  title = {Experimental {{Studies}} of {{Gas-Phase Reactivity}} in {{Relation}} to {{Complex Organic Molecules}} in {{Star-Forming Regions}}},
  author = {Cooke, Ilsa R. and Sims, Ian R.},
  year = {2019},
  month = jul,
  journal = {ACS Earth Space Chem.},
  volume = {3},
  number = {7},
  pages = {1109--1134},
  publisher = {American Chemical Society},
  doi = {10.1021/acsearthspacechem.9b00064},
  urldate = {2024-11-08}
}

@article{Cooke2020,
  title = {Benzonitrile as a {{Proxy}} for {{Benzene}} in the {{Cold ISM}}: {{Low-temperature Rate Coefficients}} for {{CN}} + {{C}} {\textsubscript{6}} {{H}} {\textsubscript{6}}},
  shorttitle = {Benzonitrile as a {{Proxy}} for {{Benzene}} in the {{Cold ISM}}},
  author = {Cooke, Ilsa R. and Gupta, Divita and Messinger, Joseph P. and Sims, Ian R.},
  year = {2020},
  month = mar,
  journal = {ApJL},
  volume = {891},
  number = {2},
  pages = {L41},
  issn = {2041-8205, 2041-8213},
  doi = {10.3847/2041-8213/ab7a9c},
  urldate = {2024-07-17},
  langid = {english}
}

@article{Dunning1989,
  title = {Gaussian Basis Sets for Use in Correlated Molecular Calculations. {{I}}. {{The}} Atoms Boron through Neon and Hydrogen},
  author = {Dunning, Jr., Thom H.},
  year = {1989},
  month = jan,
  journal = {J. Chem. Phys.},
  volume = {90},
  number = {2},
  pages = {1007--1023},
  issn = {0021-9606},
  doi = {10.1063/1.456153},
  urldate = {2024-08-19}
}

@article{Frank1985,
  title = {High {{Temperature Reaction Rate}} for {{H}} + {{O2}} = {{OH}} + {{O}} and {{OH}} + {{H2}} = {{H2O}} + {{H}}},
  author = {Frank, P. and Just, {\relax Th}.},
  year = {1985},
  month = feb,
  journal = {Ber. Bunsenges. Phys. Chem.},
  volume = {89},
  number = {2},
  pages = {181--187},
  issn = {00059021},
  doi = {10.1002/bbpc.19850890218},
  urldate = {2023-07-19},
  langid = {english}
}

@article{Garrod2007,
  title = {Non-Thermal Desorption from Interstellar Dust Grains via Exothermic Surface Reactions},
  author = {Garrod, R. T. and Wakelam, V. and Herbst, E.},
  year = {2007},
  month = jun,
  journal = {A\&A},
  volume = {467},
  number = {3},
  pages = {1103--1115},
  publisher = {EDP Sciences},
  issn = {0004-6361, 1432-0746},
  doi = {10.1051/0004-6361:20066704},
  urldate = {2025-04-23},
  copyright = {{\copyright} ESO, 2007},
  langid = {english}
}

@article{Greiner1969,
  title = {Hydroxyl {{Radical Kinetics}} by {{Kinetic Spectroscopy}}. {{V}}. {{Reactions}} with {{H2}} and {{CO}} in the {{Range}} 300--500{$^\circ$}{{K}}},
  author = {Greiner, N. R.},
  year = {1969},
  month = dec,
  journal = {J. Chem. Phys.},
  volume = {51},
  number = {11},
  pages = {5049--5051},
  issn = {0021-9606, 1089-7690},
  doi = {10.1063/1.1671902},
  urldate = {2023-07-19},
  langid = {english}
}

@article{Hasegawa1993,
  title = {New Gas--Grain Chemical Models of Quiescent Dense Interstellar Clouds: The Effects of {{H2}} Tunnelling Reactions and Cosmic Ray Induced Desorption},
  shorttitle = {New Gas--Grain Chemical Models of Quiescent Dense Interstellar Clouds},
  author = {Hasegawa, Tatsuhiko I. and Herbst, Eric},
  year = {1993},
  month = mar,
  journal = {Mon. Not. R. Astron. Soc.},
  volume = {261},
  number = {1},
  pages = {83--102},
  issn = {0035-8711},
  doi = {10.1093/mnras/261.1.83},
  urldate = {2025-04-23}
}

@article{Heard2018,
  title = {Rapid {{Acceleration}} of {{Hydrogen Atom Abstraction Reactions}} of {{OH}} at {{Very Low Temperatures}} through {{Weakly Bound Complexes}} and {{Tunneling}}},
  author = {Heard, Dwayne E.},
  year = {2018},
  month = nov,
  journal = {Acc. Chem. Res.},
  volume = {51},
  number = {11},
  pages = {2620--2627},
  publisher = {American Chemical Society},
  issn = {0001-4842},
  doi = {10.1021/acs.accounts.8b00304},
  urldate = {2024-07-29}
}

@article{Hebrard2009,
  title = {How {{Measurements}} of {{Rate Coefficients}} at {{Low Temperature Increase}} the {{Predictivity}} of {{Photochemical Models}} of {{Titan}}'s {{Atmosphere}}},
  author = {H{\'e}brard, E. and Dobrijevic, M. and Pernot, P. and Carrasco, N. and Bergeat, A. and Hickson, K. M. and Canosa, A. and Le Picard, S. D. and Sims, I. R.},
  year = {2009},
  month = oct,
  journal = {J. Phys. Chem. A},
  volume = {113},
  number = {42},
  pages = {11227--11237},
  publisher = {American Chemical Society},
  issn = {1089-5639},
  doi = {10.1021/jp905524e},
  urldate = {2024-08-21}
}

@article{Herbst1994,
  title = {Tunneling in the {{C2H}}{{H2}} Reaction at Low Temperature},
  author = {Herbst, Eric},
  year = {1994},
  month = may,
  journal = {Chem. Phys. Lett.},
  volume = {222},
  number = {3},
  pages = {297--301},
  issn = {0009-2614},
  doi = {10.1016/0009-2614(94)00333-5},
  urldate = {2025-01-31}
}

@article{Heyl2023,
  title = {Understanding Molecular Abundances in Star-Forming Regions Using Interpretable Machine Learning},
  author = {Heyl, Johannes and Butterworth, Joshua and Viti, Serena},
  year = {2023},
  month = nov,
  journal = {Mon. Not. R. Astron. Soc.},
  volume = {526},
  number = {1},
  pages = {404--422},
  issn = {0035-8711},
  doi = {10.1093/mnras/stad2814},
  urldate = {2024-10-29}
}

@article{Hincelin2011,
  title = {Oxygen Depletion in Dense Molecular Clouds: A Clue to a Low {{O2}} Abundance?},
  shorttitle = {Oxygen Depletion in Dense Molecular Clouds},
  author = {Hincelin, U. and Wakelam, V. and Hersant, F. and Guilloteau, S. and Loison, J. C. and Honvault, P. and Troe, J.},
  year = {2011},
  month = jun,
  journal = {A\&A},
  volume = {530},
  pages = {A61},
  publisher = {EDP Sciences},
  issn = {0004-6361, 1432-0746},
  doi = {10.1051/0004-6361/201016328},
  urldate = {2024-08-16},
  copyright = {{\copyright} ESO, 2011},
  langid = {english}
}

@article{Kocheril2025a,
  title = {Termination of Bottom-up Interstellar Aromatic Ring Formation at {{C6H5}}+},
  author = {Kocheril, G. S. and {Zagorec-Marks}, C. and Lewandowski, H. J.},
  year = {2025},
  month = mar,
  journal = {Nat. Astron.},
  pages = {1--7},
  publisher = {Nature Publishing Group},
  issn = {2397-3366},
  doi = {10.1038/s41550-025-02504-y},
  urldate = {2025-04-10},
  copyright = {2025 The Author(s), under exclusive licence to Springer Nature Limited},
  langid = {english}
}

@article{Loison2013,
  title = {The Gas-Phase Chemistry of Carbon Chains in Dark Cloud Chemical Models},
  author = {Loison, Jean-Christophe and Wakelam, Valentine and Hickson, Kevin M. and Bergeat, Astrid and Mereau, Raphael},
  year = {2013},
  month = nov,
  journal = {Mon. Not. R. Astron. Soc.},
  volume = {437},
  number = {1},
  pages = {930--945},
  issn = {0035-8711, 1365-2966},
  doi = {10.1093/mnras/stt1956},
  urldate = {2025-04-05},
  langid = {english}
}

@article{Loomis1995,
  title = {Stabilization of Reactants in a Weakly Bound Complex: {{OH}}--{{H2}} and {{OH}}--{{D2}}},
  shorttitle = {Stabilization of Reactants in a Weakly Bound Complex},
  author = {Loomis, Richard A. and Lester, Marsha I.},
  year = {1995},
  month = sep,
  journal = {J. Chem. Phys.},
  volume = {103},
  number = {10},
  pages = {4371--4374},
  issn = {0021-9606, 1089-7690},
  doi = {10.1063/1.470678},
  urldate = {2023-07-18},
  langid = {english}
}

@article{Loomis1997,
  title = {{{OH-H2}} Entrance Channel Complexes},
  author = {Loomis, R A and Lester, M I},
  year = {1997},
  journal = {Annu. Rev. Phys. Chem.},
  volume = {48},
  pages = {643--673},
  publisher = {Annual Reviews},
  address = {United States},
  issn = {0066-426X},
  doi = {10.1146/annurev.physchem.48.1.643},
  urldate = {2023-07-18}
}

@article{Loomis2021,
  title = {An Investigation of Spectral Line Stacking Techniques and Application to the Detection of {{HC11N}}},
  author = {Loomis, Ryan A. and Burkhardt, Andrew M. and Shingledecker, Christopher N. and Charnley, Steven B. and Cordiner, Martin A. and Herbst, Eric and Kalenskii, Sergei and Lee, Kin Long Kelvin and Willis, Eric R. and Xue, Ci and Remijan, Anthony J. and McCarthy, Michael C. and McGuire, Brett A.},
  year = {2021},
  month = feb,
  journal = {Nat. Astron.},
  volume = {5},
  number = {2},
  pages = {188--196},
  publisher = {Nature Publishing Group},
  issn = {2397-3366},
  doi = {10.1038/s41550-020-01261-4},
  urldate = {2023-05-04},
  copyright = {2021 The Author(s), under exclusive licence to Springer Nature Limited},
  langid = {english}
}

@article{Majumdar2016,
  title = {Chemistry of {{TMC-1}} with Multiply Deuterated Species and Spin Chemistry of {{H}} {\textsubscript{2}} , {{H}} {\textsubscript{2}} {\textsuperscript{+}} , {{H}} {\textsubscript{3}} {\textsuperscript{+}} and Their Isotopologues},
  author = {Majumdar, L. and Gratier, P. and Ruaud, M. and Wakelam, V. and Vastel, C. and Sipil{\"a}, O. and Hersant, F. and Dutrey, A. and Guilloteau, S.},
  year = {2016},
  month = dec,
  journal = {Mon. Not. R. Astron. Soc.},
  pages = {stw3360},
  issn = {0035-8711, 1365-2966},
  doi = {10.1093/mnras/stw3360},
  urldate = {2024-02-29},
  langid = {english}
}

@article{McGuire2018c,
  title = {Detection of the Aromatic Molecule Benzonitrile (c-{{C6H5CN}}) in the Interstellar Medium},
  author = {McGuire, Brett A. and Burkhardt, Andrew M. and Kalenskii, Sergei and Shingledecker, Christopher N. and Remijan, Anthony J. and Herbst, Eric and McCarthy, Michael C.},
  year = {2018},
  month = jan,
  journal = {Science},
  volume = {359},
  number = {6372},
  pages = {202--205},
  publisher = {American Association for the Advancement of Science},
  doi = {10.1126/science.aao4890},
  urldate = {2025-05-03}
}

@article{McGuire2021,
  title = {Detection of Two Interstellar Polycyclic Aromatic Hydrocarbons via Spectral Matched Filtering},
  author = {McGuire, Brett A. and Loomis, Ryan A. and Burkhardt, Andrew M. and Lee, Kin Long Kelvin and Shingledecker, Christopher N. and Charnley, Steven B. and Cooke, Ilsa R. and Cordiner, Martin A. and Herbst, Eric and Kalenskii, Sergei and Siebert, Mark A. and Willis, Eric R. and Xue, Ci and Remijan, Anthony J. and McCarthy, Michael C.},
  year = {2021},
  month = mar,
  journal = {Science},
  volume = {371},
  number = {6535},
  pages = {1265--1269},
  issn = {0036-8075, 1095-9203},
  doi = {10.1126/science.abb7535},
  urldate = {2022-11-26},
  langid = {english}
}

@article{Meisner2019,
  title = {The Role of Atom Tunneling in Gas-Phase Reactions in Planet-Forming Disks},
  author = {Meisner, J. and Kamp, I. and Thi, W.-F. and K{\"a}stner, J.},
  year = {2019},
  month = jul,
  journal = {A\&A},
  volume = {627},
  pages = {A45},
  issn = {0004-6361, 1432-0746},
  doi = {10.1051/0004-6361/201834974},
  urldate = {2024-11-08},
  copyright = {https://www.edpsciences.org/en/authors/copyright-and-licensing},
  langid = {english}
}

@article{Millar2023b,
  title = {The {{UMIST}} Database for Astrochemistry 2022},
  author = {Millar, T. J. and Walsh, C. and Van De Sande, M. and Markwick, A. J.},
  year = {2023},
  month = nov,
  journal = {A\&A},
  issn = {0004-6361, 1432-0746},
  doi = {10.1051/0004-6361/202346908},
  urldate = {2024-11-09},
  copyright = {https://www.edpsciences.org/en/authors/copyright-and-licensing},
  langid = {english}
}

@article{Miller1979,
  title = {Tunneling Corrections to Unimolecular Rate Constants, with Application to Formaldehyde},
  author = {Miller, William H.},
  year = {1979},
  month = nov,
  journal = {J. Am. Chem. Soc.},
  volume = {101},
  number = {23},
  pages = {6810--6814},
  publisher = {American Chemical Society},
  issn = {0002-7863},
  doi = {10.1021/ja00517a004},
  urldate = {2024-07-29}
}

@article{Miller1994,
  title = {Rotationally Inelastic and Bound State Dynamics of {{H}} {\textsubscript{2}} -{{OH}}({{X}} {\textsuperscript{2}} {{$\Pi$}})},
  author = {Miller, S.M. and Clary, D.C. and Kliesch, A. and Werner, H.-J.},
  year = {1994},
  month = oct,
  journal = {Mol. Phys.},
  volume = {83},
  number = {3},
  pages = {405--428},
  issn = {0026-8976, 1362-3028},
  doi = {10.1080/00268979400101341},
  urldate = {2023-07-19},
  langid = {english}
}

@article{Oldenborg1992,
  title = {Kinetic Study of the Hydrogel + Hydrogen Reaction from 800 to 1550 {{K}}},
  author = {Oldenborg, R. C. and Loge, G. W. and Harradine, D. M. and Winn, K. R.},
  year = {1992},
  month = oct,
  journal = {J. Phys. Chem.},
  volume = {96},
  number = {21},
  pages = {8426--8430},
  issn = {0022-3654, 1541-5740},
  doi = {10.1021/j100200a041},
  urldate = {2024-08-13},
  langid = {english}
}

@article{Orkin2006,
  title = {Rate {{Constant}} for the {{Reaction}} of {{OH}} with {{H}} {\textsubscript{2}} between 200 and 480 {{K}}},
  author = {Orkin, Vladimir L. and Kozlov, Sergey N. and Poskrebyshev, Gregory A. and Kurylo, Michael J.},
  year = {2006},
  month = jun,
  journal = {J. Phys. Chem. A},
  volume = {110},
  number = {21},
  pages = {6978--6985},
  issn = {1089-5639, 1520-5215},
  doi = {10.1021/jp057035b},
  urldate = {2023-07-18},
  langid = {english}
}

@article{Pratap1997,
  title = {A {{Study}} of the {{Physics}} and {{Chemistry}} of {{TMC-1}}},
  author = {Pratap, P. and Dickens, J. E. and Snell, R. L. and Miralles, M. P. and Bergin, E. A. and Irvine, W. M. and Schloerb, F. P.},
  year = {1997},
  month = sep,
  journal = {ApJ},
  volume = {486},
  number = {2},
  pages = {862},
  issn = {0004-637X},
  doi = {10.1086/304553},
  urldate = {2024-02-15},
  langid = {english}
}

@article{Ravishankara1981,
  title = {Kinetic Study of the Reaction of Hydroxyl with Hydrogen and Deuterium from 250 to 1050 {{K}}},
  author = {Ravishankara, A. R. and Nicovich, J. M. and Thompson, R. L. and Tully, F. P.},
  year = {1981},
  month = aug,
  journal = {J. Phys. Chem.},
  volume = {85},
  number = {17},
  pages = {2498--2503},
  issn = {0022-3654, 1541-5740},
  doi = {10.1021/j150617a018},
  urldate = {2024-08-13},
  langid = {english}
}

@article{Rodriguez-Baras2021,
  title = {Gas Phase {{Elemental}} Abundances in {{Molecular cloudS}} ({{GEMS}}) - {{IV}}. {{Observational}} Results and Statistical Trends},
  author = {{Rodr{\'i}guez-Baras}, M. and Fuente, A. and {Rivi{\'e}re-Marichalar}, P. and {Navarro-Almaida}, D. and Caselli, P. and Gerin, M. and Kramer, C. and Roueff, E. and Wakelam, V. and Esplugues, G. and {Garc{\'i}a-Burillo}, S. and Gal, R. Le and Spezzano, S. and {Alonso-Albi}, T. and Bachiller, R. and Cazaux, S. and Commercon, B. and Goicoechea, J. R. and Loison, J. C. and {Trevi{\~n}o-Morales}, S. P. and Roncero, O. and {Jim{\'e}nez-Serra}, I. and Laas, J. and Hacar, A. and Kirk, J. and Lattanzi, V. and {Mart{\'i}n-Dom{\'e}nech}, R. and {Mu{\~n}oz-Caro}, G. and Pineda, J. E. and Tercero, B. and {Ward-Thompson}, D. and Tafalla, M. and Marcelino, N. and Malinen, J. and Friesen, R. and Giuliano, B. M.},
  year = {2021},
  month = apr,
  journal = {A\&A},
  volume = {648},
  pages = {A120},
  publisher = {EDP Sciences},
  issn = {0004-6361, 1432-0746},
  doi = {10.1051/0004-6361/202040112},
  urldate = {2025-04-23},
  copyright = {{\copyright} ESO 2021},
  langid = {english}
}

@article{Ruaud2015,
  title = {Modelling Complex Organic Molecules in Dense Regions: {{Eley}}--{{Rideal}} and Complex Induced Reaction},
  shorttitle = {Modelling Complex Organic Molecules in Dense Regions},
  author = {Ruaud, M. and Loison, J. C. and Hickson, K. M. and Gratier, P. and Hersant, F. and Wakelam, V.},
  year = {2015},
  month = mar,
  journal = {Mon. Not. R. Astron. Soc.},
  volume = {447},
  number = {4},
  pages = {4004--4017},
  issn = {1365-2966, 0035-8711},
  doi = {10.1093/mnras/stu2709},
  urldate = {2025-04-22},
  langid = {english}
}

@article{Ruaud2016,
  title = {Gas and Grain Chemical Composition in Cold Cores as Predicted by the {{Nautilus}} Three-Phase Model},
  author = {Ruaud, Maxime and Wakelam, Valentine and Hersant, Franck},
  year = {2016},
  month = jul,
  journal = {Mon. Not. Roy. Astron. Soc.},
  volume = {459},
  number = {4},
  pages = {3756--3767},
  issn = {0035-8711, 1365-2966},
  doi = {10.1093/mnras/stw887},
  urldate = {2024-02-29},
  langid = {english}
}

@article{Shannon2013,
  title = {Accelerated Chemistry in the Reaction between the Hydroxyl Radical and Methanol at Interstellar Temperatures Facilitated by Tunnelling},
  author = {Shannon, Robin J. and Blitz, Mark A. and Goddard, Andrew and Heard, Dwayne E.},
  year = {2013},
  month = sep,
  journal = {Nat. Chem.},
  volume = {5},
  number = {9},
  pages = {745--749},
  publisher = {Nature Publishing Group},
  issn = {1755-4349},
  doi = {10.1038/nchem.1692},
  urldate = {2025-01-31},
  copyright = {2013 Springer Nature Limited},
  langid = {english}
}

@article{Sims2013b,
  title = {Tunnelling in Space},
  author = {Sims, Ian R.},
  year = {2013},
  month = sep,
  journal = {Nat. Chem.},
  volume = {5},
  number = {9},
  pages = {734--736},
  publisher = {Nature Publishing Group},
  issn = {1755-4349},
  doi = {10.1038/nchem.1736},
  urldate = {2025-04-04},
  copyright = {2013 Springer Nature Limited},
  langid = {english}
}

@article{Sita2022,
  title = {Discovery of {{Interstellar}} 2-{{Cyanoindene}} (2-{{C}} {\textsubscript{9}} {{H}} {\textsubscript{7}} {{CN}}) in {{GOTHAM Observations}} of {{TMC-1}}},
  author = {Sita, Madelyn L. and Changala, P. Bryan and Xue, Ci and Burkhardt, Andrew M. and Shingledecker, Christopher N. and Kelvin Lee, Kin Long and Loomis, Ryan A. and Momjian, Emmanuel and Siebert, Mark A. and Gupta, Divita and Herbst, Eric and Remijan, Anthony J. and McCarthy, Michael C. and Cooke, Ilsa R. and McGuire, Brett A.},
  year = {2022},
  month = oct,
  journal = {ApJL},
  volume = {938},
  number = {2},
  pages = {L12},
  issn = {2041-8205, 2041-8213},
  doi = {10.3847/2041-8213/ac92f4},
  urldate = {2024-07-17},
  langid = {english}
}

@article{Snell1982,
  title = {Determination of Density Structure in Dark Clouds from {{CS}} Observations},
  author = {Snell, R. L. and Langer, W. D. and Frerking, M. A.},
  year = {1982},
  month = apr,
  journal = {ApJ},
  volume = {255},
  pages = {149},
  issn = {0004-637X, 1538-4357},
  doi = {10.1086/159813},
  urldate = {2024-08-12},
  langid = {english}
}

@article{Spitzer1968,
  title = {Heating of {{H}} i {{Regions}} by {{Energetic Particles}}},
  author = {Spitzer, Jr., Lyman and Tomasko, Martin G.},
  year = {1968},
  month = jun,
  journal = {ApJ},
  volume = {152},
  pages = {971},
  publisher = {IOP},
  issn = {0004-637X},
  doi = {10.1086/149610},
  urldate = {2025-04-23}
}

@article{Talukdar1996,
  title = {Kinetics of Hydroxyl Radical Reactions with Isotopically Labeled Hydrogen},
  author = {Talukdar, Ranajit K. and Gierczak, Tomasz and Goldfarb, Leah and Rudich, Yinon and Madhava Rao, B. S. and Ravishankara, A. R.},
  year = {1996},
  month = feb,
  journal = {J. Phys. Chem.},
  volume = {100},
  number = {8},
  pages = {3037--3043},
  issn = {1520-5207},
  doi = {10.1021/jp9518724},
  urldate = {2024-08-13}
}

@article{Tully1980,
  title = {Flash Photolysis-Resonance Fluorescence Kinetic Study of the Reactions Hydroxyl + Molecular Hydrogen .Fwdarw. Water + Atomic Hydrogen and Hydroxyl + Methane .Fwdarw. Water + Methyl from 298 to 1020 {{K}}},
  author = {Tully, F. P. and Ravishankara, A. R.},
  year = {1980},
  month = nov,
  journal = {J. Phys. Chem.},
  volume = {84},
  number = {23},
  pages = {3126--3130},
  issn = {0022-3654, 1541-5740},
  doi = {10.1021/j100460a031},
  urldate = {2024-08-14},
  langid = {english}
}

@article{Vasyunin2004,
  title = {Influence of Uncertainties in the Rate Constants of Chemical Reactions on Astrochemical Modeling Results},
  author = {Vasyunin, A. I. and Sobolev, A. M. and Wiebe, D. S. and Semenov, D. A.},
  year = {2004},
  month = aug,
  journal = {Astron. Lett.},
  volume = {30},
  number = {8},
  pages = {566--576},
  issn = {1562-6873},
  doi = {10.1134/1.1784498},
  urldate = {2023-12-20},
  langid = {english}
}

@article{Wakelam2005,
  title = {Estimation and Reduction of the Uncertainties in Chemical Models: Application to Hot Core Chemistry},
  shorttitle = {Estimation and Reduction of the Uncertainties in Chemical Models},
  author = {Wakelam, V. and Selsis, F. and Herbst, E. and Caselli, P.},
  year = {2005},
  month = dec,
  journal = {A\&A},
  volume = {444},
  number = {3},
  pages = {883--891},
  publisher = {EDP Sciences},
  issn = {0004-6361, 1432-0746},
  doi = {10.1051/0004-6361:20053673},
  urldate = {2024-10-29},
  copyright = {{\copyright} ESO, 2005},
  langid = {english}
}

@article{Wakelam2006,
  title = {The Effect of Uncertainties on Chemical Models of Dark Clouds},
  author = {Wakelam, V. and Herbst, E. and Selsis, F.},
  year = {2006},
  month = may,
  journal = {A\&A},
  volume = {451},
  number = {2},
  pages = {551--562},
  publisher = {EDP Sciences},
  issn = {0004-6361, 1432-0746},
  doi = {10.1051/0004-6361:20054682},
  urldate = {2024-10-29},
  copyright = {{\copyright} ESO, 2006},
  langid = {english}
}

@article{Wakelam2009,
  title = {A Sensitivity Study of the Neutral-Neutral Reactions {{C}} + {{C}} and {{C}} + {{C}} in Cold Dense Interstellar Clouds},
  author = {Wakelam, V. and Loison, J.-C. and Herbst, E. and Talbi, D. and Quan, D. and Caralp, F.},
  year = {2009},
  month = feb,
  journal = {A\&A},
  volume = {495},
  number = {2},
  pages = {513--521},
  publisher = {EDP Sciences},
  issn = {0004-6361, 1432-0746},
  doi = {10.1051/0004-6361:200810967},
  urldate = {2024-10-29},
  copyright = {{\copyright} ESO, 2009},
  langid = {english}
}

@article{Wakelam2010,
  title = {Sensitivity Analyses of Dense Cloud Chemical Models},
  author = {Wakelam, V. and Herbst, E. and Bourlot, J. Le and Hersant, F. and Selsis, F. and Guilloteau, S.},
  year = {2010},
  month = jul,
  journal = {A\&A},
  volume = {517},
  pages = {A21},
  publisher = {EDP Sciences},
  issn = {0004-6361, 1432-0746},
  doi = {10.1051/0004-6361/200913856},
  urldate = {2024-10-29},
  copyright = {{\copyright} ESO, 2010},
  langid = {english}
}

@article{Wakelam2015,
  title = {{{THE}} 2014 {{KIDA NETWORK FOR INTERS}}TEL{{LAR CHEMISTRY}}},
  author = {Wakelam, V. and Loison, J.-C. and Herbst, E. and Pavone, B. and Bergeat, A. and B{\'e}roff, K. and Chabot, M. and Faure, A. and Galli, D. and Geppert, W. D. and Gerlich, D. and Gratier, P. and Harada, N. and Hickson, K. M. and Honvault, P. and Klippenstein, S. J. and Picard, S. D. Le and Nyman, G. and Ruaud, M. and Schlemmer, S. and Sims, I. R. and Talbi, D. and Tennyson, J. and Wester, R.},
  year = {2015},
  month = mar,
  journal = {ApJS},
  volume = {217},
  number = {2},
  pages = {20},
  publisher = {The American Astronomical Society},
  issn = {0067-0049},
  doi = {10.1088/0067-0049/217/2/20},
  urldate = {2024-12-03},
  langid = {english}
}

@article{Wakelam2024a,
  title = {The 2024 {{KIDA}} Network for Interstellar Chemistry},
  author = {Wakelam, V. and Gratier, P. and Loison, J.-C. and Hickson, K. M. and Penguen, J. and Mechineau, A.},
  year = {2024},
  month = sep,
  journal = {A\&A},
  volume = {689},
  pages = {A63},
  issn = {0004-6361, 1432-0746},
  doi = {10.1051/0004-6361/202450606},
  urldate = {2024-11-09},
  copyright = {https://creativecommons.org/licenses/by/4.0},
  langid = {english}
}

@article{Wenzel2024a,
  title = {Detection of Interstellar 1-Cyanopyrene: {{A}} Four-Ring Polycyclic Aromatic Hydrocarbon},
  shorttitle = {Detection of Interstellar 1-Cyanopyrene},
  author = {Wenzel, Gabi and Cooke, Ilsa R. and Changala, P. Bryan and Bergin, Edwin A. and Zhang, Shuo and Burkhardt, Andrew M. and Byrne, Alex N. and Charnley, Steven B. and Cordiner, Martin A. and Duffy, Miya and Fried, Zachary T. P. and Gupta, Harshal and Holdren, Martin S. and Lipnicky, Andrew and Loomis, Ryan A. and Shay, Hannah Toru and Shingledecker, Christopher N. and Siebert, Mark A. and Stewart, D. Archie and Willis, Reace H. J. and Xue, Ci and Remijan, Anthony J. and Wendlandt, Alison E. and McCarthy, Michael C. and McGuire, Brett A.},
  year = {2024},
  month = nov,
  journal = {Science},
  volume = {386},
  number = {6723},
  pages = {810--813},
  publisher = {American Association for the Advancement of Science},
  doi = {10.1126/science.adq6391},
  urldate = {2025-02-19}
}

@article{Wenzel2025,
  title = {Detections of Interstellar Aromatic Nitriles 2-Cyanopyrene and 4-Cyanopyrene in {{TMC-1}}},
  author = {Wenzel, Gabi and Speak, Thomas H. and Changala, P. Bryan and Willis, Reace H. J. and Burkhardt, Andrew M. and Zhang, Shuo and Bergin, Edwin A. and Byrne, Alex N. and Charnley, Steven B. and Fried, Zachary T. P. and Gupta, Harshal and Herbst, Eric and Holdren, Martin S. and Lipnicky, Andrew and Loomis, Ryan A. and Shingledecker, Christopher N. and Xue, Ci and Remijan, Anthony J. and Wendlandt, Alison E. and McCarthy, Michael C. and Cooke, Ilsa R. and McGuire, Brett A.},
  year = {2025},
  month = feb,
  journal = {Nat. Astron.},
  volume = {9},
  number = {2},
  pages = {262--270},
  publisher = {Nature Publishing Group},
  issn = {2397-3366},
  doi = {10.1038/s41550-024-02410-9},
  urldate = {2025-02-19},
  copyright = {2024 The Author(s)},
  langid = {english}
}

@article{Woon2009a,
  title = {{{QUANTUM CHEMICAL PREDICTIONS OF THE PROPERTIES OF KNOWN AND POSTULATED NEUTRAL INTERS}}TEL{{LAR MOLECULES}}},
  author = {Woon, David E. and Herbst, Eric},
  year = {2009},
  month = nov,
  journal = {ApJS},
  volume = {185},
  number = {2},
  pages = {273},
  publisher = {The American Astronomical Society},
  issn = {0067-0049},
  doi = {10.1088/0067-0049/185/2/273},
  urldate = {2025-04-05},
  langid = {english}
}

@article{Xu2007,
  title = {Theoretical Study on the Kinetics for {{OH}} Reactions with {{CH3OH}} and {{C2H5OH}}},
  author = {Xu, Shucheng and Lin, M.C.},
  year = {2007},
  month = jan,
  journal = {Proc. Combust. Inst.},
  volume = {31},
  number = {1},
  pages = {159--166},
  issn = {15407489},
  doi = {10.1016/j.proci.2006.07.132},
  urldate = {2025-01-31},
  langid = {english}
}

@article{ORCA5,
    title = {Software update: {The} {ORCA} program system—{Version} 5.0},
    volume = {12},
    issn = {1759-0884},
    shorttitle = {Software update},
    url = {https://onlinelibrary.wiley.com/doi/abs/10.1002/wcms.1606},
    doi = {10.1002/wcms.1606},
    language = {en},
    number = {5},
    urldate = {2024-06-14},
    journal = {WIREs Comput. Molec. Sci.},
    author = {Neese, Frank},
    year = {2022},
    note = {\_eprint: https://onlinelibrary.wiley.com/doi/pdf/10.1002/wcms.1606},
    pages = {e1606},
}

@article{ORCA2020,
    author = {Neese, Frank and Wennmohs, Frank and Becker, Ute and Riplinger, Christoph},
    title = "{The ORCA quantum chemistry program package}",
    journal = {J. Chem. Phys.},
    volume = {152},
    number = {22},
    pages = {224108},
    year = {2020},
    month = {06},
    issn = {0021-9606},
    doi = {10.1063/5.0004608},
    url = {https://doi.org/10.1063/5.0004608},
    eprint = {https://pubs.aip.org/aip/jcp/article-pdf/doi/10.1063/5.0004608/16740678/224108\_1\_online.pdf},
}

@article{ORCA,
author = {Neese,F.},
title = {The ORCA program system},
journal = {WIRES Comput. Molec. Sci.},
volume = {2},
number = {1},
pages = {73-78},
DOI = {10.1002/wcms.81},
year = {2012},
type = {journal Article}
}

@article{Avogadro2012,
  title={Avogadro: an advanced semantic chemical editor, visualization, and analysis platform},
  author={Hanwell, Marcus D and Curtis, Donald E and Lonie, David C and Vandermeersch, Tim and Zurek, Eva and Hutchison, Geoffrey R},
  journal={J. Cheminform.},
  volume={4},
  pages={1--17},
  year={2012},
  publisher={Springer}
}

@article{Werner2012,
author = {Werner, Hans-Joachim and Knowles, Peter J. and Knizia, Gerald and Manby, Frederick R. and Schütz, Martin},
title = {Molpro: a general-purpose quantum chemistry program package},
journal = {WIREs CMS},
volume = {2},
number = {2},
pages = {242-253},
doi = {https://doi.org/10.1002/wcms.82},
url = {https://wires.onlinelibrary.wiley.com/doi/abs/10.1002/wcms.82},
eprint = {https://wires.onlinelibrary.wiley.com/doi/pdf/10.1002/wcms.82},
year = {2012}
}

@misc{MOLPROtest,
title={MOLPRO, Version 2010.1, a package of ab initio programs},
author={H.-J. Werner and
P. J. Knowles and
P. Celani and
W. Gyorffy and
A. Hesselmann and
D. Kats and
G. Knizia and
A. Kohn and
T. Korona and
D. Kreplin and
R. Lindh and
Q. Ma and
F. R. Manby and
A. Mitrushenkov and
G. Rauhut and
M. Schutz and
K. R. Shamasundar and
T. B. Adler and
R. D. Amos and
S. J. Bennie and
A. Bernhardsson and
A. Berning and
J. A. Black and
P. J. Bygrave and
R. Cimiraglia and
D. L. Cooper and
D. Coughtrie and
M. J. O. Deegan and
A. J. Dobbyn and
K. Doll and 
M. Dornbach and
F. Eckert and
S. Erfort and
E. Goll and
C. Hampel and
G. Hetzer and
J. G. Hill and
M. Hodges and 
T. Hrenar and
G. Jansen and
C. Koppl and
C. Kollmar and
S. J. R. Lee and
Y. Liu and
A. W. Lloyd and
R. A. Mata and
A. J. May and
B. Mussard and
S. J. McNicholas and
W. Meyer and
T. F. Miller III and
M. E. Mura and
A. Nicklass and
D. P. O'Neill and
P. Palmieri and
D. Peng and
K. A. Peterson and
K. Pfluger and
R. Pitzer and
I. Polyak and
M. Reiher and
J. O. Richardson and
J. B. Robinson and
B. Schroder and
M. Schwilk and 
T. Shiozaki and
M. Sibaev and
H. Stoll and
A. J. Stone and
R. Tarroni and
T. Thorsteinsson and
J. Toulouse and
M. Wang and
M. Welborn and 
B. Ziegler
},
url = {https://www.molpro.net/},
version = {2010.1},
date = {2017-12-15},
year = {2017},
}

@misc{CCCBDB-2022,
    title = {{NIST} computational chemistry comparison and benchmark database},
    url = {http://cccbdb.nist.gov/},
    urldate = {2023-05-01},
    author = {Johnson, Russel III},
    year = {2022},
}

@article{Peterson1994,
    author = {Peterson, Kirk A. and Woon, David E. and Dunning, Thom H., Jr.},
    title = {Benchmark calculations with correlated molecular wave functions. IV. The classical barrier height of the H+H2→H2+H reaction},
    journal = {J. Chem. Phys.},
    volume = {100},
    number = {10},
    pages = {7410-7415},
    year = {1994},
    month = {05},
    issn = {0021-9606},
    doi = {10.1063/1.466884},
    url = {https://doi.org/10.1063/1.466884},
    eprint = {https://pubs.aip.org/aip/jcp/article-pdf/100/10/7410/19206380/7410\_1\_online.pdf},
}

@misc{ccREPO-2016,
    title = {ccRepo},
    author = {Hill, Grant},
    year = {2016},
    url = {http://www.grant-hill.group.shef.ac.uk/ccrepo/index.html},
    urldate  = {2024-06-14},
}

@article{Klippenstein2005,
    author = {Georgievskii, Yuri and Klippenstein, Stephen J.},
    title = "{Long-range transition state theory}",
    journal = {J. Chem. Phys.},
    volume = {122},
    number = {19},
    pages = {194103},
    year = {2005},
    month = {05},
    issn = {0021-9606},
    doi = {10.1063/1.1899603},
    url = {https://doi.org/10.1063/1.1899603},
    eprint = {https://pubs.aip.org/aip/jcp/article-pdf/doi/10.1063/1.1899603/15365677/194103\_1\_online.pdf},
}

@article{Quack1974,
author = {Quack, M. and Troe, J.},
title = {Specific Rate Constants of Unimolecular Processes II. Adiabatic Channel Model},
journal = {Ber. Bunsenges. Phys. Chem.},
volume = {78},
number = {3},
pages = {240-252},
doi = {https://doi.org/10.1002/bbpc.19740780306},
url = {https://onlinelibrary.wiley.com/doi/abs/10.1002/bbpc.19740780306},
eprint = {https://onlinelibrary.wiley.com/doi/pdf/10.1002/bbpc.19740780306},
year = {1974}
}

@article{Troe1985,
title = {Statistical adiabatic channel model of ion-neutral dipole capture rate constants},
journal = {Chem. Phys. Lett.},
volume = {122},
number = {5},
pages = {425-430},
year = {1985},
issn = {0009-2614},
doi = {https://doi.org/10.1016/0009-2614(85)87240-7},
url = {https://www.sciencedirect.com/science/article/pii/0009261485872407},
author = {J. Troe},
}

@Article{Stoecklin1991,
author ="Stoecklin, T. and Dateo, C. E. and Clary, D. C.",
title  ="Rate constant calculations on fast diatom–diatom reactions",
journal  ="J. Chem. Soc.{,} Faraday Trans.",
year  ="1991",
volume  ="87",
issue  ="11",
pages  ="1667-1679",
publisher  ="The Royal Society of Chemistry",
doi  ="10.1039/FT9918701667",
url  ="http://dx.doi.org/10.1039/FT9918701667",
}

@article{Clary1994,
    author = {Clary, D. C.},
    title = "{Rate constant formulae for fast reactions}",
    journal = {AIP Conf. Proc.},
    volume = {312},
    number = {1},
    pages = {405-421},
    year = {1994},
    month = {07},
    issn = {0094-243X},
    doi = {10.1063/1.46565},
    url = {https://doi.org/10.1063/1.46565},
    eprint = {https://pubs.aip.org/aip/acp/article-pdf/312/1/405/11438773/405\_1\_online.pdf},
}

@Article{Clary1993,
author ="Clary, David C. and Stoecklin, Thierry S. and Wickham, Andrew G.",
title  ="Rate constants for chemical reactions of radicals at low temperatures",
journal  ="J. Chem. Soc.{,} Faraday Trans.",
year  ="1993",
volume  ="89",
issue  ="13",
pages  ="2185-2191",
publisher  ="The Royal Society of Chemistry",
doi  ="10.1039/FT9938902185",
url  ="http://dx.doi.org/10.1039/FT9938902185",
}

@article{Stoecklin1995,
title = {Fast reactions between a linear molecule and a polar symmetric top},
journal = {J. Mol. Struc. THEOCHEM},
volume = {341},
number = {1},
pages = {53-61},
year = {1995},
issn = {0166-1280},
doi = {https://doi.org/10.1016/0166-1280(95)04209-O},
url = {https://www.sciencedirect.com/science/article/pii/016612809504209O},
author = {T. Stoecklin and D.C. Clary},
}

@article{georgievskii2005,
  title = {Long-Range Transition State Theory},
  author = {Georgievskii, Yuri and Klippenstein, Stephen J.},
  year = {2005},
  month = may,
  journal = {J. Chem. Phys.},
  volume = {122},
  number = {19},
  pages = {194103},
  issn = {0021-9606},
  doi = {10.1063/1.1899603},
  urldate = {2024-06-14}
}

@article{west2019a,
  title = {Measurements of {{Low Temperature Rate Coefficients}} for the {{Reaction}} of {{CH}} with {{CH2O}} and {{Application}} to {{Dark Cloud}} and {{AGB Stellar Wind Models}}},
  author = {West, Niclas A. and Millar, Tom J. and de Sande, Marie Van and Rutter, Edward and Blitz, Mark A. and Decin, Leen and Heard, Dwayne E.},
  year = {2019},
  month = nov,
  journal = {ApJ},
  volume = {885},
  number = {2},
  pages = {134},
  publisher = {The American Astronomical Society},
  issn = {0004-637X},
  doi = {10.3847/1538-4357/ab480e},
  urldate = {2024-06-14},
  langid = {english}
}

@article{davies1986,
  title = {The Testing of Models for Unimolecular Decomposition via Inverse Laplace Transformation of Experimental Recombination Rate Data},
  author = {Davies, Joanne W. and Green, Nicholas J. B. and Pilling, Michael J.},
  year = {1986},
  month = may,
  journal = {Chem. Phys. Lett.},
  volume = {126},
  number = {3},
  pages = {373--379},
  issn = {0009-2614},
  doi = {10.1016/S0009-2614(86)80101-4},
  urldate = {2024-06-14}
}

@book{Robertson_1996,
  title = {Unimolecular Reactions},
  author = {Holbrook, Kenneth A and Pilling, Michael J and Robertson, Struan H},
  year = {1996},
  edition = {2},
  publisher = {Wiley},
  isbn = {0-471-92268-4}
}

@book{baer1996,
  title = {Unimolecular {{Reaction Dynamics}}: {{Theory}} and {{Experiments}}},
  shorttitle = {Unimolecular {{Reaction Dynamics}}},
  author = {Baer, Tomas and Hase, William L.},
  year = {1996},
  publisher = {Oxford University Pres},
  urldate = {2024-06-14},
  langid = {english},
}

@article{lourderaj2009,
  title = {Theoretical and {{Computational Studies}} of {{Non-RRKM Unimolecular Dynamics}}},
  author = {Lourderaj, Upakarasamy and Hase, William L.},
  year = {2009},
  month = mar,
  journal = {J. Phys. Chem. A. A},
  volume = {113},
  number = {11},
  pages = {2236--2253},
  publisher = {American Chemical Society},
  issn = {1089-5639},
  doi = {10.1021/jp806659f},
  urldate = {2024-06-14}
}

@article{Peterson2008,
    author = {Peterson, Kirk A. and Adler, Thomas B. and Werner, Hans-Joachim},
    title = {Systematically convergent basis sets for explicitly correlated wavefunctions: The atoms H, He, B–Ne, and Al–Ar},
    journal = {J. Chem. Phys.},
    volume = {128},
    number = {8},
    pages = {084102},
    year = {2008},
    month = {02},
    issn = {0021-9606},
    doi = {10.1063/1.2831537},
    url = {https://doi.org/10.1063/1.2831537},
    eprint = {https://pubs.aip.org/aip/jcp/article-pdf/doi/10.1063/1.2831537/15412333/084102\_1\_online.pdf},
}

@article{Adler2007,
    author = {Adler, Thomas B. and Knizia, Gerald and Werner, Hans-Joachim},
    title = {A simple and efficient CCSD(T)-F12 approximation},
    journal = {J. Chem. Phys.},
    volume = {127},
    number = {22},
    pages = {221106},
    year = {2007},
    month = {12},
    issn = {0021-9606},
    doi = {10.1063/1.2817618},
    url = {https://doi.org/10.1063/1.2817618},
    eprint = {https://pubs.aip.org/aip/jcp/article-pdf/doi/10.1063/1.2817618/13499829/221106\_1\_online.pdf},
}

@article{Kendall1992,
    author = {Kendall, Rick A. and Dunning, Thom H., Jr. and Harrison, Robert J.},
    title = {Electron affinities of the first‐row atoms revisited. Systematic basis sets and wave functions},
    journal = {J. Chem. Phys.},
    volume = {96},
    number = {9},
    pages = {6796-6806},
    year = {1992},
    month = {05},
    issn = {0021-9606},
    doi = {10.1063/1.462569},
    url = {https://doi.org/10.1063/1.462569},
    eprint = {https://pubs.aip.org/aip/jcp/article-pdf/96/9/6796/18998924/6796\_1\_online.pdf},
}

@article{Knizia2009,
    author = {Knizia, Gerald and Adler, Thomas B. and Werner, Hans-Joachim},
    title = {Simplified CCSD(T)-F12 methods: Theory and benchmarks},
    journal = {J. Chem. Phys.},
    volume = {130},
    number = {5},
    pages = {054104},
    year = {2009},
    month = {02},
    issn = {0021-9606},
    doi = {10.1063/1.3054300},
    url = {https://doi.org/10.1063/1.3054300},
    eprint = {https://pubs.aip.org/aip/jcp/article-pdf/doi/10.1063/1.3054300/15424674/054104\_1\_online.pdf},
}

@article{Rauhut2009,
    author = {Rauhut, Guntram and Knizia, Gerald and Werner, Hans-Joachim},
    title = {Accurate calculation of vibrational frequencies using explicitly correlated coupled-cluster theory},
    journal = {J. Chem. Phys.},
    volume = {130},
    number = {5},
    pages = {054105},
    year = {2009},
    month = {02},
    issn = {0021-9606},
    doi = {10.1063/1.3070236},
    url = {https://doi.org/10.1063/1.3070236},
    eprint = {https://pubs.aip.org/aip/jcp/article-pdf/doi/10.1063/1.3070236/15422793/054105\_1\_online.pdf},
}

@article{Bloino2012,
    author = {Bloino, Julien and Barone, Vincenzo},
    title = {A second-order perturbation theory route to vibrational averages and transition properties of molecules: General formulation and application to infrared and vibrational circular dichroism spectroscopies},
    journal = {J. Chem. Phys.},
    volume = {136},
    number = {12},
    pages = {124108},
    year = {2012},
    month = {03},
    issn = {0021-9606},
    doi = {10.1063/1.3695210},
    url = {https://doi.org/10.1063/1.3695210},
    eprint = {https://pubs.aip.org/aip/jcp/article-pdf/doi/10.1063/1.3695210/13468897/124108\_1\_online.pdf},
}

@article{Canneaux2014,
author = {Canneaux, Sébastien and Bohr, Frédéric and Henon, Eric},
title = {KiSThelP: A program to predict thermodynamic properties and rate constants from quantum chemistry results†},
journal = {J. Comput. Chem.},
volume = {35},
number = {1},
pages = {82-93},
doi = {https://doi.org/10.1002/jcc.23470},
url = {https://onlinelibrary.wiley.com/doi/abs/10.1002/jcc.23470},
eprint = {https://onlinelibrary.wiley.com/doi/pdf/10.1002/jcc.23470},
year = {2014}
}

@misc{Loison2025,
  title = {Evidence for {{Phenylium Reactivity}} under {{Interstellar Relevant Conditions}}},
  author = {Loison, Jean-Christophe and Rossi, Corentin and Solem, Nicolas and Thissen, Roland and Romanzin, Claire and Alcaraz, Christian and Jacovella, Ugo},
  year = {2025},
  month = jun,
  number = {arXiv:2506.13290},
  eprint = {2506.13290},
  primaryclass = {astro-ph},
  publisher = {arXiv},
  doi = {10.48550/arXiv.2506.13290},
  urldate = {2025-06-17},
  archiveprefix = {arXiv}
}

@article{Endres2016,
  title = {The {{Cologne Database}} for {{Molecular Spectroscopy}}, {{CDMS}}, in the {{Virtual Atomic}} and {{Molecular Data Centre}}, {{VAMDC}}},
  author = {Endres, Christian P. and Schlemmer, Stephan and Schilke, Peter and Stutzki, J{\"u}rgen and M{\"u}ller, Holger S. P.},
  year = {2016},
  month = sep,
  journal = {J. Mol. Spectrosc.},
  series = {New {{Visions}} of {{Spectroscopic Databases}}, {{Volume II}}},
  volume = {327},
  pages = {95--104},
  issn = {0022-2852},
  doi = {10.1016/j.jms.2016.03.005},
  urldate = {2025-08-14}
}

@misc{Muller2025,
  title = {Molecules [{{CDMS}} Classic Documentation]},
  author = {M{\"u}ller, Holger S. P.},
  year = {2025},
  month = aug,
  journal = {The Cologne Database for Molecular Spectroscopy CDMS},
  urldate = {2025-08-14},
  howpublished = {https://cdms.astro.uni-koeln.de/classic/molecules}
}

@article{Garrod2008,
  title = {Complex {{Chemistry}} in {{Star-forming Regions}}: {{An Expanded Gas-Grain Warm-up Chemical Model}}},
  shorttitle = {Complex {{Chemistry}} in {{Star-forming Regions}}},
  author = {Garrod, Robin T. and Widicus Weaver, Susanna L. and Herbst, Eric},
  year = {2008},
  month = jul,
  journal = {ApJ},
  volume = {682},
  pages = {283--302},
  publisher = {IOP},
  issn = {0004-637X},
  doi = {10.1086/588035},
  urldate = {2025-08-14}
}

@article{Caravan2015,
  title = {Measurements of {{Rate Coefficients}} for {{Reactions}} of {{OH}} with {{Ethanol}} and {{Propan-2-ol}} at {{Very Low Temperatures}}},
  author = {Caravan, Rebecca L. and Shannon, Robin J. and Lewis, Thomas and Blitz, Mark A. and Heard, Dwayne E.},
  year = {2015},
  month = jul,
  journal = {J. Phys. Chem. A},
  volume = {119},
  number = {28},
  pages = {7130--7137},
  publisher = {American Chemical Society},
  issn = {1089-5639},
  doi = {10.1021/jp505790m},
  urldate = {2025-08-15}
}

@article{J.Shannon2014,
  title = {A Combined Experimental and Theoretical Study of Reactions between the Hydroxyl Radical and Oxygenated Hydrocarbons Relevant to Astrochemical Environments},
  author = {J.~Shannon, R. and L.~Caravan, R. and A.~Blitz, M. and E.~Heard, D.},
  year = {2014},
  journal = {Phys. Chem. Chem. Phys.},
  volume = {16},
  number = {8},
  pages = {3466--3478},
  publisher = {Royal Society of Chemistry},
  doi = {10.1039/C3CP54664K},
  urldate = {2025-08-15},
  langid = {english}
}

@article{Jimenez2016,
  title = {First Evidence of the Dramatic Enhancement of the Reactivity of Methyl Formate ({{HC}}({{O}}){{OCH3}}) with {{OH}} at Temperatures of the Interstellar Medium: A Gas-Phase Kinetic Study between 22 {{K}} and 64 {{K}}},
  shorttitle = {First Evidence of the Dramatic Enhancement of the Reactivity of Methyl Formate ({{HC}}({{O}}){{OCH3}}) with {{OH}} at Temperatures of the Interstellar Medium},
  author = {Jim{\'e}nez, E. and Anti{\~n}olo, M. and Ballesteros, B. and Canosa, A. and Albaladejo, J.},
  year = {2016},
  month = jan,
  journal = {Phys. Chem. Chem. Phys.},
  volume = {18},
  number = {3},
  pages = {2183--2191},
  publisher = {The Royal Society of Chemistry},
  issn = {1463-9084},
  doi = {10.1039/C5CP06369H},
  urldate = {2025-08-15},
  langid = {english}
}

@article{Wenzel2025a,
  title = {Discovery of the {{Seven-ring Polycyclic Aromatic Hydrocarbon Cyanocoronene}} ({{C24H11CN}}) in {{GOTHAM Observations}} of {{TMC-1}}},
  author = {Wenzel, Gabi and Gong, Siyuan and Xue, Ci and Changala, P. Bryan and Holdren, Martin S. and Speak, Thomas H. and Stewart, D. Archie and Fried, Zachary T. P. and Willis, Reace H. J. and Bergin, Edwin A. and Burkhardt, Andrew M. and Byrne, Alex N. and Charnley, Steven B. and Lipnicky, Andrew and Loomis, Ryan A. and Shingledecker, Christopher N. and Cooke, Ilsa R. and McCarthy, Michael C. and Remijan, Anthony J. and Wendlandt, Alison E. and McGuire, Brett A.},
  year = {2025},
  month = apr,
  journal = {ApJL},
  volume = {984},
  number = {1},
  pages = {L36},
  publisher = {The American Astronomical Society},
  issn = {2041-8205},
  doi = {10.3847/2041-8213/adc911},
  urldate = {2025-06-16},
  langid = {english}
}

@article{Kamp2017,
  title = {Consistent Dust and Gas Models for Protoplanetary Disks: {{II}}. {{Chemical}} Networks and Rates},
  shorttitle = {Consistent Dust and Gas Models for Protoplanetary Disks},
  author = {Kamp, I. and Thi, W.-F. and Woitke, P. and Rab, C. and Bouma, S. and M{\'e}nard, F.},
  year = {2017},
  month = nov,
  journal = {A\&A},
  volume = {607},
  pages = {A41},
  issn = {0004-6361, 1432-0746},
  doi = {10.1051/0004-6361/201730388},
  urldate = {2025-09-07},
  langid = {english}
}

@article{Demissy1980,
  title = {Kinetics of Hydrogen Abstraction by Amino Radicals from Alkanes in the Gas Phase. {{A}} Flash Photolysis-Laser Resonance Absorption Study},
  author = {Demissy, M. and Lesclaux, R.},
  year = {1980},
  month = apr,
  journal = {J. Am. Chem. Soc.},
  volume = {102},
  number = {9},
  pages = {2897--2902},
  publisher = {American Chemical Society},
  issn = {0002-7863},
  doi = {10.1021/ja00529a005},
  urldate = {2025-10-08}
}

@article{Friedrichs2000,
  title = {Direct {{Measurements}} of the {{Reaction NH2}} + {{H2}} {$\rightarrow$} {{NH3}} + {{H}} at {{Temperatures}} from 1360 to 2130 {{K}}},
  author = {Friedrichs, G. and Wagner, H. Gg},
  year = {2000},
  month = aug,
  journal = {Z. Phys. Chem.},
  volume = {214},
  number = {8},
  pages = {1151},
  publisher = {De Gruyter (O)},
  issn = {2196-7156},
  doi = {10.1524/zpch.2000.214.8.1151},
  urldate = {2025-10-08},
  chapter = {Zeitschrift f{\"u}r Physikalische Chemie},
  copyright = {De Gruyter expressly reserves the right to use all content for commercial text and data mining within the meaning of Section 44b of the German Copyright Act.},
  langid = {english}
}

@article{Hack1986,
  title = {Direct Measurements of the Reactions Amidogen + Molecular Hydrogen .Dblarw. Ammonia + Atomic Hydrogen at Temperatures from 670 to 1000 {{K}}},
  author = {Hack, W. and Rouveirolles, P. and Wagner, H. G.},
  year = {1986},
  month = may,
  journal = {J. Phys. Chem.},
  volume = {90},
  number = {11},
  pages = {2505--2511},
  publisher = {American Chemical Society},
  issn = {0022-3654},
  doi = {10.1021/j100402a048},
  urldate = {2025-10-08}
}

@article{Lesclaux1978,
  title = {The Kinetics of the Gas Phase Reactions of {{NH2}} Radicals with Alkane and Alkyl Radicals},
  author = {Lesclaux, R. and Demissy, M.},
  year = {1978},
  month = jan,
  journal = {J. Photochem.},
  series = {{{IXthe}} International Conference on Photochemistry},
  volume = {9},
  number = {2},
  pages = {110--112},
  issn = {0047-2670},
  doi = {10.1016/0047-2670(78)80075-6},
  urldate = {2025-10-08}
}

@article{Sutherland1988,
  title = {The Kinetics and Thermodynamics of the Reaction {{H}}+{{NH3}}⇄{{NH2}}+{{H2}} by the Flash Photolysis--Shock Tube Technique: {{Determination}} of the Equilibrium Constant, the Rate Constant for the Back Reaction, and the Enthalpy of Formation of the Amidogen Radical},
  shorttitle = {The Kinetics and Thermodynamics of the Reaction {{H}}+{{NH3}}⇄{{NH2}}+{{H2}} by the Flash Photolysis--Shock Tube Technique},
  author = {Sutherland, J. W. and Michael, J. V.},
  year = {1988},
  month = jan,
  journal = {J. Chem. Phys.},
  volume = {88},
  number = {2},
  pages = {830--834},
  issn = {0021-9606},
  doi = {10.1063/1.454162},
  urldate = {2025-10-08}
}

@article{Baulch2005,
  title = {Evaluated {{Kinetic Data}} for {{Combustion Modeling}}: {{Supplement II}}},
  shorttitle = {Evaluated {{Kinetic Data}} for {{Combustion Modeling}}},
  author = {Baulch, D. L. and Bowman, C. T. and Cobos, C. J. and Cox, R. A. and Just, {\relax Th}. and Kerr, J. A. and Pilling, M. J. and Stocker, D. and Troe, J. and Tsang, W. and Walker, R. W. and Warnatz, J.},
  year = {2005},
  month = sep,
  journal = {J. Phys. Chem. Ref. Data.},
  volume = {34},
  number = {3},
  pages = {757--1397},
  issn = {0047-2689, 1529-7845},
  doi = {10.1063/1.1748524},
  urldate = {2025-10-08},
  langid = {english}
}

@article{Bott1989,
  title = {A Shock Tube Study of the Reaction of the Hydroxyl Radical with {{H2}}, {{CH4}}, c-{{C5H10}}, and i-{{C4H10}}},
  author = {Bott, J. F. and Cohen, N.},
  year = {1989},
  journal = {Int. J. Chem. Kinet.},
  volume = {21},
  number = {7},
  pages = {485--498},
  issn = {1097-4601},
  doi = {10.1002/kin.550210702},
  urldate = {2025-10-08},
  copyright = {Copyright {\copyright} 1989 John Wiley \& Sons, Inc.},
  langid = {english}
}

@article{Brabbs1971,
  title = {Shock-Tube Measurements of Specific Reaction Rates in the Branched-Chain {{H2-CO-O2}} System},
  author = {Brabbs, T. A. and Belles, F. E. and Brokaw, R. S.},
  year = {1971},
  month = jan,
  journal = { Symp. Combust. Proc.},
  series = {Thirteenth Symposium ({{International}}) on {{Combustion}}},
  volume = {13},
  number = {1},
  pages = {129--136},
  issn = {0082-0784},
  doi = {10.1016/S0082-0784(71)80017-6},
  urldate = {2025-10-08}
}

@article{Eberius1971,
  title = {Experimental and Mathematical Study of a Hydrogen-Oxygen Flame},
  author = {Eberius, K. H. and Hoyermann, K. and Wagner, H. {\relax GG}.},
  year = {1971},
  month = jan,
  journal = { Symp. Combust. Proc.},
  series = {Thirteenth Symposium ({{International}}) on {{Combustion}}},
  volume = {13},
  number = {1},
  pages = {713--721},
  issn = {0082-0784},
  doi = {10.1016/S0082-0784(71)80074-7},
  urldate = {2025-10-08}
}

@article{Krasnoperov2004,
  title = {Shock {{Tube Studies Using}} a {{Novel Multipass Absorption Cell}}:\, {{Rate Constant Results For OH}} + {{H2}} and {{OH}} + {{C2H6}}},
  shorttitle = {Shock {{Tube Studies Using}} a {{Novel Multipass Absorption Cell}}},
  author = {Krasnoperov, L. N. and Michael, J. V.},
  year = {2004},
  month = jul,
  journal = {J. Phys. Chem. A},
  volume = {108},
  number = {26},
  pages = {5643--5648},
  publisher = {American Chemical Society},
  issn = {1089-5639},
  doi = {10.1021/jp040186e},
  urldate = {2025-10-08}
}

@article{Ripley1966,
  title = {Shock-{{Tube Study}} of the {{Hydrogen}}---{{Oxygen Reaction}}. {{II}}. {{Role}} of {{Exchange Initiation}}},
  author = {Ripley, Dennis L. and Gardiner, Jr., W. C.},
  year = {1966},
  month = mar,
  journal = {J. Chem. Phys.},
  volume = {44},
  number = {6},
  pages = {2285--2296},
  issn = {0021-9606},
  doi = {10.1063/1.1727036},
  urldate = {2025-10-08}
}

@article{Roth1985,
  title = {Kinetics of the High Temperature, Low Concentration {{CH4}} Oxidation Verified by {{H}} and {{O}} Atom Measurements},
  author = {Roth, P. and Just, {\relax Th}.},
  year = {1985},
  month = jan,
  journal = { Symp. Combust. Proc.},
  series = {Twentieth {{Symposium}} ({{International}}) on {{Combustion}}},
  volume = {20},
  number = {1},
  pages = {807--818},
  issn = {0082-0784},
  doi = {10.1016/S0082-0784(85)80571-3},
  urldate = {2025-10-08}
}

@inproceedings{Schmidt1984,
  title = {Absolute {{Rate Constant Measurements}} of {{OH Reactions Under Atmospheric Conditions}} by {{Laser Photolysis}}/{{Dye Laser Fluorescence}}},
  booktitle = {Physico-{{Chemical Behaviour}} of {{Atmospheric Pollutants}}},
  author = {Schmidt, V. and Zhu, Gui-Yun and Becker, K. H. and Fink, E. H.},
  editor = {Versino, B. and Angeletti, G.},
  year = {1984},
  pages = {177--187},
  publisher = {Springer Netherlands},
  address = {Dordrecht},
  doi = {10.1007/978-94-009-6505-8_19},
  isbn = {978-94-009-6505-8},
  langid = {english}
}

@article{Smith1973,
  title = {Rate Measurements of Reactions of {{OH}} by Resonance Absorption. {{Part}} 2.---{{Reactions}} of {{OH}} with {{CO}}, {{C2H4}} and {{C2H2}}},
  author = {Smith, Ian W. M. and Zellner, Reinhard},
  year = {1973},
  month = jan,
  journal = {J. Chem. Soc., Faraday Trans. 2},
  volume = {69},
  number = {0},
  pages = {1617--1627},
  publisher = {The Royal Society of Chemistry},
  issn = {0300-9238},
  doi = {10.1039/F29736901617},
  urldate = {2025-10-08},
  langid = {english}
}

@article{Vandooren1975,
  title = {Rate Constant of the Elementary Reaction of Carbon Monoxide with Hydroxyl Radical},
  author = {Vandooren, J. and Peeters, J. and Tiggelen, P. J. Van},
  year = {1975},
  month = jan,
  journal = { Symp. Combust. Proc.},
  series = {Fifteenth {{Symposium}} ({{International}}) on {{Combustion}}},
  volume = {15},
  number = {1},
  pages = {745--753},
  issn = {0082-0784},
  doi = {10.1016/S0082-0784(75)80343-2},
  urldate = {2025-10-08}
}

@article{W.C.Gardiner1973,
  title = {Elementary Reaction Rates from Post-Induction-Period Profiles in Shock-Initiated Combustion},
  author = {{W. C. Gardiner} and Mallard, W. G. and McFarland, M. and Morinaga, K. and Owen, J. H. and Rawlins, W. T. and Takeyama, T. and Walker, B. F.},
  year = {1973},
  month = jan,
  journal = {Symp. Combust. Proc.},
  series = {Fourteenth {{Symposium}} ({{International}}) on {{Combustion}}},
  volume = {14},
  number = {1},
  pages = {61--75},
  issn = {0082-0784},
  doi = {10.1016/S0082-0784(73)80009-8},
  urldate = {2025-10-08}
}

@article{Westenberg1973,
  title = {Rates of {{CO}} + {{OH}} and {{H2}} + {{OH}} over an Extended Temperature Range},
  author = {Westenberg, A. A. and {deHaas}, N.},
  year = {1973},
  month = may,
  journal = {J. Chem. Phys.},
  volume = {58},
  number = {10},
  pages = {4061--4065},
  issn = {0021-9606},
  doi = {10.1063/1.1678961},
  urldate = {2025-10-08}
}

@article{Lam2013,
  title = {A {{Shock Tube Study}} of {{H2}} + {{OH}} {$\rightarrow$} {{H2O}} + {{H Using OH Laser Absorption}}},
  author = {Lam, King-Yiu and Davidson, David F. and Hanson, Ronald K.},
  year = {2013},
  journal = {Int. J. Chem. Kinet.},
  volume = {45},
  number = {6},
  pages = {363--373},
  issn = {1097-4601},
  doi = {10.1002/kin.20771},
  urldate = {2025-10-08},
  copyright = {{\copyright} 2013 Wiley Periodicals, Inc.},
  langid = {english}
}

@article{Carty2001a,
  title = {Low Temperature Rate Coefficients for the Reactions of {{CN}} and {{C2H}} Radicals with Allene ({{CH2}}{{C}}{{CH2}}) and Methyl Acetylene ({{CH3C}}{{CH}})},
  author = {Carty, David and Le Page, Valery and Sims, Ian R. and Smith, Ian W. M.},
  year = {2001},
  month = aug,
  journal = {Chem. Phys. Lett.},
  volume = {344},
  number = {3},
  pages = {310--316},
  issn = {0009-2614},
  doi = {10.1016/S0009-2614(01)00682-0},
  urldate = {2025-10-10}
}

@article{Farhat1993,
  title = {Temperature Dependence of the Rate of Reaction of Ethynyl Radical with Hydrogen},
  author = {Farhat, S. K. and Morter, C. L. and Glass, Graham P.},
  year = {1993},
  month = dec,
  journal = {J. Phys. Chem.},
  volume = {97},
  number = {49},
  pages = {12789--12792},
  publisher = {American Chemical Society},
  issn = {0022-3654},
  doi = {10.1021/j100151a026},
  urldate = {2025-10-16}
}

@article{Koshi1992,
  title = {Temperature Dependence of the Rate Constants for the Reactions of Ethynyl Radical with Acetylene, Hydrogen, and Deuterium},
  author = {Koshi, Mitsuo and Fukuda, Koichi and Kamiya, Kenshu and Matsui, Hiroyuki},
  year = {1992},
  month = nov,
  journal = {J. Phys. Chem.},
  volume = {96},
  number = {24},
  pages = {9839--9843},
  publisher = {American Chemical Society},
  issn = {0022-3654},
  doi = {10.1021/j100203a048},
  urldate = {2025-10-16}
}

@article{Koshi1992a,
  title = {Kinetics of the Reactions of Ethynyl Radical with Acetylene, Hydrogen, and Deuterium},
  author = {Koshi, Mitsuo and Nishida, Nobuhiro and Matsui, Hiroyuki},
  year = {1992},
  month = jul,
  journal = {J. Phys. Chem.},
  volume = {96},
  number = {14},
  pages = {5875--5880},
  publisher = {American Chemical Society},
  issn = {0022-3654},
  doi = {10.1021/j100193a043},
  urldate = {2025-10-16}
}

@article{Kovacs2010,
  title = {H-{{Atom Yields}} from the {{Photolysis}} of {{Acetylene}} and from the {{Reaction}} of {{C2H}} with {{H2}}, {{C2H2}}, and {{C2H4}}},
  author = {Kov{\'a}cs, Tam{\'a}s and Blitz, Mark A. and Seakins, Paul W.},
  year = {2010},
  month = apr,
  journal = {J. Phys. Chem. A},
  volume = {114},
  number = {14},
  pages = {4735--4741},
  publisher = {American Chemical Society},
  issn = {1089-5639},
  doi = {10.1021/jp908285t},
  urldate = {2025-10-16}
}

@article{Kruse1997,
  title = {Kinetics of {{C2 Reactions}} during {{High-Temperature Pyrolysis}} of {{Acetylene}}},
  author = {Kruse, T. and Roth, P.},
  year = {1997},
  month = mar,
  journal = {J. Phys. Chem. A},
  volume = {101},
  number = {11},
  pages = {2138--2146},
  publisher = {American Chemical Society},
  issn = {1089-5639},
  doi = {10.1021/jp963373o},
  urldate = {2025-10-16}
}

@article{Matsugi2011,
  title = {Deuterium Kinetic Isotope Effects on the Gas-Phase Reactions of {{C2H}} with {{H2}}({{D2}}) and {{CH4}}({{CD4}})},
  author = {Matsugi, Akira and Suma, Kohsuke and Miyoshi, Akira},
  year = {2011},
  month = feb,
  journal = {Phys. Chem. Chem. Phys.},
  volume = {13},
  number = {9},
  pages = {4022--4031},
  publisher = {The Royal Society of Chemistry},
  issn = {1463-9084},
  doi = {10.1039/C0CP02056G},
  urldate = {2025-10-16},
  langid = {english}
}

@article{Opansky1996,
  title = {Rate {{Coefficients}} of {{C2H}} with {{C2H4}}, {{C2H6}}, and {{H2}} from 150 to 359 {{K}}},
  author = {Opansky, Brian J. and Leone, Stephen R.},
  year = {1996},
  month = jan,
  journal = {J. Phys. Chem.},
  volume = {100},
  number = {51},
  pages = {19904--19910},
  publisher = {American Chemical Society},
  issn = {0022-3654},
  doi = {10.1021/jp9619604},
  urldate = {2025-10-16}
}

@article{Peeters1996,
  title = {Absolute {{Rate Coefficients}} of the {{Reactions}} of {{C2H}} with {{NO}} and {{H2}} between 295 and 440 {{K}}},
  author = {Peeters, Jozef and Van Look, Hilde and Ceursters, Benny},
  year = {1996},
  month = jan,
  journal = {J. Phys. Chem.},
  volume = {100},
  number = {37},
  pages = {15124--15129},
  publisher = {American Chemical Society},
  issn = {0022-3654},
  doi = {10.1021/jp960201i},
  urldate = {2025-10-16}
}

@article{Peeters2002,
  title = {The Reaction of {{C2H}} with {{H2}}: {{Absolute}} Rate Coefficient Measurements and Ab Initio Study},
  shorttitle = {The Reaction of {{C2H}} with {{H2}}},
  author = {Peeters, Jozef and Ceursters, Benny and Nguyen, Hue Minh Thi and Nguyen, Minh Tho},
  year = {2002},
  month = mar,
  journal = {J. Chem. Phys.},
  volume = {116},
  number = {9},
  pages = {3700--3709},
  publisher = {AIP Publishing},
  issn = {0021-9606},
  doi = {10.1063/1.1436481},
  urldate = {2025-10-16},
  langid = {english}
}

@article{Stephens1987,
  title = {Rate Constant Measurements of Reactions of Ethynyl Radical with Hydrogen, Oxygen, Acetylene and Nitric Oxide Using Color Center Laser Kinetic Spectroscopy},
  author = {Stephens, J. W. and Hall, Jeffrey L. and Solka, H. and Yan, W. B. and Curl, Robert F. and Glass, G. P.},
  year = {1987},
  month = oct,
  journal = {J. Phys. Chem.},
  volume = {91},
  number = {22},
  pages = {5740--5743},
  publisher = {American Chemical Society},
  issn = {0022-3654},
  doi = {10.1021/j100306a044},
  urldate = {2025-10-16}
}

@article{Xue2025,
  title = {The {{Molecular Inventory}} of {{TMC-1}} with {{GOTHAM Observations}}},
  author = {Xue, Ci and Byrne, Alex N. and Morgan, Larry and Wenzel, Gabi and Changala, P. Bryan and Fried, Zachary T. P. and Loomis, Ryan A. and Remijan, Anthony and Bergin, Edwin A. and Cooke, Ilsa R. and Frayer, David and Burkhardt, Andrew M. and Charnley, Steven B. and Cordiner, Martin A. and Lipnicky, Andrew and McCarthy, Michael C. and McGuire, Brett A.},
  year = 2025,
  month = oct,
  journal = {ApJS},
  volume = {281},
  number = {1},
  pages = {9},
  publisher = {The American Astronomical Society},
  issn = {0067-0049},
  doi = {10.3847/1538-4365/ae04e5},
  urldate = {2025-11-04},
  langid = {english}
}

@article{Albers1975,
  title = {Absolute Rate Coefficients for the Reaction of {{H-atoms}} with {{N2O}} and Some Reactions of {{CN}} Radicals},
  author = {Albers, E. A. and Hoyermann, K. and Schacke, H. and Schmatjko, K. J. and Wagner, H. {\relax Gg}. and Wolfrum, J.},
  year = 1975,
  month = jan,
  journal = { Symp. Combust. Proc.},
  series = {Fifteenth {{Symposium}} ({{International}}) on {{Combustion}}},
  volume = {15},
  number = {1},
  pages = {765--773},
  issn = {0082-0784},
  doi = {10.1016/S0082-0784(75)80345-6},
  urldate = {2025-11-06}
}

@article{Balla1987,
  title = {Kinetics of Gas-Phase Cyanogen by Diode Laser Absorption},
  author = {Balla, R. {\relax Jeffrey}. and Pasternack, L.},
  year = 1987,
  month = jan,
  journal = {J. Phys. Chem.},
  volume = {91},
  number = {1},
  pages = {73--78},
  publisher = {American Chemical Society},
  issn = {0022-3654},
  doi = {10.1021/j100285a019},
  urldate = {2025-11-06}
}

@article{Boden1997,
  title = {Kinetics of Reactions Involving {{CN}} Emission {{IV}}. {{Study}} of the Reactions of {{CN}} by Electronic Absorption Spectroscopy},
  author = {Boden, J. C. and Thrush, Brian Arthur},
  year = 1997,
  month = jan,
  journal = {Proc. R. Soc. Lond. A. Math. Phys. Sci.},
  volume = {305},
  number = {1480},
  pages = {107--123},
  publisher = {Royal Society},
  doi = {10.1098/rspa.1968.0108},
  urldate = {2025-11-06}
}

@article{Choi2004,
  title = {H Atom Branching Ratios from the Reactions of {{CN}} Radicals with {{C2H2}} and {{C2H4}}},
  author = {Choi, N. and Blitz, M. A. and McKee, K. and Pilling, M. J. and Seakins, P. W.},
  year = 2004,
  month = jan,
  journal = {Chem. Phys. Lett.},
  volume = {384},
  number = {1},
  pages = {68--72},
  issn = {0009-2614},
  doi = {10.1016/j.cplett.2003.11.100},
  urldate = {2025-11-06}
}

@article{DeJuan1987,
  title = {Pulsed Photolysis-Laser-Induced Fluorescence Measurements of the Rate Constants for Reactions of the Cyanogen Radical with Hydrogen, Oxygen, Ammonia, Hydrogen Chloride, Hydrogen Bromide, and Hydrogen Iodide},
  author = {De Juan, {\relax Julian}. and Smith, Ian W. M. and Veyret, B.},
  year = 1987,
  month = jan,
  journal = {J. Phys. Chem.},
  volume = {91},
  number = {1},
  pages = {69--72},
  publisher = {American Chemical Society},
  issn = {0022-3654},
  doi = {10.1021/j100285a018},
  urldate = {2025-11-06}
}

@article{He1998,
  title = {Thermal {{Rate Constant}} for {{CN}} + {{H2}}/{{D2}} {$\rightarrow$} {{HCN}}/{{DCN}} + {{H}}/{{D Reaction}} from {{T}} = 293 to 380 {{K}}},
  author = {He, G. and Tokue, I. and Macdonald, R. Glen},
  year = 1998,
  month = jun,
  journal = {J. Phys. Chem. A},
  volume = {102},
  number = {24},
  pages = {4585--4591},
  publisher = {American Chemical Society},
  issn = {1089-5639},
  doi = {10.1021/jp980875o},
  urldate = {2025-11-06}
}

@article{Jacobs1989,
  title = {Measurements of Absolute Rate Coefficients for the Reactions of {{CN}} Radicals with {{H2O}}, {{H2}}, and {{CO2}} in the Temperature Range 295 {{K}}{$\leq$}{{T}}{$\leq$}1027 {{K}}},
  author = {Jacobs, A. and Wahl, M. and Weller, R. and Wolfrum, J.},
  year = 1989,
  month = jan,
  journal = { Symp. Combust. Proc.},
  volume = {22},
  number = {1},
  pages = {1093--1100},
  issn = {0082-0784},
  doi = {10.1016/S0082-0784(89)80119-5},
  urldate = {2025-11-06}
}

@article{Li1984,
  title = {Laser Measurements of the Effects of Vibrational Energy on the Reactions of {{CN}}},
  author = {Li, Xuechu and Sayah, Nahid and Jackson, William M.},
  year = 1984,
  month = jul,
  journal = {J. Chem. Phys.},
  volume = {81},
  number = {2},
  pages = {833--840},
  issn = {0021-9606},
  doi = {10.1063/1.447717},
  urldate = {2025-11-06}
}

@article{Lichtin1985,
  title = {Kinetics of {{CN}} Radical Reactions with Selected Molecules at Room Temperature},
  author = {Lichtin, D. A. and Lin, M. C.},
  year = 1985,
  month = jul,
  journal = {Chem. Phys.},
  volume = {96},
  number = {3},
  pages = {473--482},
  issn = {0301-0104},
  doi = {10.1016/0301-0104(85)85109-0},
  urldate = {2025-11-06}
}

@article{Natarajan1988,
  title = {A Shock Tube Study of {{CN}} Radical Reactions with {{H2}} and {{NO}} Verified by {{H}}, {{N}} and {{O}} Atom Measurements},
  author = {Natarajan, K. and Roth, P.},
  year = 1988,
  month = jan,
  journal = { Symp. Combust. Proc.},
  series = {Twenty-{{First Symposuim}} ({{International}} on {{Combustion}})},
  volume = {21},
  number = {1},
  pages = {729--737},
  issn = {0082-0784},
  doi = {10.1016/S0082-0784(88)80305-9},
  urldate = {2025-11-06}
}

@article{Schacke1977,
  title = {{Reaktionen von Molek\"ulen in definierten Schwingungszust\"anden (IV). Reaktionen schwingungsangeregter Cyan-Radikale mit Wasserstoff und einfachen Kohlenwasserstoffen}},
  author = {Schacke, H. and Wagner, H. {\relax Gg}. and Wolfrum, J.},
  year = 1977,
  journal = {Ber. Bunsenges. Phys. Chem.},
  volume = {81},
  number = {7},
  pages = {670--676},
  issn = {0005-9021},
  doi = {10.1002/bbpc.19770810709},
  urldate = {2025-11-06},
  copyright = {Copyright \copyright{} 1977 Wiley-VCH Verlag GmbH \& Co. KGaA, Weinheim},
  langid = {ngerman}
}

@article{Sims1988b,
  title = {Rate Constants for the Reactions {{CN}}({$\nu$}=0), {{CN}}({$\nu$}=1)+{{H2}}, {{D2}}{$\rightarrow$}{{HCN}}, {{DCN}}+{{H}}, {{D}} between 295 and 768 {{K}}, and Comparisons with Transition State Theory Calculations},
  author = {Sims, Ian R. and Smith, Ian W. M.},
  year = 1988,
  month = sep,
  journal = {Chem. Phys. Lett.},
  volume = {149},
  number = {5},
  pages = {565--571},
  issn = {0009-2614},
  doi = {10.1016/0009-2614(88)80384-1},
  urldate = {2025-11-06}
}

@article{Sun1990,
  title = {Experimental and Reduced Dimensionality Quantum Rate Coefficients for {{H2}}({{D2}})+{{CN}}{$\rightarrow$}{{H}}({{D}}){{CN}}+{{H}}({{D}})},
  author = {Sun, Q. and Yang, D. L. and Wang, N. S. and Bowman, J. M. and Lin, M. C.},
  year = 1990,
  month = oct,
  journal = {J. Chem. Phys.},
  volume = {93},
  number = {7},
  pages = {4730--4739},
  issn = {0021-9606},
  doi = {10.1063/1.458663},
  urldate = {2025-11-06}
}

@article{Anderson1998,
  title = {Infrared Spectroscopy and Time-Resolved Dynamics of the Ortho-{{H2}}--{{OH}} Entrance Channel Complex},
  author = {Anderson, David T. and Schwartz, Rebecca L. and Todd, Michael W. and Lester, Marsha I.},
  year = 1998,
  month = sep,
  journal = {J. Chem. Phys.},
  volume = {109},
  number = {9},
  pages = {3461--3473},
  issn = {0021-9606},
  doi = {10.1063/1.476941},
  urldate = {2025-11-27}
}

@article{Hernandez1995,
  title = {Electronic Spectra of the {{OH}}({{A2$\Sigma$}}+)-{{H2}} and {{OH}}({{A2$\Sigma$}}+)-{{D2}} Complexes},
  author = {Hern{\'a}ndez, Ram{\'o}n and Clary, David C.},
  year = 1995,
  month = oct,
  journal = {Chem. Phys. Lett.},
  volume = {244},
  number = {5},
  pages = {421--426},
  issn = {0009-2614},
  doi = {10.1016/0009-2614(95)00944-Y},
  urldate = {2025-11-27}
}

@article{Krause1998,
  title = {Time-Resolved Dissociation of the {{H2}}--{{OH}} Entrance Channel Complex},
  author = {Krause, Paul J. and Clary, David C. and Anderson, David T. and Todd, Michael W. and Schwartz, Rebecca L. and Lester, Marsha I.},
  year = 1998,
  month = sep,
  journal = {Chem. Phys. Lett.},
  volume = {294},
  number = {6},
  pages = {518--522},
  issn = {0009-2614},
  doi = {10.1016/S0009-2614(98)00869-0},
  urldate = {2025-11-27}
}

@article{Lester1997,
  title = {Electronic {{Quenching}} of {{OH A 2$\Sigma$}}+ (v` = 0, 1) in {{Complexes}} with {{Hydrogen}} and {{Nitrogen}}},
  author = {Lester, Marsha I. and Loomis, Richard A. and Schwartz, Rebecca L. and Walch, Stephen P.},
  year = 1997,
  month = dec,
  journal = {J. Phys. Chem. A},
  volume = {101},
  number = {49},
  pages = {9195--9206},
  publisher = {American Chemical Society},
  issn = {1089-5639},
  doi = {10.1021/jp9727557},
  urldate = {2025-11-27}
}

@article{Loomis1996a,
  title = {Electronic Spectroscopy and Quenching Dynamics of {{OH}}--{{H2}}/{{D2}} Pre-reactive Complexes},
  author = {Loomis, Richard A. and Schwartz, Rebecca L. and Lester, Marsha I.},
  year = 1996,
  month = may,
  journal = {J. Chem. Phys.},
  volume = {104},
  number = {18},
  pages = {6984--6996},
  publisher = {AIP Publishing},
  issn = {0021-9606},
  doi = {10.1063/1.471408},
  urldate = {2025-11-27},
  langid = {english}
}

@article{Schwartz1997a,
  title = {Infrared Spectroscopy of {{OH}}{{H2}} Entrance Channel Complexes},
  author = {Schwartz, Rebecca L. and Anderson, David T. and Todd, Michael W. and Lester, Marsha I.},
  year = 1997,
  month = jul,
  journal = {Chem. Phys. Lett.},
  volume = {273},
  number = {1},
  pages = {18--24},
  issn = {0009-2614},
  doi = {10.1016/S0009-2614(97)00597-6},
  urldate = {2025-11-27}
}

@article{Wheeler1999,
  title = {Stimulated {{Raman}} Excitation of the Ortho-{{H2}}--{{OH}} Entrance Channel Complex},
  author = {Wheeler, Martyn D. and Todd, Michael W. and Anderson, David T. and Lester, Marsha I.},
  year = 1999,
  month = apr,
  journal = {J. Chem. Phys.},
  volume = {110},
  number = {14},
  pages = {6732--6742},
  issn = {0021-9606},
  doi = {10.1063/1.478578},
  urldate = {2025-11-27}
}

@article{Chen1998,
  title = {Spectroscopy and Dynamics of the {{H2}}--{{CN}} van Der {{Waals}} Complex},
  author = {Chen, Yaling and Heaven, Michael C.},
  year = 1998,
  month = oct,
  journal = {J. Chem. Phys.},
  volume = {109},
  number = {13},
  pages = {5171--5174},
  issn = {0021-9606},
  doi = {10.1063/1.477132},
  urldate = {2025-11-27}
}

@article{Ruscic2005,
  title = {Active {{Thermochemical Tables}}: Thermochemistry for the 21st Century},
  shorttitle = {Active {{Thermochemical Tables}}},
  author = {Ruscic, Branko and Pinzon, Reinhardt E. and von Laszewski, Gregor and Kodeboyina, Deepti and Burcat, Alexander and Leahy, David and Montoy, David and Wagner, Albert F.},
  year = 2005,
  month = jan,
  journal = {J. Phys.: Conf. Ser.},
  volume = {16},
  number = {1},
  pages = {561},
  issn = {1742-6596},
  doi = {10.1088/1742-6596/16/1/078},
  urldate = {2025-11-27},
  langid = {english}
}

@article{Blitz2021,
  title = {Global {{Master Equation Analysis}} of {{Rate Data}} for the {{Reaction C2H4}} + {{H}} ⇄ {{C2H5}}: {{$\Delta$fH0}}{$\ominus$}{{C2H5}}},
  shorttitle = {Global {{Master Equation Analysis}} of {{Rate Data}} for the {{Reaction C2H4}} + {{H}} ⇄ {{C2H5}}},
  author = {Blitz, Mark A. and Pilling, Michael J. and Robertson, Struan H. and Seakins, Paul W. and Speak, Thomas H.},
  year = 2021,
  month = nov,
  journal = {J. Phys. Chem. A},
  volume = {125},
  number = {43},
  pages = {9548--9565},
  publisher = {American Chemical Society},
  issn = {1089-5639},
  doi = {10.1021/acs.jpca.1c05911},
  urldate = {2025-11-28}
}

@article{Blitz2025,
  title = {First {{Direct Observation}} of {{Equilibrium Involving Cl Atoms}}: {{Cl}} + {{C2H4}} {$\Leftrightarrow$} {{ClCH2CH2}} by {{VUV Monitoring}}},
  shorttitle = {First {{Direct Observation}} of {{Equilibrium Involving Cl Atoms}}},
  author = {Blitz, Mark A. and Speak, Thomas Henry and Seakins, Paul W.},
  year = 2025,
  month = oct,
  journal = {J. Phys. Chem. A},
  volume = {129},
  number = {42},
  pages = {9733--9744},
  publisher = {American Chemical Society},
  issn = {1089-5639},
  doi = {10.1021/acs.jpca.5c05430},
  urldate = {2025-11-28}
}

@article{Medeiros2018,
  title = {Kinetics of the {{Reaction}} of {{OH}} with {{Isoprene}} over a {{Wide Range}} of {{Temperature}} and {{Pressure Including Direct Observation}} of {{Equilibrium}} with the {{OH Adducts}}},
  author = {Medeiros, D. J. and Blitz, M. A. and James, L. and Speak, T. H. and Seakins, P. W.},
  year = 2018,
  month = sep,
  journal = {J. Phys. Chem. A},
  volume = {122},
  number = {37},
  pages = {7239--7255},
  publisher = {American Chemical Society},
  issn = {1089-5639},
  doi = {10.1021/acs.jpca.8b04829},
  urldate = {2025-11-28}
}

@article{Medeiros2020,
  title = {Direct {{Trace Fitting}} of {{Experimental Data Using}} the {{Master Equation}}: {{Testing Theory}} and {{Experiments}} on the {{OH}} + {{C2H4 Reaction}}},
  shorttitle = {Direct {{Trace Fitting}} of {{Experimental Data Using}} the {{Master Equation}}},
  author = {Medeiros, D. J. and Robertson, S. H. and Blitz, M. A. and Seakins, P. W.},
  year = 2020,
  month = may,
  journal = {J. Phys. Chem. A},
  volume = {124},
  number = {20},
  pages = {4015--4024},
  publisher = {American Chemical Society},
  issn = {1089-5639},
  doi = {10.1021/acs.jpca.0c02132},
  urldate = {2025-11-28}
}

@article{Cabezas2025a,
  title = {Discovery of Interstellar Phenalene (c-{{C13H10}}): {{A}} New Piece in the Chemical Puzzle of {{PAHs}} in Space},
  shorttitle = {Discovery of Interstellar Phenalene (c-{{C13H10}})},
  author = {Cabezas, C. and Ag{\'u}ndez, M. and P{\'e}rez, C. and {Villar-Castro}, D. and Molpeceres, G. and P{\'e}rez, D. and Steber, A. L. and Fuentetaja, R. and Tercero, B. and Marcelino, N. and Lesarri, A. and de Vicente, P. and Cernicharo, J.},
  year = 2025,
  month = sep,
  journal = {A\&A},
  volume = {701},
  pages = {L8},
  publisher = {EDP Sciences},
  issn = {0004-6361, 1432-0746},
  doi = {10.1051/0004-6361/202556687},
  urldate = {2025-11-29},
  copyright = {\copyright{} The Authors 2025},
  langid = {english}
}

@article{Zhang2012,
  title = {Prediction of {{Reaction Barriers}} and {{Thermochemical Properties}} with {{Explicitly Correlated Coupled-Cluster Methods}}: {{A Basis Set Assessment}}},
  shorttitle = {Prediction of {{Reaction Barriers}} and {{Thermochemical Properties}} with {{Explicitly Correlated Coupled-Cluster Methods}}},
  author = {Zhang, Jinmei and Valeev, Edward F.},
  year = 2012,
  month = sep,
  journal = {J. Chem. Theory Comput.},
  volume = {8},
  number = {9},
  pages = {3175--3186},
  publisher = {American Chemical Society},
  issn = {1549-9618},
  doi = {10.1021/ct3005547},
  urldate = {2026-01-27}
}

@article{Wheeler1999a,
author = {Wheeler, Martyn D. and Anderson, David T. and Todd, Michael W. and Lester, Marsha I. and Krause, Paul J. and Clary, David C.},
title = {Mode-selective decay dynamics of the ortho-H2—OH complex: experiment and theory},
journal = {Mol. Phys.},
volume = {97},
number = {1-2},
pages = {151--158},
year = {1999},
publisher = {Taylor \& Francis},
doi = {10.1080/00268979909482817},
URL = {https://doi.org/10.1080/00268979909482817},
eprint = {https://doi.org/10.1080/00268979909482817}
}

@article{Wheeler1999b,
    author = {Wheeler, Martyn D. and Todd, Michael W. and Anderson, David T. and Lester, Marsha I.},
    title = {Stimulated Raman excitation of the ortho-H2–OH entrance channel complex},
    journal = {J. Chem. Phys.},
    volume = {110},
    number = {14},
    pages = {6732-6742},
    year = {1999},
    month = {04},
    issn = {0021-9606},
    doi = {10.1063/1.478578},
    url = {https://doi.org/10.1063/1.478578},
    eprint = {https://pubs.aip.org/aip/jcp/article-pdf/110/14/6732/19184373/6732_1_online.pdf},
}

@article{Yaling1998,
    author = {Chen, Yaling and Heaven, Michael C.},
    title = {Spectroscopy and dynamics of the H2–CN van der Waals complex},
    journal = {J. Chem. Phys.},
    volume = {109},
    number = {13},
    pages = {5171-5174},
    year = {1998},
    month = {10},
    issn = {0021-9606},
    doi = {10.1063/1.477132},
    url = {https://doi.org/10.1063/1.477132},
    eprint = {https://pubs.aip.org/aip/jcp/article-pdf/109/13/5171/19256433/5171_1_online.pdf},
}

@article{Brutschy2000,
author = {Brutschy, Bernhard and Hobza, Pavel},
title = {van der Waals Molecules III:  Introduction},
journal = {Chem. Rev.},
volume = {100},
number = {11},
pages = {3861-3862},
year = {2000},
doi = {10.1021/cr990074x},
    note ={PMID: 11749331},
URL = {  
        https://doi.org/10.1021/cr990074x
},
eprint = { 
          https://doi.org/10.1021/cr990074x
}
}

@article{Nesbitt1988,
author = {Nesbitt, David J.},
title = {High-resolution infrared spectroscopy of weakly bound molecular complexes},
journal = {Chem. Rev.},
volume = {88},
number = {6},
pages = {843-870},
year = {1988},
doi = {10.1021/cr00088a003},
URL = {  
        https://doi.org/10.1021/cr00088a003
},
eprint = {    
        https://doi.org/10.1021/cr00088a003
}
}

@article{Fawzy2005,
    author = {Fawzy, Wafaa M. and Kerenskaya, Galina and Heaven, Michael C.},
    title = {Experimental detection and theoretical characterization of the H2–NH(X) van der Waals complex},
    journal = {J. Chem. Phys.},
    volume = {122},
    number = {14},
    pages = {144318},
    year = {2005},
    month = {04},
    issn = {0021-9606},
    doi = {10.1063/1.1879932},
    url = {https://doi.org/10.1063/1.1879932},
    eprint = {https://pubs.aip.org/aip/jcp/article-pdf/doi/10.1063/1.1879932/15363856/144318_1_online.pdf},
}

@article{Tarabukin2021,
title = {Rotational spectroscopy and bound state calculations of deuterated NH3–H2 van der Waals complexes},
journal = {J. Mol. Spectrosc.},
volume = {377},
pages = {111442},
year = {2021},
issn = {0022-2852},
doi = {https://doi.org/10.1016/j.jms.2021.111442},
url = {https://www.sciencedirect.com/science/article/pii/S0022285221000266},
author = {I.V. Tarabukin and L.A. Surin and M. Hermanns and B. Heyne and S. Schlemmer and K.L.K. Lee and M.C. McCarthy and A. {van der Avoird}}
}

@article{Surin2017,
doi = {10.3847/1538-4357/aa615a},
url = {https://doi.org/10.3847/1538-4357/aa615a},
year = {2017},
month = {mar},
publisher = {The American Astronomical Society},
volume = {838},
number = {1},
pages = {27},
author = {Surin, L. A. and Tarabukin, I. V. and Schlemmer, S. and Breier, A. A. and Giesen, T. F. and McCarthy, M. C. and Avoird, A. van der},
title = {Rotational Spectroscopy of the NH3–H2 Molecular Complex},
journal = {ApJ}
}

@article{Fuentetaja2026a,
  title = {Discovery of {{1H-cyclopent}}[Cd]Indene (c-{{C}}\_{{11H}}\_8) in {{TMC-1}} with the {{QUIJOTE}} Line Survey: {{A}} New Three-Ringed Polycyclic Aromatic Hydrocarbon},
  shorttitle = {Discovery of {{1H-cyclopent}}[Cd]Indene (c-{{C}}\_{{11H}}\_8) in {{TMC-1}} with the {{QUIJOTE}} Line Survey},
  author = {Fuentetaja, R. and Cabezas, C. and Ag{\'u}ndez, M. and Tercero, B. and Marcelino, N. and de Vicente, P. and Cernicharo, J.},
  year = 2026,
  month = jan,
  journal = {A\&A},
  publisher = {EDP Sciences},
  issn = {0004-6361, 1432-0746},
  doi = {10.1051/0004-6361/202558398},
  urldate = {2026-02-02},
  copyright = {\copyright{} 2026, ESO},
  langid = {english}
}

@article{Cernicharo2026a,
  title = {Discovery of Two New Isomers of Cyanoacenaphthylene ({{C}}{\textsubscript{12}} {{H}}{\textsubscript{7}} {{CN}}) in the {{Taurus}} Molecular Cloud 1 with the {{QUIJOTE}} Line Survey},
  author = {Cernicharo, J. and Tercero, B. and Marcelino, N. and {L{\'o}pez-P{\'e}rez}, J. A. and Gallego, J. D. and Tercero, F. and Esplugues, G. and Cabezas, C. and Ag{\'u}ndez, M. and Limeres, C. and Steber, A. L. and P{\'e}rez, D. and P{\'e}rez, C. and Lesarri, A. and De Vicente, P.},
  year = 2026,
  month = jan,
  journal = {A\&A},
  volume = {705},
  pages = {L7},
  issn = {0004-6361, 1432-0746},
  doi = {10.1051/0004-6361/202557893},
  urldate = {2026-02-10},
  copyright = {https://creativecommons.org/licenses/by/4.0},
  langid = {english}
}

\appendix
\section{Calculations of Collision Limits and Enthalpies of formation}\label{CollisionLimit-RCs_SI}
The dipole moments and polarizabilities which were necessary for calculating the collision limit rate coefficient values in Table \ref{tab:Reactions} were taken from the CCCBDB. If experimental values were available then they were used, but if not then a calculated value was, where either B2PLYP, B2PLYP=Full, or B2PLYP=Full and ultrafine in combination with either cc-pVDZ, cc-pVTZ, cc-pVQZ, aug-cc-pVDZ, aug-cc-pVTZ, or aug-cc-pVQZ were the method and basis set. If no such calculated value existed in CCCBDB, then these were calculated using B2PLYP-D3BJ with the aug-cc-pVTZ basis set. Ionization energies were also necessary for collision limit rate coefficients, and these were taken from the National Institute of Standards and Technology (NIST) Chemistry WebBook. Where ionization energies were unavailable in the NIST Chemistry WebBook then the vertical ionization energy was calculated by doing a single point energy correction on the B2PLYP-D3BJ, aug-cc-pVTZ geometry optimized molecule and ionized molecule with CCSD(T) and aug-cc-pVTZ. If this was unachievable then CCSD(T) with cc-pVDZ-F12 was used instead. 

As described in Section \ref{Sec:Screening}, some chemical species included in reactions within Table \ref{tab:Reactions} currently have no thermochemical data listed in ATcT. To determine the enthalpy of formation for these species, quantum chemical calculations were conducted. For the molecules: C$_{6}$, C$_{8}$, C$_{8}$H, and C$_{8}$H$_{2}$, structures and zero point energies at the B2PLYP-D3/aug-cc-pVTZ level with single point energy corrections using CCSD(T)-F12/cc-pVTZ-F12 were calculated to determine their enthalpies of formation. For the molecules \ce{C10}, and \ce{C10H} CCSD(T)-F12/cc-pVDZ-F12 was used instead of CCSD(T)-F12/cc-pVTZ-F12. Table \ref{tab:FormationEnthalpies} lists the calculated formation enthalpies for each of these species.

\begin{deluxetable}{cc}[H]
\tablenum{3}
\tablecaption{Calculated enthalpies of formation for chemical species contained within Table \ref{tab:Reactions} which do not have such data listed in ATcT (version 1.220).}
\label{tab:FormationEnthalpies}
\tablehead{\colhead{Molecule} & \colhead{$\Delta$H$^{\circ}$$_{\text{f}}$ (0 K) (kJ mol$^{-1}$)}}
\startdata
C$_{6}$ & 1267.9 \\
C$_{8}$ & 1498.0 \\
C$_{8}$H & 1251.9 \\
C$_{8}$H$_{2}$ & 904.2 \\
C$_{10}$ & 1851.5 \\
C$_{10}$H & 1614.1 \\
\enddata
\tablecomments{These values were used in the calculation of reaction enthalpies for reactions 37--39, and 54 in Table \ref{tab:Reactions} to determine if a reaction was exothermic or endothermic.}
\end{deluxetable}

\section{Aromatic Abundance Uncertainties Associated with Tunneling}\label{AAUAWT}
\begin{figure}
    \centering  \includegraphics[width=0.7\linewidth]{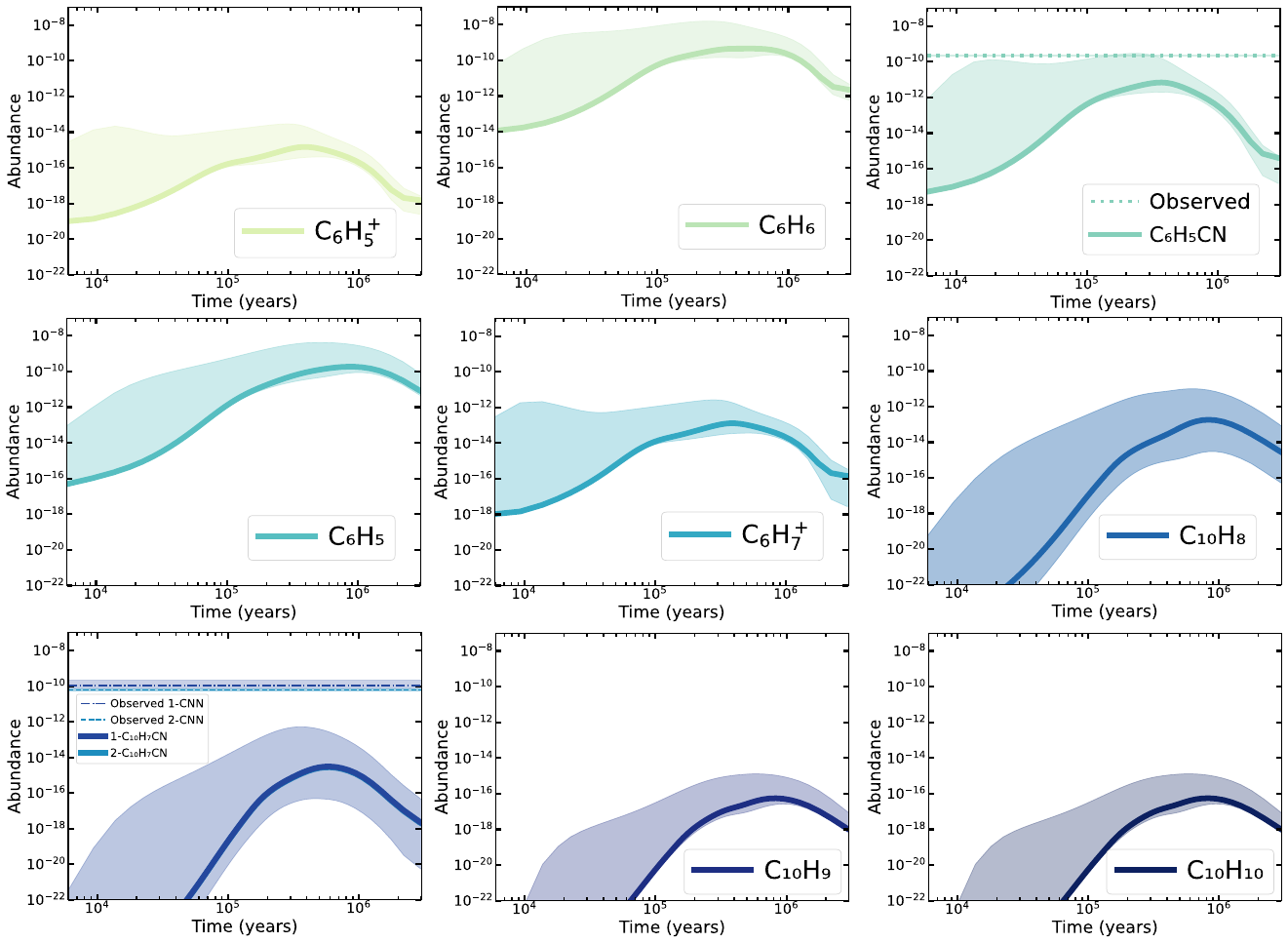}
    \caption{The abundance time profiles of all 10 aromatic molecules analyzed in this work using Network 1. The shaded regions on each plot represent the estimated abundance uncertainty due to dormant hydrogen atom transfer reactions found in kida.uva.2024 and Network 1.}
    \label{Grouped-Plots1}
\end{figure}

\begin{figure}
    \centering  \includegraphics[width=0.7\linewidth]{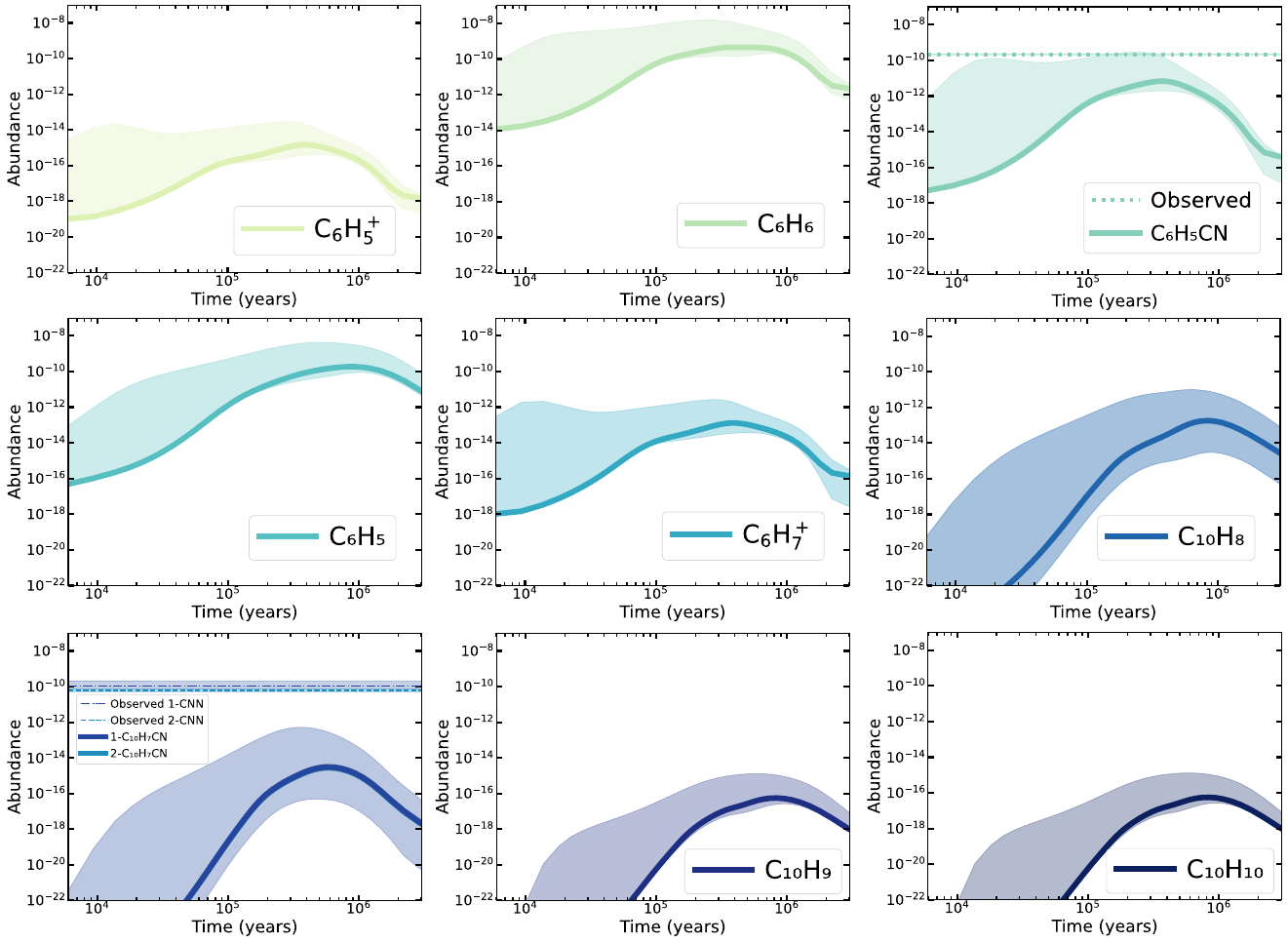}
    \caption{The same plots as displayed in Figure \ref{Grouped-Plots1}, but using Network 2, where the shaded regions also represent the estimated abundance uncertainty due to dormant hydrogen atom transfer reactions.}
    \label{Grouped-Plots2}
\end{figure}

Unlike \ce{C6H5CN}, 1-CNN, and 2-CNN, the molecules: c-C$_{6}$H$_{5}$$^{+}$, \ce{C6H6}, C$_{6}$H$_{7}$$^{+}$, \ce{C6H5}, \ce{C10H8}, C$_{10}$H$_{9}$, and C$_{10}$H$_{10}$ have not been observed in the ISM and therefore have no observed abundance to compare with their modeled abundances. Consequently, the level of agreement between their modeled abundances and their actual abundances in the ISM is unknown. Given the deviation between the modeling of \ce{C6H5CN}, 1-CNN, and 2-CNN and their observed abundance values, there could be similar or even larger discrepancies for c-C$_{6}$H$_{5}$$^{+}$, \ce{C6H6}, C$_{6}$H$_{7}$$^{+}$, \ce{C6H5}, \ce{C10H8}, C$_{10}$H$_{9}$, and C$_{10}$H$_{10}$ due to the various chemical pathways which link them to \ce{C6H5CN}, 1-CNN, and 2-CNN. In spite of their currently unknown modeled--observed abundance agreement, the current modeled abundances of c-C$_{6}$H$_{5}$$^{+}$, \ce{C6H6}, C$_{6}$H$_{7}$$^{+}$, \ce{C6H5}, \ce{C10H8}, C$_{10}$H$_{9}$, and C$_{10}$H$_{10}$ are vulnerable to dormant hydrogen atom tunneling reactions in Network 1 and kida.uva.2024. Figures \ref{Grouped-Plots1} and \ref{Grouped-Plots2} display the notable uncertainties in the modeled abundances of these molecules which exist solely from these hydrogen atom tunneling reactions. Between the time period of 1--5 $\times$ 10$^{5}$ years the maximal increase and decrease to these molecules using both Networks 1 and 2 are shown in Table \ref{tab:Abundance-changes}. Additionally, the direct, maximal effect that each reaction shown in Table \ref{tab:Reactions} has on the molecules: \ce{C6H5CN}, 1-CNN, 2-CNN, c-C$_{6}$H$_{5}$$^{+}$, \ce{C6H6}, C$_{6}$H$_{7}$$^{+}$, \ce{C6H5}, \ce{C10H8}, C$_{10}$H$_{9}$, and C$_{10}$H$_{10}$ is shown as well in Table \ref{tab:Table1-Impacts}. As described in Section \ref{subsec:"Tunnellable Reactions"}, the reaction numbers 7, 10, 11, 13, 35, 36, 52, and 57 in Table \ref{tab:Reactions} have the most notable impact on the modeled molecular abundances of \ce{C6H5CN}, 1-CNN, 2-CNN, c-C$_{6}$H$_{5}$$^{+}$, \ce{C6H6}, C$_{6}$H$_{7}$$^{+}$, \ce{C6H5}, \ce{C10H8}, C$_{10}$H$_{9}$, and C$_{10}$H$_{10}$ when either Network 1 or 2 is used.

\begin{deluxetable}{ccccc}[H]
\tablenum{4}
\renewcommand{\arraystretch}{0.7} 
\setlength{\tabcolsep}{18pt}       
\tablecaption{Maximal modeled molecular abundance increases and decreases for c-C$_{6}$H$_{5}$$^{+}$, \ce{C6H6}, C$_{6}$H$_{7}$$^{+}$, \ce{C6H5}, \ce{C10H8}, C$_{10}$H$_{9}$, and C$_{10}$H$_{10}$ during the period of 1--5 $\times$ 10$^{5}$ years.}
\label{tab:Abundance-changes}
\tablehead{\colhead{Molecule} & 
\colhead{\makecell{Maximal Increase \\ ($\times$ more)}} & \colhead{\makecell{Time of Maximal \\ Increase (years)}} & \colhead{\makecell{Maximal Decrease \\ ($\times$ less)}} & \colhead{\makecell{Time of Maximal \\ Decrease (years)}}
}
\startdata
\cutinhead{Network 1}
\ce{C6H6} & 100 & 1 $\times$ 10$^{5}$ & 3 & 4 $\times$ 10$^{5}$ \\
\ce{C6H5+} & 70 & 1 $\times$ 10$^{5}$ & 4 & 3 $\times$ 10$^{5}$ \\
\ce{C6H7+} & 80 & 1 $\times$ 10$^{5}$ & 4 & 3 $\times$ 10$^{5}$ \\
\ce{C6H5} & 300 & 1 $\times$ 10$^{5}$ & 3 & 5 $\times$ 10$^{5}$ \\
\ce{C10H8} & 20000 & 1 $\times$ 10$^{5}$ & 50 & 5 $\times$ 10$^{5}$ \\
\ce{C10H9} & 4000 & 1 $\times$ 10$^{5}$ & 3 & 5 $\times$ 10$^{5}$ \\
\ce{C10H10} & 4000 & 1 $\times$ 10$^{5}$ & 3 & 5 $\times$ 10$^{5}$ \\
\cutinhead{Network 2}
\ce{C6H6} & 200 & 1 $\times$ 10$^{5}$ & 3 & 4 $\times$ 10$^{5}$ \\
\ce{C6H5+} & 90 & 1 $\times$ 10$^{5}$ & 4 & 3 $\times$ 10$^{5}$ \\
\ce{C6H7+} & 100 & 1 $\times$ 10$^{5}$ & 4 & 3 $\times$ 10$^{5}$ \\
\ce{C6H5} & 500 & 1 $\times$ 10$^{5}$ & 3 & 5 $\times$ 10$^{5}$ \\
\ce{C10H8} & 50000 & 1 $\times$ 10$^{5}$ & 70 & 5 $\times$ 10$^{5}$ \\
\ce{C10H9} & 7000 & 1 $\times$ 10$^{5}$ & 3 & 5 $\times$ 10$^{5}$ \\
\ce{C10H10} & 7000 & 1 $\times$ 10$^{5}$ & 3 & 5 $\times$ 10$^{5}$
\enddata
\tablecomments{Abundance changes are rounded to one significant figure, as are the simulated chemical ages of TMC-1. The maximal changes to the modeled abundances of \ce{C6H5CN}, 1-CNN, and 2-CNN during 1--5 $\times$ 10$^{5}$ years are discussed in Section \ref{AromaticImpact}.}
\end{deluxetable}
\vspace{-2cm}
\startlongtable
\begin{deluxetable*}{ccccc}
\tablenum{5}
\tablecaption{Of the 59 simulations conducted for each reaction in Table \ref{tab:Reactions} (using the \textit{k}\textsubscript{Coll, 10 K} values also found in Table \ref{tab:Reactions}, asides from Reactions \ref{eq:1}-\ref{eq:4}, where MESMER values were used instead), the maximally impacted species, the factor they are changed by, and the time of maximal change are shown here for both Networks 1 and 2. The impacted species here are limited to c-C$_{6}$H$_{5}$$^{+}$, \ce{C6H6}, C$_{6}$H$_{7}$$^{+}$, \ce{C6H5}, \ce{C6H5CN}, \ce{C10H8}, \ce{1-C10H7CN}, \ce{2-C10H7CN}, \ce{C10H9}, and \ce{C10H10}. \label{tab:Table1-Impacts}}
\tablehead{\colhead{Reaction} & 
\colhead{Maximally Impacted Species} & \colhead{Abundance Change ($\times$)} & 
\colhead{Increase or Decrease} & \colhead{Time of Maximal Change (years)}
}
\startdata
\cutinhead{Network 1}
2 & \ce{C10H9} & 2 & Increase & 1 $\times$ 10$^{6}$ \\
3 & \ce{1-C10H7CN} & 1 & Increase & 9 $\times$ 10$^{6}$ \\
4 & \ce{1-C10H7CN} & 1 & Decrease & 9 $\times$ 10$^{6}$ \\
5 & \ce{1-C10H7CN} & 1 & Decrease & 7 $\times$ 10$^{4}$ \\
6  & \ce{2-C10H7CN} & 1 & Decrease & 6 $\times$ 10$^{4}$ \\
7  & \ce{1-C10H7CN} & 83 & Decrease & 9 $\times$ 10$^{3}$ \\
8  & \ce{C10H8} & 1 & Decrease & 4 $\times$ 10$^{5}$ \\
9  & \ce{1-C10H7CN} & 1 & Increase & 5 $\times$ 10$^{3}$ \\
10  & \ce{1-C10H7CN} & 1107 & Increase & 5 $\times$ 10$^{3}$ \\
11  & \ce{2-C10H7CN} & 229 & Increase & 5 $\times$ 10$^{3}$ \\
12  & \ce{2-C10H7CN} & 1 & Increase & 9 $\times$ 10$^{3}$ \\
13  & \ce{1-C10H7CN} & 3 & Decrease & 5 $\times$ 10$^{3}$ \\
14  & \ce{1-C10H7CN} & 1 & Decrease & 9 $\times$ 10$^{6}$ \\
15 & \ce{2-C10H7CN} & 1 & Decrease & 9 $\times$ 10$^{5}$ \\
16 & \ce{1-C10H7CN} & 1 & Decrease & 9 $\times$ 10$^{6}$ \\
17 & N/A & N/A & N/A & N/A \\
18 & \ce{2-C10H7CN} & 1 & Decrease & 7 $\times$ 10$^{4}$ \\
19 & \ce{2-C10H7CN} & 1 & Increase & 4 $\times$ 10$^{4}$ \\
20 & \ce{1-C10H7CN} & 1 & Increase & 5 $\times$ 10$^{3}$ \\
21 & \ce{2-C10H7CN} & 1 & Decrease & 4 $\times$ 10$^{5}$ \\
22 & \ce{C10H8} & 1 & Decrease & 3 $\times$ 10$^{5}$ \\
23 & \ce{2-C10H7CN} & 1 & Increase & 9 $\times$ 10$^{3}$ \\
24 & \ce{C6H5CN} & 1 & Increase & 5 $\times$ 10$^{4}$ \\
25 & N/A & N/A & N/A & N/A \\
26 & \ce{1-C10H7CN} & 2 & Decrease & 9 $\times$ 10$^{6}$ \\
27 & \ce{C6H6} & 1 & Increase & 2 $\times$ 10$^{6}$ \\
28 & \ce{1-C10H7CN} & 1 & Decrease & 5 $\times$ 10$^{3}$ \\
29 & \ce{1-C10H7CN} & 2 & Decrease & 9 $\times$ 10$^{6}$ \\
30 & \ce{C6H7+} & 2 & Decrease & 4 $\times$ 10$^{6}$ \\
31 & \ce{C6H5CN} & 1 & Increase & 5 $\times$ 10$^{5}$ \\
32 & \ce{1-C10H7CN} & 1 & Decrease & 1 $\times$ 10$^{6}$ \\
33 & \ce{C6H6} & 1 & Increase & 1 $\times$ 10$^{6}$ \\
34 & \ce{2-C10H7CN} & 1 & Increase & 1 $\times$ 10$^{6}$ \\
35 & \ce{2-C10H7CN} & 357 & Increase & 5 $\times$ 10$^{4}$ \\
36 & \ce{1-C10H7CN} & 1449 & Increase & 5 $\times$ 10$^{4}$ \\
37 & \ce{1-C10H7CN} & 3 & Increase & 6 $\times$ 10$^{4}$ \\
38 & \ce{1-C10H7CN} & 2 & Increase & 5 $\times$ 10$^{4}$ \\
39 & \ce{1-C10H7CN} & 1 & Increase & 6 $\times$ 10$^{4}$ \\
40 & \ce{C6H5CN} & 1 & Decrease & 2 $\times$ 10$^{6}$ \\
41 & \ce{2-C10H7CN} & 1 & Increase & 1 $\times$ 10$^{5}$ \\
42 & \ce{2-C10H7CN} & 1 & Increase & 9 $\times$ 10$^{6}$ \\
43 & \ce{C10H8} & 1 & Decrease & 5 $\times$ 10$^{5}$ \\
44 & \ce{C6H7+} & 1 & Decrease & 9 $\times$ 10$^{6}$ \\
45 & \ce{C10H8} & 1 & Increase & 1 $\times$ 10$^{6}$ \\
46 & \ce{2-C10H7CN} & 1 & Decrease & 9 $\times$ 10$^{5}$ \\
47 & \ce{1-C10H7CN} & 1 & Increase & 2 $\times$ 10$^{6}$ \\
48 & \ce{1-C10H7CN} & 1 & Increase & 9 $\times$ 10$^{6}$ \\
49 & \ce{1-C10H7CN} & 1 & Decrease & 9 $\times$ 10$^{6}$ \\
50 & \ce{2-C10H7CN} & 1 & Decrease & 7 $\times$ 10$^{4}$ \\
51 & \ce{C6H7+} & 1 & Increase & 9 $\times$ 10$^{6}$ \\
52 & \ce{1-C10H7CN} & 4 & Increase & 3 $\times$ 10$^{5}$ \\
53 & \ce{C10H9} & 1 & Increase & 9 $\times$ 10$^{6}$ \\
54 & \ce{C10H8} & 1 & Increase & 2 $\times$ 10$^{5}$ \\
55 & \ce{C6H7+} & 1 & Decrease & 9 $\times$ 10$^{6}$ \\
56 & \ce{C10H8} & 1 & Decrease & 7 $\times$ 10$^{5}$ \\
57 & \ce{C10H8} & 3 & Decrease & 5 $\times$ 10$^{5}$ \\
58 & \ce{2-C10H7CN} & 1 & Decrease & 7 $\times$ 10$^{4}$ \\
59 & \ce{2-C10H7CN} & 1 & Increase & 1 $\times$ 10$^{5}$ \\
60 & \ce{C10H9} & 1 & Decrease & 2 $\times$ 10$^{5}$ \\
61 & \ce{C6H7+} & 1 & Decrease & 2 $\times$ 10$^{6}$ \\
62 & \ce{C6H7+} & 1 & Decrease & 9 $\times$ 10$^{6}$ \\
\cutinhead{Network 2}
2 & \ce{C10H9} & 2 & Increase & 1 $\times$ 10$^{6}$ \\
3 & \ce{1-C10H7CN} & 1 & Increase & 9 $\times$ 10$^{6}$ \\
4 & \ce{1-C10H7CN} & 1 & Decrease & 9 $\times$ 10$^{6}$ \\
5 & \ce{2-C10H7CN} & 1 & Increase & 6 $\times$ 10$^{4}$ \\
6  & \ce{2-C10H7CN} & 1 & Increase & 7 $\times$ 10$^{4}$ \\
7  & \ce{1-C10H7CN} & 83 & Decrease & 9 $\times$ 10$^{3}$ \\
8  & \ce{C10H8} & 1 & Decrease & 4 $\times$ 10$^{5}$ \\
9  & \ce{1-C10H7CN} & 1 & Increase & 5 $\times$ 10$^{3}$ \\
10  & \ce{1-C10H7CN} & 1071 & Increase & 5 $\times$ 10$^{3}$ \\
11  & \ce{2-C10H7CN} & 1181 & Increase & 5 $\times$ 10$^{3}$ \\
12  & \ce{2-C10H7CN} & 1 & Increase & 9 $\times$ 10$^{3}$ \\
13  & \ce{1-C10H7CN} & 3 & Decrease & 5 $\times$ 10$^{3}$ \\
14  & \ce{1-C10H7CN} & 1 & Decrease & 9 $\times$ 10$^{6}$ \\
15 & \ce{C10H10} & 1 & Decrease & 9 $\times$ 10$^{6}$ \\
16 & \ce{1-C10H7CN} & 1 & Decrease & 9 $\times$ 10$^{6}$ \\
17 & N/A & N/A & N/A & N/A \\
18 & \ce{2-C10H7CN} & 1 & Increase & 7 $\times$ 10$^{4}$ \\
19 & \ce{2-C10H7CN} & 1 & Increase & 7 $\times$ 10$^{4}$ \\
20 & \ce{1-C10H7CN} & 1 & Increase & 5 $\times$ 10$^{4}$ \\
21 & \ce{1-C10H7CN} & 1 & Decrease & 3 $\times$ 10$^{5}$ \\
22 & \ce{C10H8} & 1 & Decrease & 3 $\times$ 10$^{5}$ \\
23 & \ce{2-C10H7CN} & 1 & Increase & 9 $\times$ 10$^{3}$ \\
24 & \ce{1-C10H7CN} & 1 & Increase & 5 $\times$ 10$^{4}$ \\
25 & N/A & N/A & N/A & N/A \\
26 & \ce{1-C10H7CN} & 2 & Decrease & 9 $\times$ 10$^{6}$ \\
27 & \ce{C6H6} & 1 & Increase & 2 $\times$ 10$^{6}$ \\
28 & \ce{1-C10H7CN} & 1 & Decrease & 5 $\times$ 10$^{3}$ \\
29 & \ce{1-C10H7CN} & 2 & Decrease & 9 $\times$ 10$^{6}$ \\
30 & \ce{C6H7+} & 3 & Decrease & 4 $\times$ 10$^{6}$ \\
31 & \ce{C6H5CN} & 1 & Increase & 5 $\times$ 10$^{5}$ \\
32 & \ce{2-C10H7CN} & 1 & Increase & 6 $\times$ 10$^{4}$ \\
33 & \ce{C6H6} & 1 & Increase & 1 $\times$ 10$^{6}$ \\
34 & \ce{1-C10H7CN} & 1 & Increase & 9 $\times$ 10$^{6}$ \\
35 & \ce{2-C10H7CN} & 770 & Increase & 5 $\times$ 10$^{4}$ \\
36 & \ce{1-C10H7CN} & 3156 & Increase & 5 $\times$ 10$^{4}$ \\
37 & \ce{1-C10H7CN} & 3 & Increase & 5 $\times$ 10$^{4}$ \\
38 & \ce{1-C10H7CN} & 2 & Increase & 5 $\times$ 10$^{4}$ \\
39 & \ce{1-C10H7CN} & 1 & Increase & 5 $\times$ 10$^{4}$ \\
40 & \ce{C6H5CN} & 1 & Decrease & 2 $\times$ 10$^{6}$ \\
41 & \ce{2-C10H7CN} & 1 & Increase & 5 $\times$ 10$^{4}$ \\
42 & \ce{1-C10H7CN} & 1 & Increase & 9 $\times$ 10$^{6}$ \\
43 & \ce{C6H6} & 1 & Decrease & 4 $\times$ 10$^{5}$ \\
44 & \ce{C6H7+} & 1 & Decrease & 9 $\times$ 10$^{6}$ \\
45 & \ce{C10H8} & 1 & Increase & 1 $\times$ 10$^{6}$ \\
46 & \ce{2-C10H7CN} & 1 & Decrease & 9 $\times$ 10$^{5}$ \\
47 & \ce{1-C10H7CN} & 1 & Increase & 9 $\times$ 10$^{6}$ \\
48 & \ce{2-C10H7CN} & 1 & Increase & 9 $\times$ 10$^{6}$ \\
49 & \ce{1-C10H7CN} & 1 & Decrease & 9 $\times$ 10$^{6}$ \\
50 & \ce{2-C10H7CN} & 1 & Increase & 1 $\times$ 10$^{5}$ \\
51 & \ce{C6H7+} & 1 & Increase & 9 $\times$ 10$^{6}$ \\
52 & \ce{1-C10H7CN} & 5 & Increase & 3 $\times$ 10$^{5}$ \\
53 & \ce{1-C10H7CN} & 1 & Increase & 7 $\times$ 10$^{4}$ \\
54 & \ce{C6H7+} & 1 & Decrease & 5 $\times$ 10$^{5}$ \\
55 & \ce{C6H7+} & 1 & Decrease & 9 $\times$ 10$^{6}$ \\
56 & \ce{C10H8} & 1 & Decrease & 7 $\times$ 10$^{5}$ \\
57 & \ce{C10H8} & 3 & Decrease & 5 $\times$ 10$^{5}$ \\
58 & \ce{2-C10H7CN} & 1 & Increase & 7 $\times$ 10$^{4}$ \\
59 & \ce{2-C10H7CN} & 1 & Increase & 7 $\times$ 10$^{4}$ \\
60 & \ce{C10H9} & 1 & Decrease & 2 $\times$ 10$^{5}$ \\
61 & \ce{C6H7+} & 1 & Decrease & 2 $\times$ 10$^{6}$ \\
62 & \ce{C6H7+} & 1 & Decrease & 9 $\times$ 10$^{6}$ \\
\enddata
\tablecomments{Abundance changes and times of maximal changes are rounded to one significant figure. The considered chemical ages of TMC-1 correspond to the beginning of each simulation to 9 $\times$ 10$^{6}$ years.}
\end{deluxetable*}

\section{Rate Coefficient Calculations}\label{RCC_SI}
\begin{figure}[H]
    \centering  \includegraphics[width=0.77\linewidth]{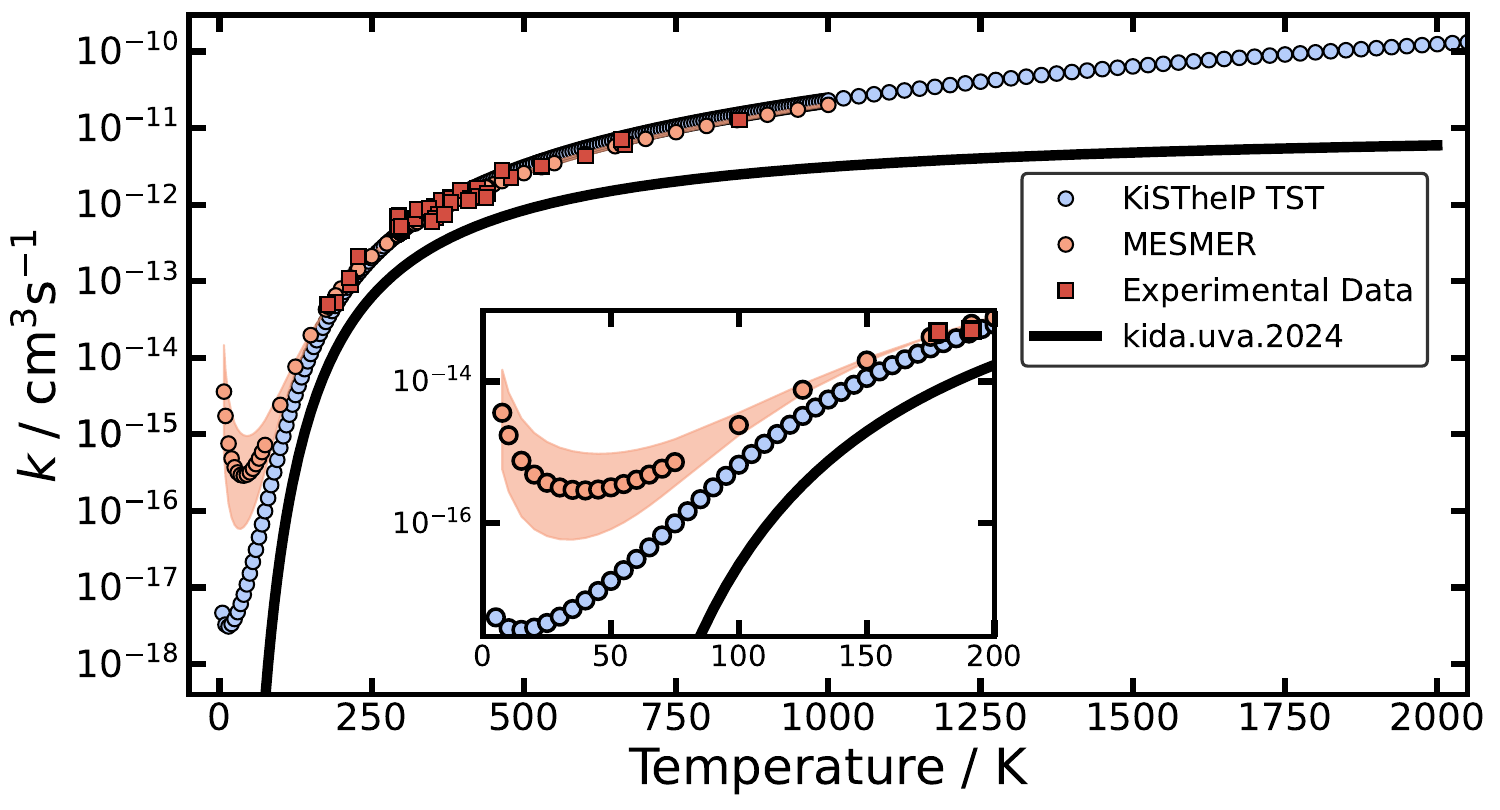}
    \caption{Displayed are the calculated KiSThelP TST, and MESMER rate coefficients for Reaction \ref{eq:1}, compared with experimental data, and kida.uva.2024—using the $\alpha$, $\beta$, and $\gamma$ values supplied within the reaction network. The inset more clearly shows the temperature range relevant to the ISM. The shaded peach region represents the uncertainty in the calculated MESMER values, which were determined as described in Section \ref{MESMER}. The MESMER data was fitted to the experimental data included in this plot, which includes data from: \citet{Stephens1987}, \citet{Koshi1992}, \citet{Koshi1992a}, \citet{Farhat1993}, \citet{Peeters1996}, \citet{Opansky1996}, \citet{Kruse1997}, \citet{Peeters2002}, \citet{Kovacs2010}, \citet{Matsugi2011}.} 
    \label{C2H+H2-RCs_SI}
\end{figure}

\begin{figure}[H]
    \centering  \includegraphics[width=0.77\linewidth]{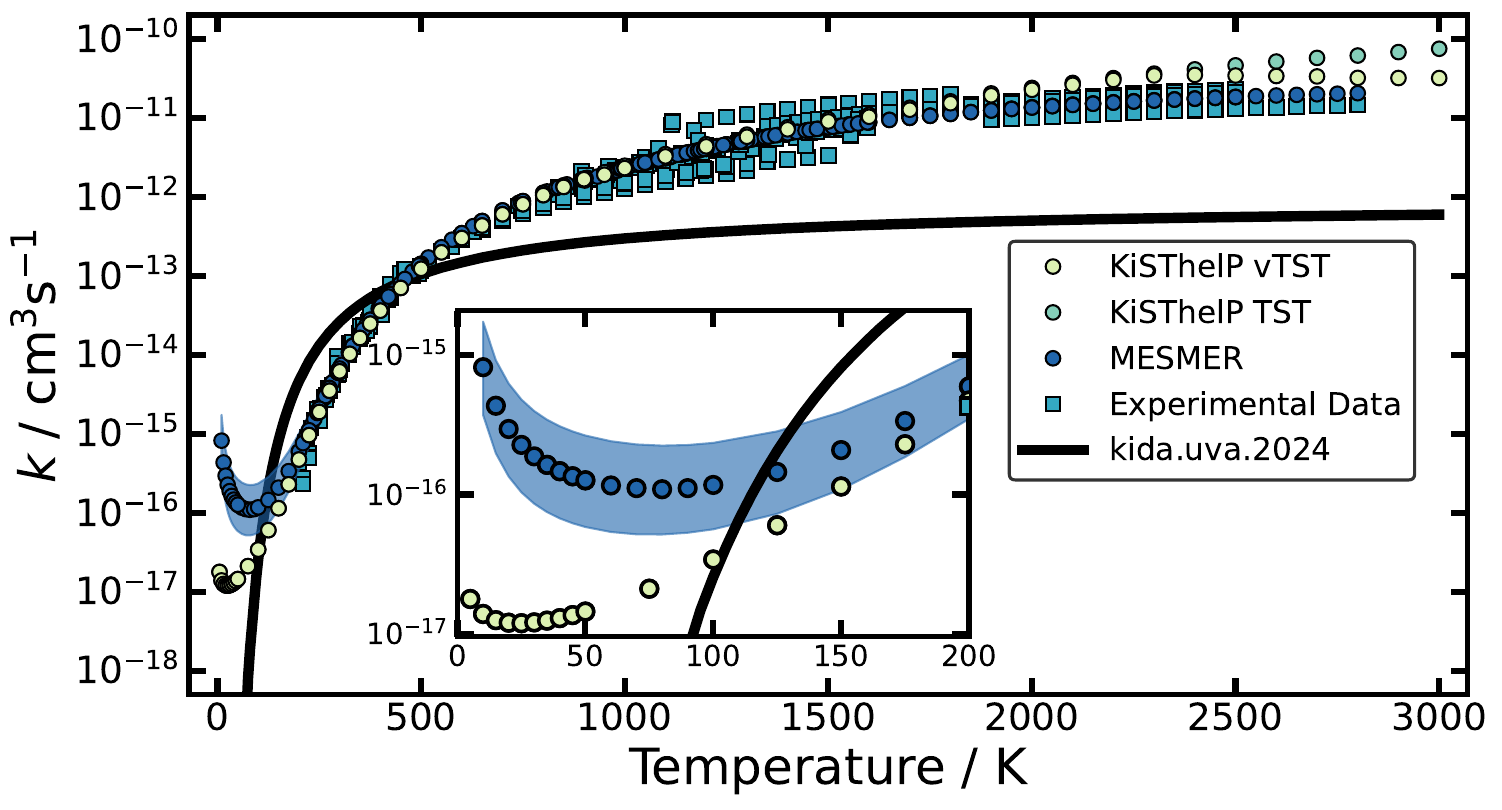}
    \caption{Displayed are the calculated KiSThelP vTST, KiSThelP TST, and MESMER rate coefficients for Reaction \ref{eq:2}, compared with experimental data, and kida.uva.2024—using the $\alpha$, $\beta$, and $\gamma$ values supplied within the reaction network. The inset more clearly shows the temperature range relevant to the ISM, and in it the KiSThelP vTST and TST values are overlaid. The shaded blue region represents the uncertainty in the calculated MESMER values, which were determined as described in Section \ref{MESMER}. The MESMER data was fitted to the experimental data included in this plot, which includes data from: \citet{Ripley1966}, \citet{Greiner1969}, \citet{Brabbs1971}, \citet{Eberius1971}, \citet{Westenberg1973}, \citet{Smith1973}, \citet{W.C.Gardiner1973}, \citet{Vandooren1975}, \citet{Atkinson1975}, \citet{Tully1980}, \citet{Ravishankara1981}, \citet{Schmidt1984}, \citet{Frank1985}, \citet{Roth1985}, \citet{Bott1989}, \citet{Oldenborg1992}, \citet{Talukdar1996}, \citet{Krasnoperov2004}, \citet{Baulch2005}, \citet{Orkin2006}, and \citet{Lam2013}.}
    \label{OH+H2-RCs_SI}
\end{figure}

\begin{figure}[H]
    \centering  \includegraphics[width=0.77\linewidth]{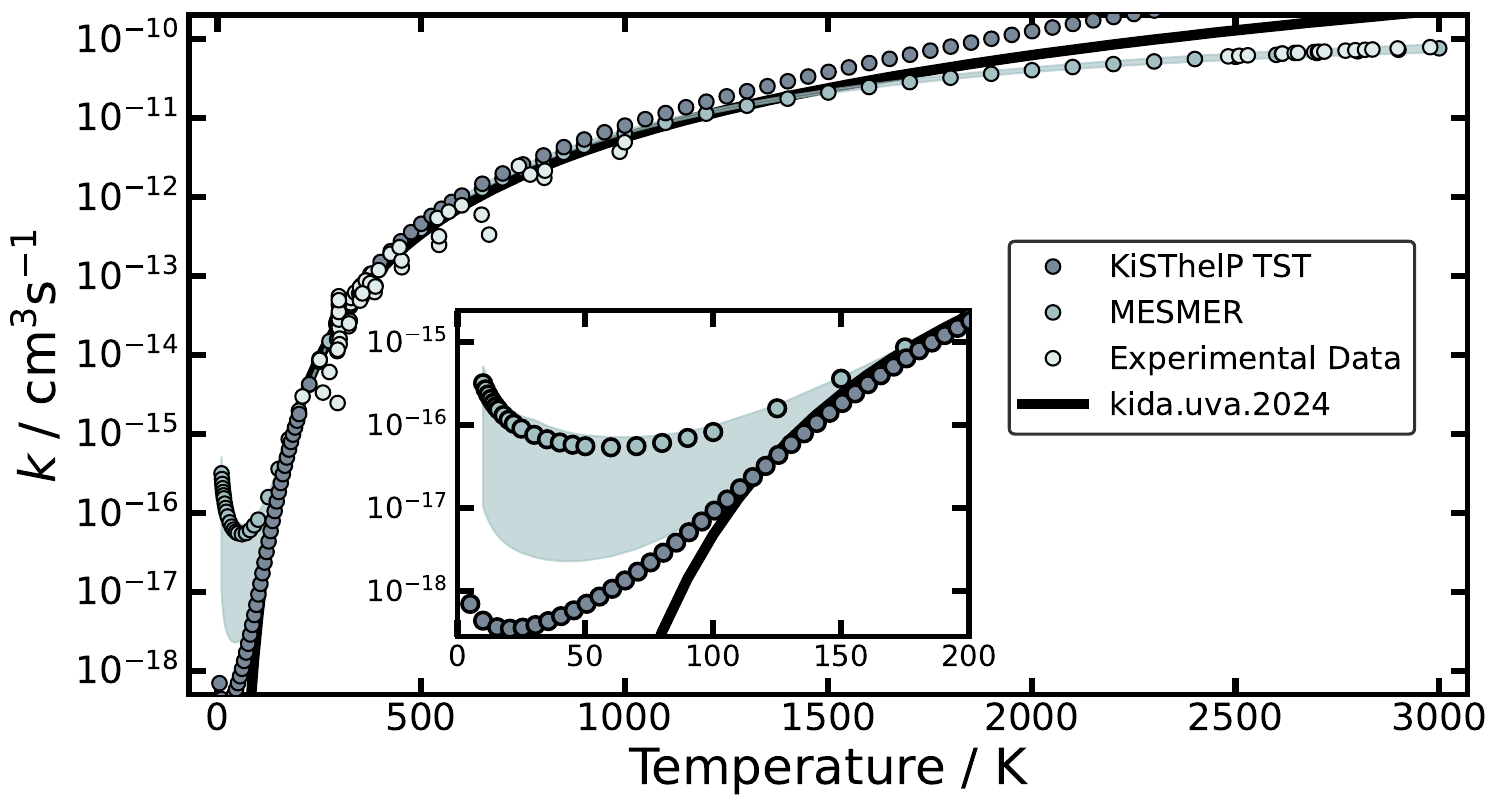}
    \caption{Displayed are the calculated KiSThelP TST (dark grey), and MESMER rate coefficients (grey) for Reaction \ref{eq:3}, compared with experimental data (light grey), and kida.uva.2024 (black line)—using the $\alpha$, $\beta$, and $\gamma$ values supplied within the reaction network. The inset more clearly shows the temperature range relevant to the ISM, and in it the KiSThelP vTST and TST values are overlaid. The shaded grey region represents the uncertainty in the calculated MESMER values, which were determined as described in Section \ref{MESMER}. The MESMER data was fitted to the experimental data included in this plot, which includes data from \citet{Albers1975}, \citet{Schacke1977}, \citet{Li1984}, \citet{Lichtin1985}, \citet{Balla1987}, \citet{DeJuan1987}, \citet{Natarajan1988}, \citet{Sims1988b}, \citet{Jacobs1989}, \citet{Sun1990}, \citet{Boden1997}, \citet{He1998}, and \citet{Choi2004}.}
    \label{CN+H2-RCs_SI}
\end{figure}

\begin{figure}[H]
    \centering  \includegraphics[width=0.75\linewidth]{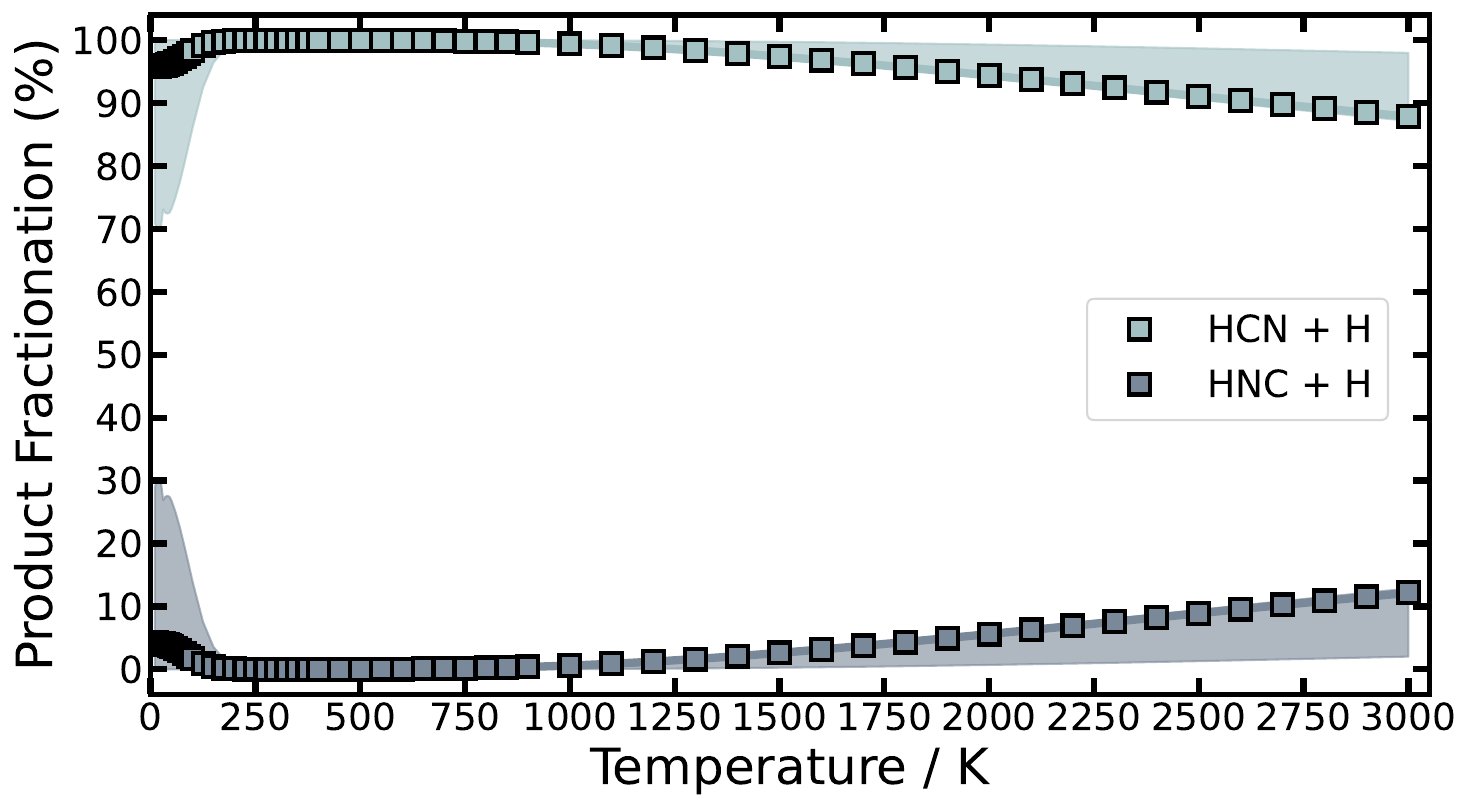}
    \caption{Shown is the MESMER calculated temperature dependence of the product branching ratio for the CN + \ce{H2} reaction from 10 to 3000 K. The predicted fraction of HCN + H formation grey, while HNC + H formation is dark grey. The shaded regions represent the associated uncertainty in the calculated product fractionation percentages.}
    \label{CN+H2-BR_SI}
\end{figure}

\begin{figure}[H]
    \centering  \includegraphics[width=0.77\linewidth]{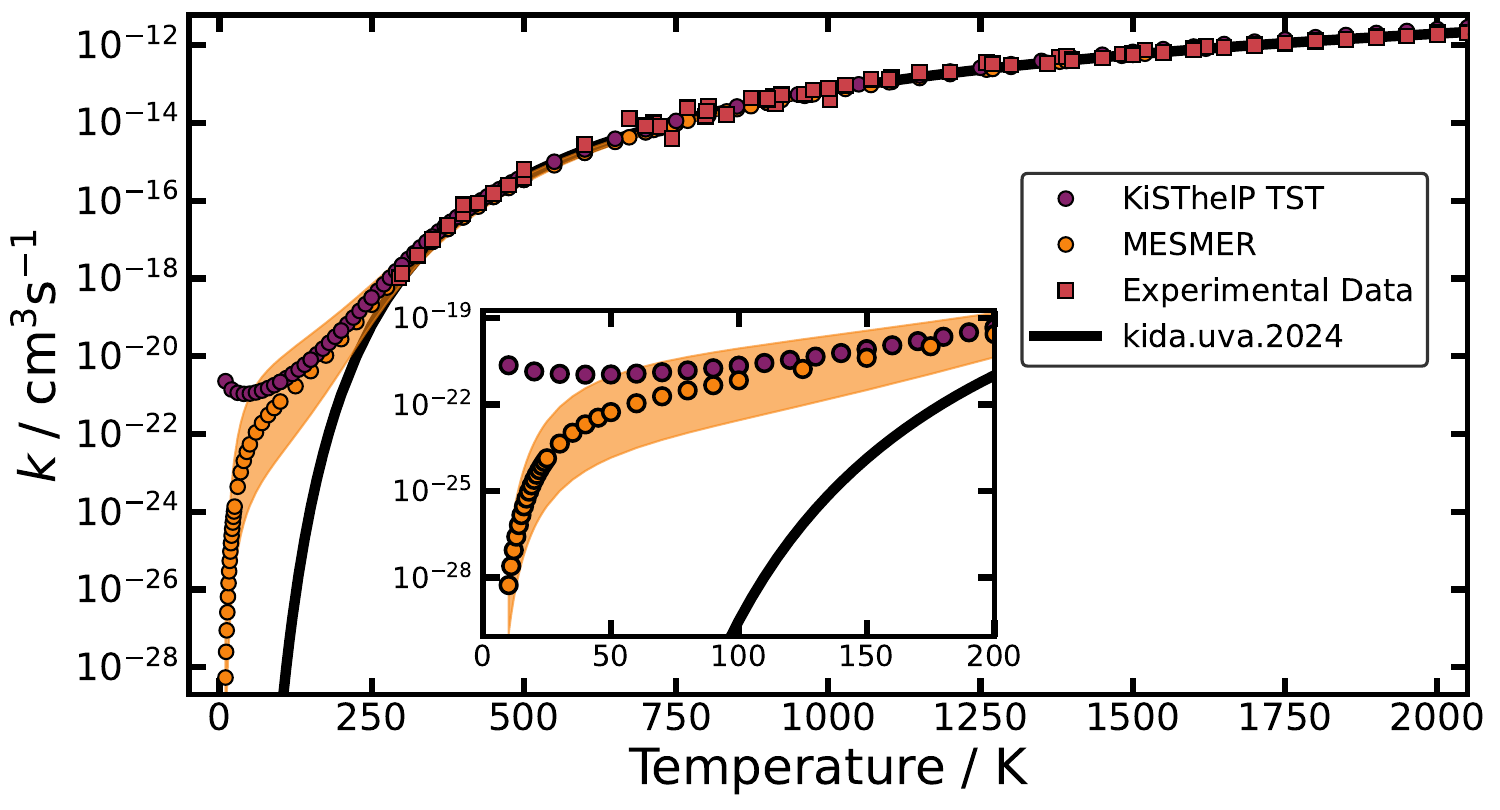}
    \caption{Displayed are the calculated KiSThelP TST, and MESMER rate coefficients for Reaction \ref{eq:4}, compared with experimental data, and kida.uva.2024—using the $\alpha$, $\beta$, and $\gamma$ values supplied within the reaction network. The inset more clearly shows the temperature range relevant to the ISM. The shaded orange region represents the uncertainty in the calculated MESMER values, which were determined as described in Section \ref{MESMER}. The MESMER data was fitted to the experimental data included in this plot, which includes data from: \citet{lesclaux1978}, \citet{Demissy1980}, \citet{Hack1986}, \citet{Sutherland1988}, and \citet{Friedrichs2000}.}
    \label{NH2+H2-RCs_SI}
\end{figure}

Included in Figures \ref{C2H+H2-RCs_SI}--\ref{CN+H2-RCs_SI}, and \ref{NH2+H2-RCs_SI} are the plotted temperature dependencies of the rate coefficients for Reactions \ref{eq:1}--\ref{eq:4}, along with the temperature dependence of the product fractionation for Reaction \ref{eq:3} in Figure \ref{CN+H2-BR_SI}. Within Figures \ref{C2H+H2-RCs_SI}--\ref{CN+H2-RCs_SI}, and \ref{NH2+H2-RCs_SI} are MESMER calculated rate coefficients, KiSThelP vTST and/or KiSThelP TST rate coefficients, experimental data from the literature (which MESMER values are fitted to), and the temperature dependencies of the $\alpha$, $\beta$, and $\gamma$ values for each reaction over a range of temperatures (from a couple thousand K to 5 K). At low temperatures ($< 500$ K) the difference between TST and vTST was found to have converged for Reactions \ref{eq:1}--\ref{eq:4}. When comparing predicted rate coefficients to experimental values using vTST, vTST showed improved accuracy at temperatures over 1500 K; for example in Figure \ref{OH+H2-RCs_SI} for Reaction \ref{eq:2}. 

The MESMER extrapolations for Reactions \ref{eq:1}--\ref{eq:4}, based on the best fit parameters provided in Table \ref{tab:MESMER_Params}, were conducted at a density of 2 $\times$ $10^{4}$ $\mathrm{cm^{-3}}$, where collisional stabilization into the Van der Waals complexes is not predicted. The experimental data which these calculations were fit to were measured over a wide range of different pressures, however showed no pressure dependence. At temperatures below 30 K for Reactions \ref{eq:1}--\ref{eq:3} and below 100 K for Reaction \ref{eq:4}, where the TST rate coefficients exhibit a mild upturn, canonical TST with a one-dimensional Eckart correction overestimates the tunneling probability. Here, a plateau region is expected instead of the observed upturn. This predicted upturn behavior is likely an artifact of the multiplicative tunneling correction where the tunneling transmission coefficient is increasing faster than the uncorrected TST rate coefficient falls, rather than a true physical acceleration of the low-temperature rate coefficients. In Reaction \ref{eq:4}, at low temperatures ($<$ 50 K) the MESMER predicted rate coefficients exhibit a sharp decrease. This decrease is linked to the microcanonical treatment of this system where at low temperatures only a small number of grains are populated. These energy grains have a low density of states and the number of accessible states in the transition state collapses. In addition, the energy level specific tunneling probabilities in the lowest energy grains become negligible. This results in the ``canonical'' rate coefficients calculated from Boltzmann averaging of the microcanonical rate coefficients for reaction \ref{eq:4} being significantly lower than the canonical TST values determined in KiSThelP. The combination of these factors, causes our low-temperature KiSThelP TST values to be larger than the MESMER values in Figure \ref{NH2+H2-RCs_SI} for temperatures below approximately 200 K.

\begin{deluxetable}{lcccc}[H]
\tablenum{6}
\tablecaption{Best fit parameters from MESMER analysis of experimental data for Reactions \ref{eq:1}--\ref{eq:4}.}
\label{tab:MESMER_Params}
\tablehead{
\colhead{} & 
\colhead{C$_{2}$H + H$_{2}$} & 
\colhead{OH + H$_{2}$} & 
\colhead{CN + H$_{2}$} & 
\colhead{NH$_{2}$ + H$_{2}$}
}
\startdata
\textbf{A (cm$^{3}$ s$^{-1}$)} & 2.43 (10.16) $\times$ 10$^{-9}$ & 1.352 (0.64) $\times$ 10$^{-10}$ & 3.18 (0.36) $\times$ 10$^{-10}$ & 6.92 (80.08) $\times$ 10$^{-10}$ \\
\textbf{n} & 0.702 (2.576) & -0.736 (0.201) & -0.256 (0.042) & 0.097 (44.15) \\
\textbf{VdW (kJ mol$^{-1}$)} & -0.923 (0.541) & -1.243 (1.346) & -0.889 (0.796) & -1.342 (0.437) \\
\textbf{TS (kJ mol$^{-1}$)} & 9.571 (0.328) & 26.347 (0.043) & 16.616 (0.021) & 47.65 (2.70) \\
\textbf{TS 2 (kJ mol$^{-1}$)} & N/A & N/A & 73.63 (3.54) & N/A \\
\textbf{$\nu$TS (cm$^{-1}$)} & 527 (44) & 1160 (17) & 688.2 (4.4) & 1557 (41) \\
\textbf{$\nu$TS 2 (cm$^{-1}$)} & N/A & N/A & 2811 (1197) & N/A \\
\tableline
$\chi$$^{2}$/n (Fitted) & 0.037 & 0.147 & 0.804 & 0.073 \\
$\chi$$^{2}$/n (Unfitted) & 0.296 & 0.557 & 1.027 & 0.100 \\
$\chi$$^{2}$/n (Modified Arrhenius) & 0.040 & 0.169 & 0.815 & 0.073 \\
\enddata
\tablecomments{Bolded are the parameters which were varied within MESMER calculations. A is the pre-exponential factor within the modified Arrhenius description of the barrierless formation of entrance channel complexes (Equation \ref{Modified_Arrhenius_Eq}), n is the temperature dependence of this A factor, VdW corresponds to $\Delta$H$_{0K}$ for Van der Waals complexes in Reactions \ref{eq:1}--\ref{eq:4} while TS is $\Delta$H$_{0K}$ for the transition states, and $\nu$TS corresponds to the imaginary frequency of the transition state. TS 2 and $\nu$TS 2 refer to the transition state associated with the formation of HNC + H for Reaction \ref{eq:3}. Uncertainties are given in parentheses. A measure of how well the MESMER simulations reproduce the experimental values ($\chi$$^{2}$/n) is provided for the fitted and unfitted surfaces with comparison to a standard modified Arrhenius fit.}
\end{deluxetable}


\section{Quantum Chemical Calculations}\label{QCC_SI}
Calculated vibrations were treated with a harmonic oscillator model and supported by individual treatment of some low frequency vibrational modes. Where torsional vibrational modes were identified these were replaced by either free or hindered rotors based on rotational barriers inferred from relaxed scans of dihedral angles representing the underlying torsional motion. Additionally, the low frequency mode (or modes) corresponding to formation (or fragmentation) of the loose complex were excluded from the zero point energies of the Van der Waals complexes. The result of these approaches was that the entrance channel complexes for Reactions \ref{eq:1}--\ref{eq:4} were assigned as -1.02, -0.21, -1.15, and -1.40 kJ mol$^{-1}$, respectively; whereas, with full harmonic zero point energy these were 1.68, 1.91, 0.72, and 1.96 kJ mol$^{-1}$, respectively. The presence of slightly submerged entrance channel complexes is in line with spectroscopy of loose complexes formed between H$_{2}$ and small molecules with dipole moments \citep{Brutschy2000,Nesbitt1988}. Such studies allowed for direct determination of the binding and dissociation energies for the ortho-H$_{2}$ and OH complex with a ground state dissociation energy of 0.65 kJ mol$^{-1}$ \citep{Loomis1995,Loomis1997,Krause1998,Wheeler1999a,Wheeler1999b,Anderson1998}, which matched Miller-Clary-Kliesch-Werner (MCKW) calculations \citep{Miller1994}. Spectroscopy was also carried out on the entrance channel complex formed in the reaction of CN and H$_{2}$ and its ground state dissociation energy was determined to be 0.45 kJ mol$^{-1}$ \citep{Yaling1998}. We know of no direct experimental study of the NH$_{2}$ and H$_{2}$ entrance channel complex but the dissociation energy of NH and H$_{2}$ has been determined to be 0.38 kJ mol$^{-1}$ \citep{Fawzy2005} and rotational spectroscopy of complexes formed between NH$_{3}$ and H$_{2}$ were found to have dissociation energies between 0.39 and 0.79 kJ mol$^{-1}$ \citep{Surin2017,Tarabukin2021}. In all the cases discussed, the dissociation energy was a much smaller value than the binding energy due to the difference in zero point energy in the complex and the reactants. Although, the choice to exclude modes representing translation into and out of the loose complexes from the zero point energies is not entirely justified, examination of the single point energy and zero point energy corrected scans of the entrance channel and approach to the abstraction barriers showed that defined minima exist relative to the reaction asymptotes in the local region, albeit emerged slightly compared to the reactants treated separately. In addition, the entrance channel well depths were floated in the fitting process in MESMER and TST rate coefficients excluding the entrance channel complexes were calculated.

For the reaction of \ce{C2H} + \ce{H2} (Reaction \ref{eq:1}), initial simulations with both KiSThelP and MESMER produced results in marked discrepancy with the experimental data. When comparing the MESMER and KiSThelP TST rate coefficient predictions at 10 K, (1.66 $\times$ 10$^{-15}$ and 3.27 $\times$ $10^{-18}$ $\mathrm{cm^{3}\,s^{-1}}$, respectively), the importance of the loose entrance channel complex is highlighted. An additional important factor to consider for this reaction was the impact of the low-frequency vibrational modes in the transition state. Upon visualization of the harmonic motion of these vibrational modes, it was apparent that a torsional mode for the \ce{H2} fragment was present. Replacing this with a hindered rotor treatment was therefore considered. The magnitude of the barrier for inclusion was determined by mapping the energy corresponding to this motion and fitting a sine curve. This generated a barrier height of $\approx$ 1 kJ mol$^{-1}$ which was included in both the MESMER and KiSThelP calculations. The resulting predictions equally well replicated the results achieved by fitting the lowest energy vibrational mode. Meanwhile, the results over the temperature range of the experimental data were relatively insensitive to the height of the hindered rotor barrier between 0.1 and 5 kJ mol$^{-1}$. The difference between a harmonic oscillator and hindered rotor treatment converges at low temperatures, where thermal energy is below the torsional barrier. The importance of including a hindered rotor treatment for torsional modes for the correct calculation of the density of states at elevated temperatures in MESMER has been demonstrated already for the reactions of OH, H and Cl with ethylene \citep{Medeiros2020, Blitz2021, Blitz2025}, and OH with isoprene \citep{Medeiros2018}. 

Included in Sections \ref{C2H+H2_SI2} to \ref{NH2+H2_SI2} are the cartesian coordinates of the optimized geometries (in Å), vibrational frequencies (in cm$^{-1}$), and rotational constants (in MHz) for each of the stationary points along the potential energy surfaces within Figures \ref{C2H+H2}, \ref{OH+H2}, \ref{CN+H2}, and \ref{NH2+H2}. The following data from the \textit{ab-initio} calculations for Reactions \ref{eq:1} and \ref{eq:4} were calculated with CCSD(T)/CBS//CCSD(T)-F12c/cc-pVTZ-F12, while the data for Reactions \ref{eq:2} and \ref{eq:3} were calculated with CCSD(T)/CBS//CCSD(T)-F12b/aug-cc-pVTZ.

\pagebreak
\subsection{\ce{C2H} + \ce{H2}}\label{C2H+H2_SI2}
\begin{table}[h!]
\tablenum{7}
\centering
\caption{Summary of the computed properties for \ce{C2H} in the CCSD(T)/CBS//CCSD(T)-F12c/cc-pVTZ-F12 set of calculations.}
\label{tab:moleculeX}
\end{table}
\vspace{-10mm}
\begin{minipage}{0.45\textwidth}
\centering
\begin{tabular}{l c c c}
\multicolumn{4}{c}{\textbf{Cartesian Coordinates (Angstroms)}} \\[3pt]
\toprule
Atom & {x} & {y} & {z} \\
\midrule
H &  0.000000 &  0.000000 &  1.540342 \\
C & -0.000000 & -0.000000 &  0.476269 \\
C &  0.000000 &  0.000000 & -0.731611 \\
\bottomrule
\end{tabular}
\end{minipage}
\hfill
\begin{minipage}{0.45\textwidth}
\centering
\begin{tabular}{c}
\multicolumn{1}{c}{\textbf{Vibrational Frequencies}} \\[3pt]
\toprule
$\nu$ (cm$^{-1}$)\\
\midrule
410.98 \\
410.98 \\
2058.76 \\
3450.47 \\
\bottomrule
\end{tabular}

\vspace{0.5cm}

\begin{tabular}{l c}
\multicolumn{2}{c}{\textbf{Rotational Constants (MHz)}} \\[3pt]
\toprule
 & Value \\
\midrule
A & 0.000000 \\
B & 44124.266940 \\
C & 44124.266940 \\
\bottomrule
\end{tabular}
\end{minipage}
\begin{table}[h!]
\tablenum{8}
\centering
\caption{Summary of the computed properties for \ce{H2} in the CCSD(T)/CBS//CCSD(T)-F12c/cc-pVTZ-F12 set of calculations.}
\label{tab:moleculeX}
\end{table}

\begin{minipage}{0.45\textwidth}
\centering
\begin{tabular}{l c c c}
\multicolumn{4}{c}{\textbf{Cartesian Coordinates (Angstroms)}} \\[3pt]
\toprule
Atom & {x} & {y} & {z} \\
\midrule
H & 0.000000 & 0.000000 &  0.370834 \\
H & 0.000000 & 0.000000 & -0.370834 \\
\bottomrule
\end{tabular}
\end{minipage}
\hfill
\begin{minipage}{0.45\textwidth}
\centering
\begin{tabular}{c}
\multicolumn{1}{c}{\textbf{Vibrational Frequencies}} \\[3pt]
\toprule
$\nu$ (cm$^{-1}$)\\
\midrule
4401.76 \\
\bottomrule
\end{tabular}

\vspace{0.5cm}

\begin{tabular}{l c}
\multicolumn{2}{c}{\textbf{Rotational Constants (MHz)}} \\[3pt]
\toprule
 & Value \\
\midrule
A & 0.000000 \\
B & 1822918.766075 \\
C & 1822918.766075 \\
\bottomrule
\end{tabular}
\end{minipage}
\pagebreak
\begin{table}[h!]
\tablenum{9}
\centering
\caption{Summary of the computed properties for the \ce{C2H + H2} Van der Waals complex in the CCSD(T)/CBS//CCSD(T)-F12c/cc-pVTZ-F12 set of calculations.}
\label{tab:moleculeX}
\end{table}

\begin{minipage}{0.45\textwidth}
\centering
\begin{tabular}{l c c c}
\multicolumn{4}{c}{\textbf{Cartesian Coordinates (Angstroms)}} \\[3pt]
\toprule
Atom & {x} & {y} & {z} \\
\midrule
C &  0.636519 & -0.679367 & -0.000281 \\
C &  0.372867 &  0.499466 &  0.000537 \\
H &  0.138511 &  1.537560 & -0.000286 \\
H & -2.680412 & -0.266110 &  0.000175 \\
H & -3.417748 & -0.179962 & -0.000142 \\
\bottomrule
\end{tabular}
\end{minipage}
\hfill
\begin{minipage}{0.45\textwidth}
\centering
\begin{tabular}{c}
\multicolumn{1}{c}{\textbf{Vibrational Frequencies}} \\[3pt]
\toprule
$\nu$ (cm$^{-1}$)\\
\midrule
  19.22 \\
  96.99 \\
 157.85 \\
 188.23 \\
 441.00 \\
 444.18 \\
2058.51 \\
3449.16 \\
4391.45 \\
\bottomrule
\end{tabular}


\begin{tabular}{l c}
\multicolumn{2}{c}{\textbf{Rotational Constants (MHz)}} \\[3pt]
\toprule
 & (MHz) \\
\midrule
A & 46435.015473 \\
B & 20805.529031 \\
C & 14367.896186 \\
\bottomrule
\end{tabular}
\end{minipage}
\vspace{-0.5cm}
\begin{table}[h!]
\tablenum{10}
\centering
\caption{Summary of the computed properties for the \ce{C2H + H2} transition state in the CCSD(T)/CBS//CCSD(T)-F12c/cc-pVTZ-F12 set of calculations.}
\label{tab:moleculeX}
\end{table}

\begin{minipage}{0.45\textwidth}
\centering
\begin{tabular}{l c c c}
\multicolumn{4}{c}{\textbf{Cartesian Coordinates (Angstroms)}} \\[3pt]
\toprule
Atom & {x} & {y} & {z} \\
\midrule
C & -0.311810 & 0.322044 & 0.115041 \\
H & -1.978848 & 2.099162 & 0.737307 \\
H & -1.465246 & 1.554719 & 0.547700 \\
C & 0.495213 & -0.527684 & -0.181092 \\
H & 1.204762 & -1.276530 & -0.442570 \\
\bottomrule
\end{tabular}
\end{minipage}
\hfill
\begin{minipage}{0.45\textwidth}
\centering
\begin{tabular}{c}
\multicolumn{1}{c}{\textbf{Vibrational Frequencies}} \\[3pt]
\toprule
$\nu$ (cm$^{-1}$)\\
\midrule
-524.57 \\
  95.00 \\
  95.00 \\
 372.53 \\
 391.53 \\
 427.99 \\
 694.51 \\
2051.81 \\
3450.67 \\
3596.24 \\
\bottomrule
\end{tabular}


\begin{tabular}{l c}
\multicolumn{2}{c}{\textbf{Rotational Constants (MHz)}} \\[3pt]
\toprule
 & (MHz) \\
\midrule
A & 0.000000 \\
B & 19144.156532 \\
C & 19144.029011 \\
\bottomrule
\end{tabular}
\end{minipage}
\begin{table}[h!]
\tablenum{11}
\centering
\caption{Summary of the computed properties for \ce{C2H2} in the CCSD(T)/CBS//CCSD(T)-F12c/cc-pVTZ-F12 set of calculations.}
\label{tab:moleculeX}
\end{table}

\begin{minipage}{0.45\textwidth}
\centering
\begin{tabular}{l c c c}
\multicolumn{4}{c}{\textbf{Cartesian Coordinates (Angstroms)}} \\[3pt]
\toprule
Atom & {x} & {y} & {z} \\
\midrule
C & -0.000000 & 0.000000 & 0.602765 \\
C & 0.000000 & -0.000000 & -0.602765 \\
H & 0.000000 & -0.000000 & 1.665867 \\
H & -0.000000 & 0.000000 & -1.665867 \\
\bottomrule
\end{tabular}
\end{minipage}
\hfill
\begin{minipage}{0.45\textwidth}
\centering
\begin{tabular}{c}
\multicolumn{1}{c}{\textbf{Vibrational Frequencies}} \\[3pt]
\toprule
$\nu$ (cm$^{-1}$)\\
\midrule
 618.49 \\
 618.49 \\
 750.36 \\
 750.36 \\
2007.49 \\
3410.87 \\
3502.95 \\
\bottomrule
\end{tabular}

\vspace{0.5cm}

\begin{tabular}{l c}
\multicolumn{2}{c}{\textbf{Rotational Constants (MHz)}} \\[3pt]
\toprule
 & (MHz) \\
\midrule
A & 0.000000 \\
B & 35285.894320 \\
C & 35285.894320 \\
\bottomrule
\end{tabular}
\end{minipage}
\subsection{\ce{OH} + \ce{H2}}\label{OH+H2_SI2}
\begin{table}[H]
\tablenum{12}
\centering
\caption{Summary of the computed properties for OH in the CCSD(T)/CBS//CCSD(T)-F12b/aug-cc-pVTZ set of calculations.}
\label{tab:moleculeX}
\end{table}

\begin{minipage}{0.45\textwidth}
\centering
\begin{tabular}{l c c c}
\multicolumn{4}{c}{\textbf{Cartesian Coordinates (Angstroms)}} \\[3pt]
\toprule
Atom & {x} & {y} & {z} \\
\midrule
O & 0.000000000 & 0.000000000 & -0.057554843 \\
H & 0.000000000 & 0.000000000 &  0.912815476 \\
\bottomrule
\end{tabular}
\end{minipage}
\hfill
\begin{minipage}{0.45\textwidth}
\centering
\begin{tabular}{c}
\multicolumn{1}{c}{\textbf{Vibrational Frequencies}} \\[3pt]
\toprule
$\nu$ (cm$^{-1}$)\\
\midrule
3739.39 \\
\bottomrule
\end{tabular}

\vspace{0.5cm}

\begin{tabular}{l c}
\multicolumn{2}{c}{\textbf{Rotational Constants (MHz)}} \\[3pt]
\toprule
 & (MHz) \\
\midrule
A & 566005.723 \\
B & 566005.723 \\
C & 0.00000000 \\
\bottomrule
\end{tabular}
\end{minipage}
\pagebreak
\begin{table}[h!]
\tablenum{13}
\centering
\caption{Summary of the computed properties for H2 in the CCSD(T)/CBS//CCSD(T)-F12b/aug-cc-pVTZ set of calculations.}
\label{tab:moleculeX}
\end{table}

\begin{minipage}{0.45\textwidth}
\centering
\begin{tabular}{l c c c}
\multicolumn{4}{c}{\textbf{Cartesian Coordinates (Angstroms)}} \\[3pt]
\toprule
Atom & {x} & {y} & {z} \\
\midrule
H & 0.000000000 & 0.000000000 & -0.370963341 \\
H & 0.000000000 & 0.000000000 &  0.370963341 \\
\bottomrule
\end{tabular}
\end{minipage}
\hfill
\begin{minipage}{0.45\textwidth}
\centering
\begin{tabular}{c}
\multicolumn{1}{c}{\textbf{Vibrational Frequencies}} \\[3pt]
\toprule
$\nu$ (cm$^{-1}$)\\
\midrule
4401.47 \\
\bottomrule
\end{tabular}

\vspace{0.5cm}

\begin{tabular}{l c}
\multicolumn{2}{c}{\textbf{Rotational Constants (MHz)}} \\[3pt]
\toprule
 & (MHz) \\
\midrule
A & 1821749.693 \\
B & 1821749.693 \\
C & 0.000000000 \\
\bottomrule
\end{tabular}
\end{minipage}
\begin{table}[h!]
\tablenum{14}
\centering
\caption{Summary of the computed properties for the \ce{OH + H2} Van der Waals complex in the CCSD(T)/CBS//CCSD(T)-F12b/aug-cc-pVTZ set of calculations.}
\label{tab:moleculeX}
\end{table}

\begin{minipage}{0.45\textwidth}
\centering
\begin{tabular}{l c c c}
\multicolumn{4}{c}{\textbf{Cartesian Coordinates (Angstroms)}} \\[3pt]
\toprule
Atom & {x} & {y} & {z} \\
\midrule
H & -0.371446684 &  0.000000000 & -2.819410983 \\
H &  0.371430438 &  0.000000000 & -2.819452449 \\
O & -0.000001132 &  0.000000000 &  0.391656347 \\
H &  0.000032609 &  0.000000000 & -0.578480071 \\
\bottomrule
\end{tabular}
\end{minipage}
\hfill
\begin{minipage}{0.45\textwidth}
\centering
\begin{tabular}{c}
\multicolumn{1}{c}{\textbf{Vibrational Frequencies}} \\[3pt]
\toprule
$\nu$ (cm$^{-1}$)\\
\midrule
125.00 \\
128.74 \\
156.67 \\
381.72 \\
3740.12 \\
4382.25 \\
\bottomrule
\end{tabular}

\vspace{0.5cm}

\begin{tabular}{l c}
\multicolumn{2}{c}{\textbf{Rotational Constants (MHz)}} \\[3pt]
\toprule
 & (MHz) \\
\midrule
A & 26826.07 \\
B & 26435.38 \\
C & 1815127.32 \\
\bottomrule
\end{tabular}
\end{minipage}
\pagebreak
\begin{table}[h!]
\tablenum{15}
\centering
\caption{Summary of the computed properties for the \ce{OH + H2} transition state in the CCSD(T)/CBS//CCSD(T)-F12b/aug-cc-pVTZ set of calculations.}
\label{tab:moleculeX}
\end{table}

\begin{minipage}{0.45\textwidth}
\centering
\begin{tabular}{l c c c}
\multicolumn{4}{c}{\textbf{Cartesian Coordinates (Angstroms)}} \\[3pt]
\toprule
Atom & {x} & {y} & {z} \\
\midrule
H & -0.000073386 & -0.123236707 &  1.174593401 \\
H &  0.000046079 &  0.106706927 &  1.959081975 \\
O &  0.000002531 & -0.056358974 & -0.181904822 \\
H & -0.000012846 &  0.911313989 & -0.248955248 \\
\bottomrule
\end{tabular}
\end{minipage}
\hfill
\begin{minipage}{0.45\textwidth}
\centering
\begin{tabular}{c}
\multicolumn{1}{c}{\textbf{Vibrational Frequencies}} \\[3pt]
\toprule
$\nu$ (cm$^{-1}$)\\
\midrule
-1224.43 \\
 493.08 \\
 580.51 \\
1042.91 \\
2650.12 \\
3746.60 \\
\bottomrule
\end{tabular}

\vspace{0.5cm}

\begin{tabular}{l c}
\multicolumn{2}{c}{\textbf{Rotational Constants (MHz)}} \\[3pt]
\toprule
 & (MHz) \\
\midrule
A & 74637.23 \\
B & 86293.91 \\
C & 552536.09 \\
\bottomrule
\end{tabular}
\end{minipage}
\begin{table}[h!]
\tablenum{16}
\centering
\caption{Summary of the computed properties for \ce{H2O} in the CCSD(T)/CBS//CCSD(T)-F12b/aug-cc-pVTZ set of calculations.}
\label{tab:moleculeX}
\end{table}

\begin{minipage}{0.45\textwidth}
\centering
\begin{tabular}{l c c c}
\multicolumn{4}{c}{\textbf{Cartesian Coordinates (Angstroms)}} \\[3pt]
\toprule
Atom & {x} & {y} & {z} \\
\midrule
O &  0.000000000 & -0.000000000 & -0.065702173 \\
H &  0.000000000 &  0.757485242 &  0.521277250 \\
H &  0.000000000 & -0.757485242 &  0.521277250 \\
\bottomrule
\end{tabular}
\end{minipage}
\hfill
\begin{minipage}{0.45\textwidth}
\centering
\begin{tabular}{c}
\multicolumn{1}{c}{\textbf{Vibrational Frequencies}} \\[3pt]
\toprule
$\nu$ (cm$^{-1}$)\\
\midrule
1646.72 \\
3831.97 \\    
3941.45 \\
\bottomrule
\end{tabular}

\vspace{0.5cm}

\begin{tabular}{l c}
\multicolumn{2}{c}{\textbf{Rotational Constants (MHz)}} \\[3pt]
\toprule
 & (MHz) \\
\midrule
A & 436667.85 \\
B & 284708.17 \\
C & 818130.84 \\
\bottomrule
\end{tabular}
\end{minipage}
\pagebreak
\subsection{\ce{CN} + \ce{H2}}\label{CN+H2_SI2}

\begin{table}[h!]
\tablenum{17}
\centering
\caption{Summary of the computed properties for \ce{CN} in the CCSD(T)/CBS//CCSD(T)-F12b/aug-cc-pVTZ set of calculations.}
\label{tab:moleculeX}
\end{table}

\begin{minipage}{0.45\textwidth}
\centering
\begin{tabular}{l c c c}
\multicolumn{4}{c}{\textbf{Cartesian Coordinates (Angstroms)}} \\[3pt]
\toprule
Atom & {x} & {y} & {z} \\
\midrule
C & 0.000000000 & 0.000000000 & -0.632117912 \\
N & 0.000000000 & 0.000000000 &  0.541842957 \\
\bottomrule
\end{tabular}
\end{minipage}
\hfill
\begin{minipage}{0.45\textwidth}
\centering
\begin{tabular}{c}
\multicolumn{1}{c}{\textbf{Vibrational Frequencies}} \\[3pt]
\toprule
$\nu$ (cm$^{-1}$)\\
\midrule
2068.14 \\
\bottomrule
\end{tabular}

\vspace{0.5cm}

\begin{tabular}{l c}
\multicolumn{2}{c}{\textbf{Rotational Constants (MHz)}} \\[3pt]
\toprule
 & (MHz) \\
\midrule
A & 56751.265 \\
B & 56751.265 \\
C & 0.0000000 \\
\bottomrule
\end{tabular}
\end{minipage}
\begin{table}[h!]
\tablenum{18}
\centering
\caption{Summary of the computed properties for \ce{H2} in the CCSD(T)/CBS//CCSD(T)-F12b/aug-cc-pVTZ set of calculations.}
\label{tab:moleculeX}
\end{table}

\begin{minipage}{0.45\textwidth}
\centering
\begin{tabular}{l c c c}
\multicolumn{4}{c}{\textbf{Cartesian Coordinates (Angstroms)}} \\[3pt]
\toprule
Atom & {x} & {y} & {z} \\
\midrule
H & 0.000000000 & 0.000000000 & -0.370963341 \\
H & 0.000000000 & 0.000000000 &  0.370963341 \\
\bottomrule
\end{tabular}
\end{minipage}
\hfill
\begin{minipage}{0.45\textwidth}
\centering
\begin{tabular}{c}
\multicolumn{1}{c}{\textbf{Vibrational Frequencies}} \\[3pt]
\toprule
$\nu$ (cm$^{-1}$)\\
\midrule
4401.47 \\
\bottomrule
\end{tabular}

\vspace{0.5cm}

\begin{tabular}{l c}
\multicolumn{2}{c}{\textbf{Rotational Constants (MHz)}} \\[3pt]
\toprule
 & (MHz) \\
\midrule
A & 1821749.689 \\
B & 1821749.689 \\
C & 0.000000000 \\
\bottomrule
\end{tabular}
\end{minipage}
\pagebreak
\begin{table}[h!]
\tablenum{19}
\centering
\caption{Summary of the computed properties for the \ce{CN + H2} Van der Waals complex in the CCSD(T)/CBS//CCSD(T)-F12b/aug-cc-pVTZ set of calculations.}
\label{tab:moleculeX}
\end{table}

\begin{minipage}{0.45\textwidth}
\centering
\begin{tabular}{l c c c}
\multicolumn{4}{c}{\textbf{Cartesian Coordinates (Angstroms)}} \\[3pt]
\toprule
Atom & {x} & {y} & {z} \\
\midrule
H & 0.241950332 &  0.000157910 & -3.535338834 \\
H & 0.889699223 & -0.000136285 & -3.169732310 \\
N & 0.242690409 &  0.000008670 &  0.719891070 \\
C & -0.378128050 & -0.000011937 & -0.276906838 \\
\bottomrule
\end{tabular}
\end{minipage}
\hfill
\begin{minipage}{0.45\textwidth}
\centering
\begin{tabular}{c}
\multicolumn{1}{c}{\textbf{Vibrational Frequencies}} \\[3pt]
\toprule
$\nu$ (cm$^{-1}$)\\
\midrule
  17.75 \\
  67.54 \\
  81.01 \\
 150.37 \\
2067.91 \\
4393.88 \\
\bottomrule
\end{tabular}

\vspace{0.5cm}

\begin{tabular}{l c}
\multicolumn{2}{c}{\textbf{Rotational Constants (MHz)}} \\[3pt]
\toprule
 & (MHz) \\
\midrule
A & 16365.27 \\
B & 14743.71 \\
C & 148798.41 \\
\bottomrule
\end{tabular}
\end{minipage}
\begin{table}[h!]
\tablenum{20}
\centering
\caption{Summary of the computed properties for the \ce{CN + H2} transition state (for the HCN forming route) in the CCSD(T)/CBS//CCSD(T)-F12b/aug-cc-pVTZ set of calculations.}
\label{tab:moleculeX}
\end{table}

\begin{minipage}{0.45\textwidth}
\centering
\begin{tabular}{l c c c}
\multicolumn{4}{c}{\textbf{Cartesian Coordinates (Angstroms)}} \\[3pt]
\toprule
Atom & {x} & {y} & {z} \\
\midrule
H & 0.000000000 & -0.000000563 & -2.076472972 \\
H & 0.000000000 & -0.000001897 & -2.867017005 \\
C & 0.000000000 &  0.000000563 & -0.438569154 \\
N & 0.000000000 & -0.000000563 &  0.731445168 \\
\bottomrule
\end{tabular}
\end{minipage}
\hfill
\begin{minipage}{0.45\textwidth}
\centering
\begin{tabular}{c}
\multicolumn{1}{c}{\textbf{Vibrational Frequencies}} \\[3pt]
\toprule
$\nu$ (cm$^{-1}$)\\
\midrule
-692.94 \\
104.91 \\
104.95 \\
522.17 \\     
522.22 \\    
2103.62 \\
\bottomrule
\end{tabular}

\vspace{0.5cm}

\begin{tabular}{l c}
\multicolumn{2}{c}{\textbf{Rotational Constants (MHz)}} \\[3pt]
\toprule
 & (MHz) \\
\midrule
A & 22542.921 \\
B & 22542.921 \\
C & 0.0000000 \\
\bottomrule
\end{tabular}
\end{minipage}
\pagebreak
\begin{table}[h!]
\tablenum{21}
\centering
\caption{Summary of the computed properties for the \ce{CN + H2} transition state (for the HNC forming route) in the CCSD(T)/CBS//CCSD(T)-F12b/aug-cc-pVTZ set of calculations.}
\label{tab:moleculeX}
\end{table}

\begin{minipage}{0.45\textwidth}
\centering
\begin{tabular}{l c c c}
\multicolumn{4}{c}{\textbf{Cartesian Coordinates (Angstroms)}} \\[3pt]
\toprule
Atom & {x} & {y} & {z} \\
\midrule
H & 0.000000000 & -0.000948149 & -1.744497869 \\
H & 0.000000000 & -0.002387627 & -2.598122632 \\
N & 0.000000000 &  0.000640187 & -0.378478934 \\
C & 0.000000000 & -0.000466944 &  0.805708358 \\
\bottomrule
\end{tabular}
\end{minipage}
\hfill
\begin{minipage}{0.45\textwidth}
\centering
\begin{tabular}{c}
\multicolumn{1}{c}{\textbf{Vibrational Frequencies}} \\[3pt]
\toprule
$\nu$ (cm$^{-1}$)\\
\midrule
-2810.27 \\
  21.47 \\
 782.21 \\
 782.31 \\
1833.87 \\
2113.07 \\
\bottomrule
\end{tabular}

\vspace{0.5cm}

\begin{tabular}{l c}
\multicolumn{2}{c}{\textbf{Rotational Constants (MHz)}} \\[3pt]
\toprule
 & (MHz) \\
\midrule
A & 25665.5316 \\
B & 25665.5317 \\
C & 0.00000000 \\
\bottomrule
\end{tabular}
\end{minipage}
\begin{table}[h!]
\tablenum{22}
\centering
\caption{Summary of the computed properties for HCN in the CCSD(T)/CBS//CCSD(T)-F12b/aug-cc-pVTZ set of calculations.}
\label{tab:moleculeX}
\end{table}

\begin{minipage}{0.45\textwidth}
\centering
\begin{tabular}{l c c c}
\multicolumn{4}{c}{\textbf{Cartesian Coordinates (Angstroms)}} \\[3pt]
\toprule
Atom & {x} & {y} & {z} \\
\midrule
C & 0.000000000 & 0.000000000 &  0.559347968 \\
H & 0.000000000 & 0.000000000 &  1.627973293 \\
N & 0.000000000 & 0.000000000 & -0.596405882 \\
\bottomrule
\end{tabular}
\end{minipage}
\hfill
\begin{minipage}{0.45\textwidth}
\centering
\begin{tabular}{c}
\multicolumn{1}{c}{\textbf{Vibrational Frequencies}} \\[3pt]
\toprule
$\nu$ (cm$^{-1}$)\\
\midrule
 727.68 \\
 727.68 \\
2123.41 \\
3435.30 \\
\bottomrule
\end{tabular}

\vspace{0.5cm}

\begin{tabular}{l c}
\multicolumn{2}{c}{\textbf{Rotational Constants (MHz)}} \\[3pt]
\toprule
 & (MHz) \\
\midrule
A & 44310.641 \\
B & 44310.641 \\
C & 0.0000000 \\
\bottomrule
\end{tabular}
\end{minipage}
\pagebreak
\begin{table}[h!]
\tablenum{23}
\centering
\caption{Summary of the computed properties for HNC in the CCSD(T)/CBS//CCSD(T)-F12b/aug-cc-pVTZ set of calculations.}
\label{tab:moleculeX}
\end{table}

\begin{minipage}{0.45\textwidth}
\centering
\begin{tabular}{l c c c}
\multicolumn{4}{c}{\textbf{Cartesian Coordinates (Angstroms)}} \\[3pt]
\toprule
Atom & {x} & {y} & {z} \\
\midrule
N & 0.000000000 & 0.000000000 & -0.586074419 \\
C & 0.000000000 & 0.000000000 &  0.553143085 \\
H & 0.000000000 & 0.000000000 &  1.552241363 \\
\bottomrule
\end{tabular}
\end{minipage}
\hfill
\begin{minipage}{0.45\textwidth}
\centering
\begin{tabular}{c}
\multicolumn{1}{c}{\textbf{Vibrational Frequencies}} \\[3pt]
\toprule
$\nu$ (cm$^{-1}$)\\
\midrule
463.22 \\
463.22 \\
2049.80 \\
3810.57 \\
\bottomrule
\end{tabular}

\vspace{0.5cm}

\begin{tabular}{l c}
\multicolumn{2}{c}{\textbf{Rotational Constants (MHz)}} \\[3pt]
\toprule
 & (MHz) \\
\midrule
A & 45231.709 \\
B & 45231.709 \\
C & 0.0000000 \\
\bottomrule
\end{tabular}
\end{minipage}

\subsection{\ce{NH2} + \ce{H2}}\label{NH2+H2_SI2}

\begin{table}[h!]
\tablenum{24}
\centering
\caption{Summary of the computed properties for NH2 in the CCSD(T)/CBS//CCSD(T)-F12c/cc-pVTZ-F12 set of calculations.}
\label{tab:moleculeX}
\end{table}

\begin{minipage}{0.45\textwidth}
\centering
\begin{tabular}{l c c c}
\multicolumn{4}{c}{\textbf{Cartesian Coordinates (Angstroms)}} \\[3pt]
\toprule
Atom & {x} & {y} & {z} \\
\midrule
N &  0.000000 & -0.000000 &  0.142169 \\
H &  0.000000 &  0.802129 & -0.496084 \\
H &  0.000000 & -0.802129 & -0.496084 \\
\bottomrule
\end{tabular}
\end{minipage}
\hfill
\begin{minipage}{0.45\textwidth}
\centering
\begin{tabular}{c}
\multicolumn{1}{c}{\textbf{Vibrational Frequencies}} \\[3pt]
\toprule
$\nu$ (cm$^{-1}$)\\
\midrule
1544.75 \\
3375.76 \\
3471.31 \\
\bottomrule
\end{tabular}

\vspace{0.5cm}

\begin{tabular}{l c}
\multicolumn{2}{c}{\textbf{Rotational Constants (MHz)}} \\[3pt]
\toprule
 & (MHz) \\
\midrule
A & 703947.784792 \\
B & 389617.969178 \\
C & 250804.037455 \\
\bottomrule
\end{tabular}
\end{minipage}
\pagebreak
\begin{table}[h!]
\tablenum{25}
\centering
\caption{Summary of the computed properties for H2 in the CCSD(T)/CBS//CCSD(T)-F12c/cc-pVTZ-F12 set of calculations.}
\label{tab:moleculeX}
\end{table}

\begin{minipage}{0.45\textwidth}
\centering
\begin{tabular}{l c c c}
\multicolumn{4}{c}{\textbf{Cartesian Coordinates (Angstroms)}} \\[3pt]
\toprule
Atom & {x} & {y} & {z} \\
\midrule
H & 0.000000 & 0.000000 &  0.370834 \\
H & 0.000000 & 0.000000 & -0.370834 \\
\bottomrule
\end{tabular}
\end{minipage}
\hfill
\begin{minipage}{0.45\textwidth}
\centering
\begin{tabular}{c}
\multicolumn{1}{c}{\textbf{Vibrational Frequencies}} \\[3pt]
\toprule
$\nu$ (cm$^{-1}$)\\
\midrule
4401.76 \\
\bottomrule
\end{tabular}

\vspace{0.5cm}

\begin{tabular}{l c}
\multicolumn{2}{c}{\textbf{Rotational Constants (MHz)}} \\[3pt]
\toprule
 & (MHz) \\
\midrule
A & 1822918.833783 \\
B & 1822918.833783 \\
C & 0.000000 \\
\bottomrule
\end{tabular}
\end{minipage}
\begin{table}[h!]
\tablenum{26}
\centering
\caption{Summary of the computed properties for the \ce{NH2 + H2} Van der Waals complex in the CCSD(T)/CBS//CCSD(T)-F12c/cc-pVTZ-F12 set of calculations.}
\label{tab:moleculeX}
\end{table}

\begin{minipage}{0.45\textwidth}
\centering
\begin{tabular}{l c c c}
\multicolumn{4}{c}{\textbf{Cartesian Coordinates (Angstroms)}} \\[3pt]
\toprule
Atom & {x} & {y} & {z} \\
\midrule
N & -0.438314 & -0.008712 &  0.076926 \\
H & -1.080074 &  0.782974 &  0.186598 \\
H & -1.055044 & -0.821866 &  0.172019 \\
H &  3.134859 &  0.010341 &  0.035211 \\
H &  2.391764 &  0.009450 &  0.017877 \\
\bottomrule
\end{tabular}
\end{minipage}
\hfill
\begin{minipage}{0.45\textwidth}
\centering
\begin{tabular}{c}
\multicolumn{1}{c}{\textbf{Vibrational Frequencies}} \\[3pt]
\toprule
$\nu$ (cm$^{-1}$)\\
\midrule
  36.18 \\
  59.53 \\
 128.94 \\
 283.20 \\
 287.23 \\
1544.83 \\
3376.95 \\
3472.63 \\
4376.20 \\
\bottomrule
\end{tabular}

\vspace{0.5cm}

\begin{tabular}{l c}
\multicolumn{2}{c}{\textbf{Rotational Constants (MHz)}} \\[3pt]
\toprule
 & (MHz) \\
\midrule
A & 384929.947448 \\
B & 24923.472601 \\
C & 23439.309159 \\
\bottomrule
\end{tabular}
\end{minipage}
\pagebreak
\begin{table}[h!]
\tablenum{27}
\centering
\caption{Summary of the computed properties for the \ce{NH2 + H2} transition state in the CCSD(T)/CBS//CCSD(T)-F12c/cc-pVTZ-F12 set of calculations.}
\label{tab:moleculeX}
\end{table}

\begin{minipage}{0.45\textwidth}
\centering
\begin{tabular}{l c c c}
\multicolumn{4}{c}{\textbf{Cartesian Coordinates (Angstroms)}} \\[3pt]
\toprule
Atom & {x} & {y} & {z} \\
\midrule
N &  0.002381 &  0.326274 &  0.000000 \\
H &  0.619048 &  0.187659 &  0.805277 \\
H &  0.619047 &  0.187661 & -0.805277 \\
H & -0.575452 & -0.860675 & -0.000003 \\
H & -0.665024 & -1.739322 &  0.000003 \\
\bottomrule
\end{tabular}
\end{minipage}
\hfill
\begin{minipage}{0.45\textwidth}
\centering
\begin{tabular}{c}
\multicolumn{1}{c}{\textbf{Vibrational Frequencies}} \\[3pt]
\toprule
$\nu$ (cm$^{-1}$)\\
\midrule 
-1557.46 \\
  647.01 \\
  674.59 \\
 1087.07 \\
 1210.08 \\
 1551.16 \\
 1951.52 \\
 3394.22 \\
 3491.31 \\
\bottomrule
\end{tabular}

\vspace{0.5cm}

\begin{tabular}{l c}
\multicolumn{2}{c}{\textbf{Rotational Constants (MHz)}} \\[3pt]
\toprule
 & (MHz) \\
\midrule
A & 249854.650171 \\
B & 76455.881889 \\
C & 70171.881623 \\
\bottomrule
\end{tabular}
\end{minipage}
\begin{table}[h!]
\tablenum{28}
\centering
\caption{Summary of the computed properties for \ce{NH3} in the CCSD(T)/CBS//CCSD(T)-F12c/cc-pVTZ-F12 set of calculations.}
\label{tab:moleculeX}
\end{table}

\begin{minipage}{0.45\textwidth}
\centering
\begin{tabular}{l c c c}
\multicolumn{4}{c}{\textbf{Cartesian Coordinates (Angstroms)}} \\[3pt]
\toprule
Atom & {x} & {y} & {z} \\
\midrule
N & -0.000000 & -0.000004 &  0.115437 \\
H &  0.000000 &  0.937193 & -0.267592 \\
H & -0.811638 & -0.468595 & -0.267590 \\
H &  0.811638 & -0.468595 & -0.267590 \\
\bottomrule
\end{tabular}
\end{minipage}
\hfill
\begin{minipage}{0.45\textwidth}
\centering
\begin{tabular}{c}
\multicolumn{1}{c}{\textbf{Vibrational Frequencies}} \\[3pt]
\toprule
$\nu$ (cm$^{-1}$)\\
\midrule 
1056.71 \\
1674.97 \\
1675.21 \\
3477.52 \\
3609.19 \\
3609.21 \\
\bottomrule
\end{tabular}

\vspace{0.5cm}

\begin{tabular}{l c}
\multicolumn{2}{c}{\textbf{Rotational Constants (MHz)}} \\[3pt]
\toprule
 & (MHz) \\
\midrule
A & 298526.750086 \\
B & 298523.267032 \\
C & 190272.222464 \\
\bottomrule
\end{tabular}
\end{minipage}
\pagebreak
\subsection{Modeling Impacts of MESMER Rate Coefficients}\label{H2+Radical-modelling_SI}
\begin{figure}
    \centering  \includegraphics[width=0.7\linewidth]{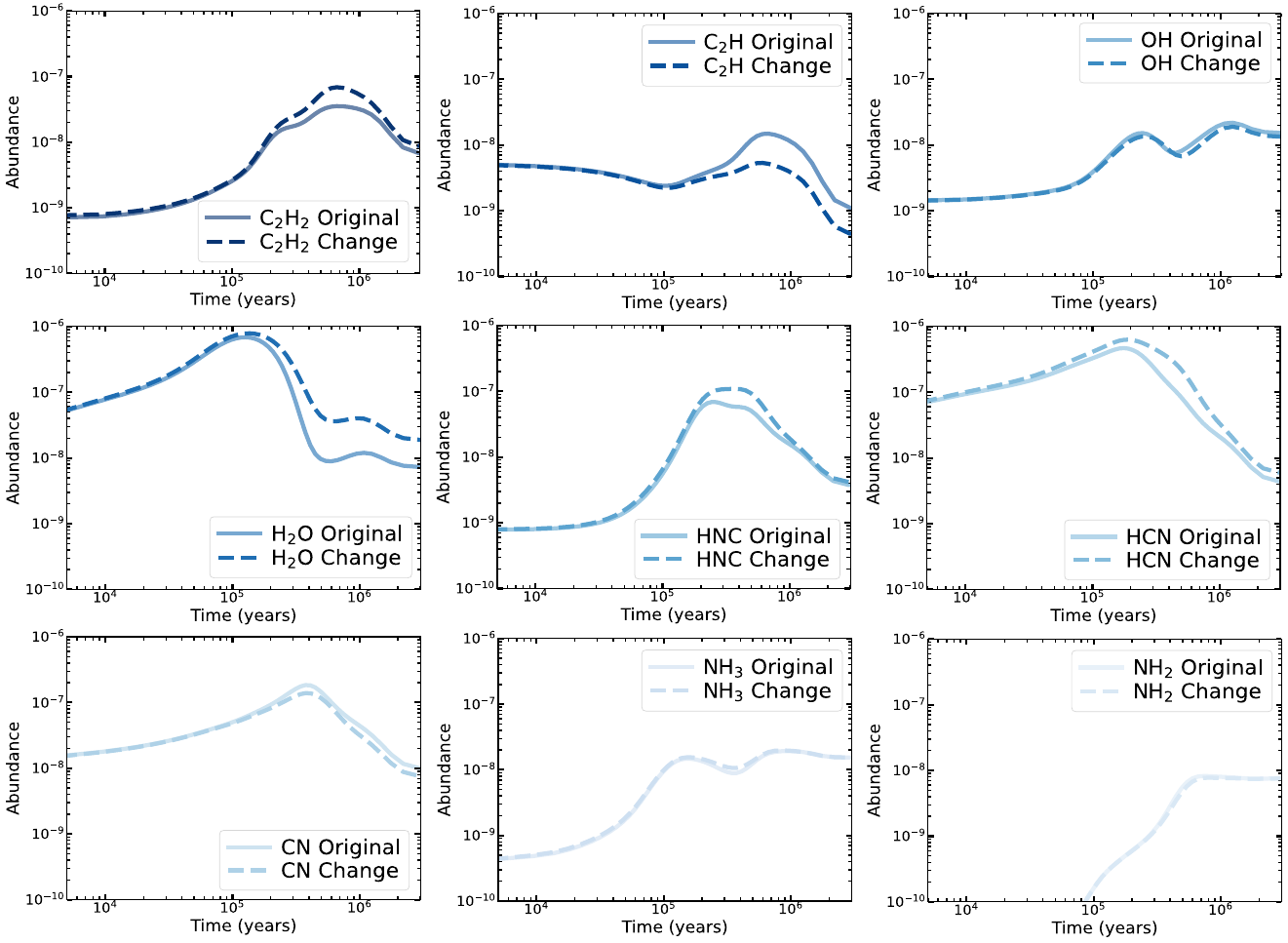}
    \caption{The effect on the abundances of species involved in Reactions \ref{eq:1}--\ref{eq:4}, when the new 10 K, 2 $\times$ 10$^{4}$ cm$^{-3}$ rate coefficients for these four reactions are used in Network 1. Here, Original refers to Network 1, and Change is when the newly calculated MESMER rate coefficients for these four reactions are used instead.}
    \label{Grouped-Plot_RNs}
\end{figure}

\begin{figure}
    \centering  \includegraphics[width=0.7\linewidth]{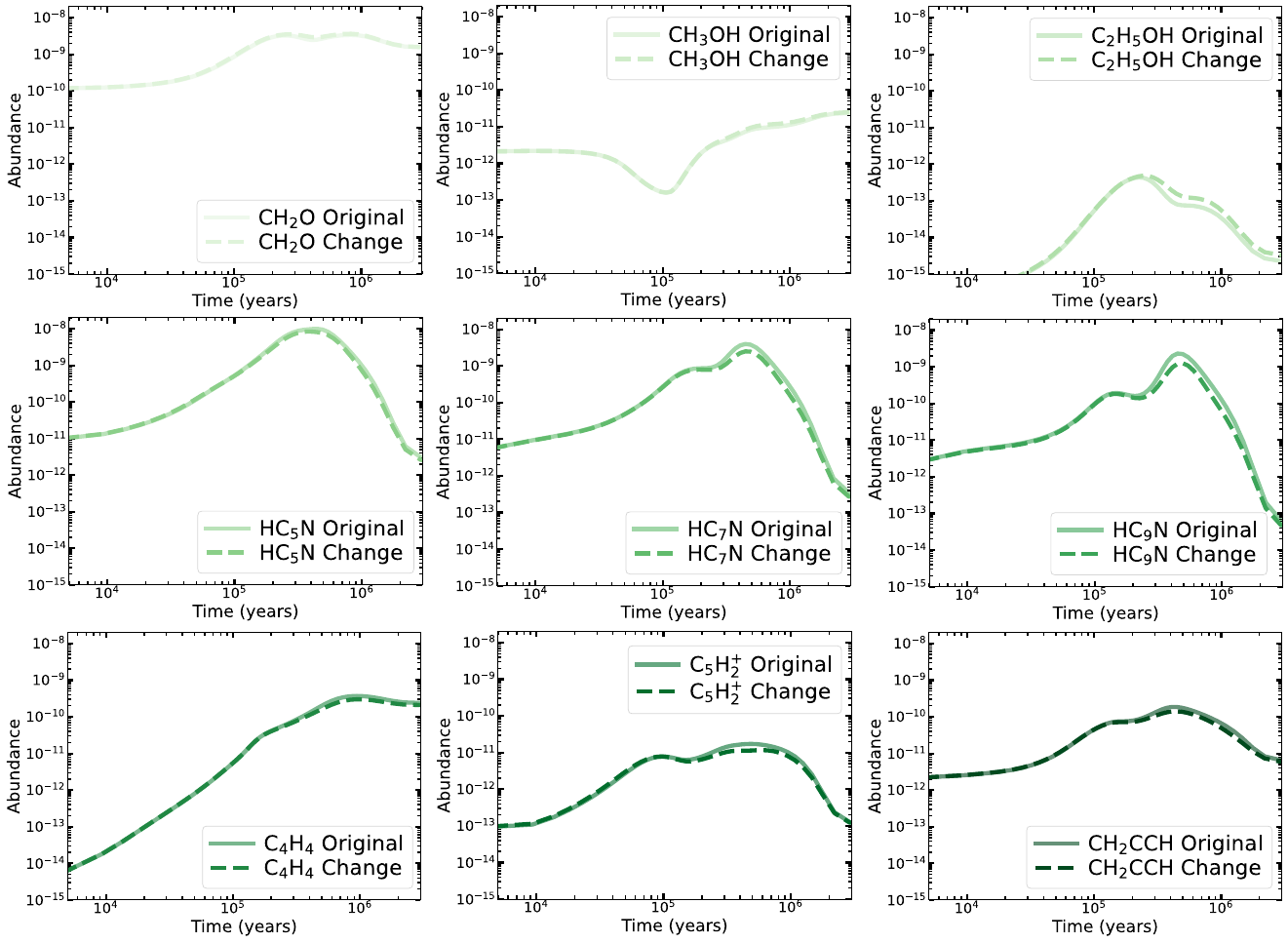}
    \caption{The abundance changes in \ce{CH2O}, \ce{CH3OH}, \ce{C2H5OH}, \ce{HC5N}, \ce{HC7N}, \ce{HC9N}, \ce{C4H4}, \ce{C5H2+}, and \ce{CH2CCH} (from top left to bottom right)—where Original refers to the use of Network 1, and Change is when Network 1 and the newly calculated MESMER rate coefficients for Reactions \ref{eq:1}--\ref{eq:4} are used instead of the ones listed in kida.uva.2024.}
    \label{Grouped-Plot-Randos}
\end{figure}

Modifying Network 1 to include the five, 10 K, MESMER calculated rate coefficients for Reactions \ref{eq:1}--\ref{eq:4}, instead of the 10 K values from kida.uva.2024 (along with adding the HNC reaction pathway for Reaction \ref{eq:3}) produces the most notable changes to the modeled molecular abundances of \ce{H2O}, \ce{HCN}, \ce{C2H}, \ce{HNC}, and \ce{C2H2}. Lesser impacts to the OH, CN, \ce{NH2} and \ce{NH3} modeled abundances were also found. Figure \ref{Grouped-Plot_RNs} shows the modeled abundance changes to all nine of these species when using this modified version of Network 1. During the period of 1--5 $\times$ 10$^{5}$ years, the abundances of \ce{H2O}, \ce{HCN}, \ce{HNC}, and \ce{C2H2} maximally increased by $\sim$5$\times$, $\sim$3$\times$, and $\sim$2$\times$, and \ce{C2H} maximally decreased by $\sim$2$\times$. Meanwhile the maximal abundance changes to OH, CN, and \ce{NH2}, and \ce{NH3} were all less than 1.7$\times$ different. In addition to these nine species, Figure \ref{Grouped-Plot-Randos} shows the impacts to the modeled abundances of formaldehyde (\ce{CH2O}), methanol (\ce{CH3OH}), ethanol (\ce{C2H5OH}), cyanodiacetylene (\ce{HC5N}), cyanohexatriyne (\ce{HC7N}), cyanooctatetrayne (\ce{HC9N}), \ce{C4H4}, \ce{C5H2+}, and \ce{CH2CCH}. The maximal increase to any of the modeled abundances of these molecules is a $\sim$2$\times$ increase to the abundance of \ce{HC9N}. Although the changes to the abundances of the species in Figure \ref{Grouped-Plot-Randos} are moderate to minor, they are evidence that the changing of the five rate coefficients for Reactions \ref{eq:1}--\ref{eq:4} impact more than just the species directly involved in Reactions \ref{eq:1}--\ref{eq:4}.

\end{document}